\documentclass[longauth]{aa} 

\usepackage{natbib}
\usepackage{graphicx}
\usepackage{color}
\usepackage{txfonts}
\usepackage{hyperref}
\usepackage{tabularx}
\usepackage{amsmath}
\usepackage{amssymb}
\usepackage{mathtools}
\usepackage{txfonts}
\usepackage{subcaption}
\usepackage[flushleft]{threeparttable} 

\providecommand{\gmag}{\ensuremath{G}}
\providecommand{\gbp}{\ensuremath{G_{\rm BP}}}
\providecommand{\grp}{\ensuremath{G_{\rm RP}}}

\newcommand\gaia{\textit{Gaia~}}
\newcommand\gdrone{\gaia~DR1~}
\newcommand\gdrtwo{\gaia~DR2~}
\newcommand\gdrthree{\gaia~DR3~}
\newcommand\egdr[1]{\gaia~EDR#1}

\newcommand\secref[1]{Sect.~\ref{#1}}
\newcommand\secrefalt[1]{Section~\ref{#1}}
\newcommand\figref[1]{Fig.~\ref{#1}}

\newcommand\figrefalt[1]{Figure~\ref{#1}}

\newcommand\equref[1]{Eq.~\eqref{#1}}

\newcommand\tabref[1]{Table~\ref{#1}}

\newcommand{\kms}{km~s$^{-1}$}
\newcommand{\sr}{$\sigma_R$}
\newcommand{\st}{$\sigma_\phi$}
\newcommand{\vr}{$v_R$}
\newcommand{\vt}{$v_\phi$}

\newcommand{\orcit}[1]{\protect\href{https://orcid.org/#1}{\protect\includegraphics[width=8pt]{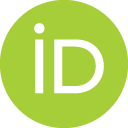}}}

\makeatletter
\renewcommand*\maketitle{%
  \thispagestyle{firstpage}
\begingroup
    \if@wideboxfn
    \setlength\bibindent{1.4\parindent}
    \else
    \setlength\bibindent{\parindent}
    \fi
    \renewcommand*\thefootnote{\@fnsymbol\c@footnote}%
    \renewcommand\@makefntext[1]{%
    \ifaa@longfn\hsize\textwidth\fi
    \noindent
    \hb@xt@\bibindent{\hss\@makefnmark\enspace}##1}
  \ifaa@twocolumn
  \begingroup
    \begin{aa@strip}
          \aa@maketitle
    \end{aa@strip}
    \@thanks            
  \endgroup
  \else
    \begingroup
      \let\thanks\footnote
      \aa@maketitle
    \endgroup
  \fi
\endgroup
  \setcounter{footnote}{0}%
}
\makeatother
\begin{document}

%----------------------------------------------------------------
% Title, authors and abstract
%----------------------------------------------------------------
\title{\gaia Early Data Release 3: Structure and properties of the Magellanic Clouds}
\subtitle{}

\author{
{\it Gaia} Collaboration
\and X.        ~Luri                          \orcit{0000-0001-5428-9397}\inst{\ref{inst:0001}}
\and L.        ~Chemin                        \orcit{0000-0002-3834-7937}\inst{\ref{inst:0002}}
\and G.        ~Clementini                    \orcit{0000-0001-9206-9723}\inst{\ref{inst:0003}}
\and H.E.      ~Delgado                       \orcit{0000-0003-1409-4282}\inst{\ref{inst:0004}}
\and P.J.      ~McMillan                      \orcit{0000-0002-8861-2620}\inst{\ref{inst:0005}}
\and M.        ~Romero-G\'{o}mez              \orcit{0000-0003-3936-1025}\inst{\ref{inst:0001}}
\and E.        ~Balbinot                      \orcit{0000-0002-1322-3153}\inst{\ref{inst:0007}}
\and A.        ~Castro-Ginard                 \orcit{0000-0002-9419-3725}\inst{\ref{inst:0001}}
\and R.        ~Mor                           \orcit{0000-0002-8179-6527}\inst{\ref{inst:0001}}
\and V.        ~Ripepi                        \orcit{0000-0003-1801-426X}\inst{\ref{inst:0010}}
\and L.M.      ~Sarro                         \orcit{0000-0002-5622-5191}\inst{\ref{inst:0004}}
\and M.-R. L.  ~Cioni                         \orcit{0000-0002-6797-696x}\inst{\ref{inst:0012}}
\and C.        ~Fabricius                     \orcit{0000-0003-2639-1372}\inst{\ref{inst:0001}}
\and A.        ~Garofalo                      \orcit{0000-0002-5907-0375}\inst{\ref{inst:0003}}
\and A.        ~Helmi                         \orcit{0000-0003-3937-7641}\inst{\ref{inst:0007}}
\and T.        ~Muraveva                      \orcit{0000-0002-0969-1915}\inst{\ref{inst:0003}}
\and A.G.A.    ~Brown                         \orcit{0000-0002-7419-9679}\inst{\ref{inst:0017}}
\and A.        ~Vallenari                     \orcit{0000-0003-0014-519X}\inst{\ref{inst:0018}}
\and T.        ~Prusti                        \orcit{0000-0003-3120-7867}\inst{\ref{inst:0019}}
\and J.H.J.    ~de Bruijne                    \orcit{0000-0001-6459-8599}\inst{\ref{inst:0019}}
\and C.        ~Babusiaux                     \orcit{0000-0002-7631-348X}\inst{\ref{inst:0021},\ref{inst:0022}}
\and M.        ~Biermann                      \inst{\ref{inst:0023}}
\and O.L.      ~Creevey                       \orcit{0000-0003-1853-6631}\inst{\ref{inst:0024}}
\and D.W.      ~Evans                         \orcit{0000-0002-6685-5998}\inst{\ref{inst:0025}}
\and L.        ~Eyer                          \orcit{0000-0002-0182-8040}\inst{\ref{inst:0026}}
\and A.        ~Hutton                        \inst{\ref{inst:0027}}
\and F.        ~Jansen                        \inst{\ref{inst:0019}}
\and C.        ~Jordi                         \orcit{0000-0001-5495-9602}\inst{\ref{inst:0001}}
\and S.A.      ~Klioner                       \orcit{0000-0003-4682-7831}\inst{\ref{inst:0030}}
\and U.        ~Lammers                       \orcit{0000-0001-8309-3801}\inst{\ref{inst:0031}}
\and L.        ~Lindegren                     \orcit{0000-0002-5443-3026}\inst{\ref{inst:0005}}
\and F.        ~Mignard                       \inst{\ref{inst:0024}}
\and C.        ~Panem                         \inst{\ref{inst:0034}}
\and D.        ~Pourbaix                      \orcit{0000-0002-3020-1837}\inst{\ref{inst:0035},\ref{inst:0036}}
\and S.        ~Randich                       \orcit{0000-0003-2438-0899}\inst{\ref{inst:0037}}
\and P.        ~Sartoretti                    \inst{\ref{inst:0022}}
\and C.        ~Soubiran                      \orcit{0000-0003-3304-8134}\inst{\ref{inst:0039}}
\and N.A.      ~Walton                        \orcit{0000-0003-3983-8778}\inst{\ref{inst:0025}}
\and F.        ~Arenou                        \orcit{0000-0003-2837-3899}\inst{\ref{inst:0022}}
\and C.A.L.    ~Bailer-Jones                  \inst{\ref{inst:0042}}
\and U.        ~Bastian                       \orcit{0000-0002-8667-1715}\inst{\ref{inst:0023}}
\and M.        ~Cropper                       \orcit{0000-0003-4571-9468}\inst{\ref{inst:0044}}
\and R.        ~Drimmel                       \orcit{0000-0002-1777-5502}\inst{\ref{inst:0045}}
\and D.        ~Katz                          \orcit{0000-0001-7986-3164}\inst{\ref{inst:0022}}
\and M.G.      ~Lattanzi                      \orcit{0000-0003-0429-7748}\inst{\ref{inst:0045},\ref{inst:0048}}
\and F.        ~van Leeuwen                   \inst{\ref{inst:0025}}
\and J.        ~Bakker                        \inst{\ref{inst:0031}}
\and J.        ~Casta\~{n}eda                 \orcit{0000-0001-7820-946X}\inst{\ref{inst:0051}}
\and F.        ~De Angeli                     \inst{\ref{inst:0025}}
\and C.        ~Ducourant                     \orcit{0000-0003-4843-8979}\inst{\ref{inst:0039}}
\and M.        ~Fouesneau                     \orcit{0000-0001-9256-5516}\inst{\ref{inst:0042}}
\and Y.        ~Fr\'{e}mat                    \orcit{0000-0002-4645-6017}\inst{\ref{inst:0055}}
\and R.        ~Guerra                        \orcit{0000-0002-9850-8982}\inst{\ref{inst:0031}}
\and A.        ~Guerrier                      \inst{\ref{inst:0034}}
\and J.        ~Guiraud                       \inst{\ref{inst:0034}}
\and A.        ~Jean-Antoine Piccolo          \inst{\ref{inst:0034}}
\and E.        ~Masana                        \orcit{0000-0002-4819-329X}\inst{\ref{inst:0001}}
\and R.        ~Messineo                      \inst{\ref{inst:0061}}
\and N.        ~Mowlavi                       \inst{\ref{inst:0026}}
\and C.        ~Nicolas                       \inst{\ref{inst:0034}}
\and K.        ~Nienartowicz                  \orcit{0000-0001-5415-0547}\inst{\ref{inst:0064},\ref{inst:0065}}
\and F.        ~Pailler                       \inst{\ref{inst:0034}}
\and P.        ~Panuzzo                       \orcit{0000-0002-0016-8271}\inst{\ref{inst:0022}}
\and F.        ~Riclet                        \inst{\ref{inst:0034}}
\and W.        ~Roux                          \inst{\ref{inst:0034}}
\and G.M.      ~Seabroke                      \inst{\ref{inst:0044}}
\and R.        ~Sordo                         \orcit{0000-0003-4979-0659}\inst{\ref{inst:0018}}
\and P.        ~Tanga                         \orcit{0000-0002-2718-997X}\inst{\ref{inst:0024}}
\and F.        ~Th\'{e}venin                  \inst{\ref{inst:0024}}
\and G.        ~Gracia-Abril                  \inst{\ref{inst:0074},\ref{inst:0023}}
\and J.        ~Portell                       \orcit{0000-0002-8886-8925}\inst{\ref{inst:0001}}
\and D.        ~Teyssier                      \orcit{0000-0002-6261-5292}\inst{\ref{inst:0077}}
\and M.        ~Altmann                       \orcit{0000-0002-0530-0913}\inst{\ref{inst:0023},\ref{inst:0079}}
\and R.        ~Andrae                        \inst{\ref{inst:0042}}
\and I.        ~Bellas-Velidis                \inst{\ref{inst:0081}}
\and K.        ~Benson                        \inst{\ref{inst:0044}}
\and J.        ~Berthier                      \orcit{0000-0003-1846-6485}\inst{\ref{inst:0083}}
\and R.        ~Blomme                        \orcit{0000-0002-2526-346X}\inst{\ref{inst:0055}}
\and E.        ~Brugaletta                    \orcit{0000-0003-2598-6737}\inst{\ref{inst:0085}}
\and P.W.      ~Burgess                       \inst{\ref{inst:0025}}
\and G.        ~Busso                         \orcit{0000-0003-0937-9849}\inst{\ref{inst:0025}}
\and B.        ~Carry                         \orcit{0000-0001-5242-3089}\inst{\ref{inst:0024}}
\and A.        ~Cellino                       \orcit{0000-0002-6645-334X}\inst{\ref{inst:0045}}
\and N.        ~Cheek                         \inst{\ref{inst:0090}}
\and Y.        ~Damerdji                      \inst{\ref{inst:0091},\ref{inst:0092}}
\and M.        ~Davidson                      \inst{\ref{inst:0093}}
\and L.        ~Delchambre                    \inst{\ref{inst:0091}}
\and A.        ~Dell'Oro                      \orcit{0000-0003-1561-9685}\inst{\ref{inst:0037}}
\and J.        ~Fern\'{a}ndez-Hern\'{a}ndez   \inst{\ref{inst:0096}}
\and L.        ~Galluccio                     \orcit{0000-0002-8541-0476}\inst{\ref{inst:0024}}
\and P.        ~Garc\'{i}a-Lario              \inst{\ref{inst:0031}}
\and M.        ~Garcia-Reinaldos              \inst{\ref{inst:0031}}
\and J.        ~Gonz\'{a}lez-N\'{u}\~{n}ez    \orcit{0000-0001-5311-5555}\inst{\ref{inst:0090},\ref{inst:0101}}
\and E.        ~Gosset                        \inst{\ref{inst:0091},\ref{inst:0036}}
\and R.        ~Haigron                       \inst{\ref{inst:0022}}
\and J.-L.     ~Halbwachs                     \orcit{0000-0003-2968-6395}\inst{\ref{inst:0105}}
\and N.C.      ~Hambly                        \orcit{0000-0002-9901-9064}\inst{\ref{inst:0093}}
\and D.L.      ~Harrison                      \orcit{0000-0001-8687-6588}\inst{\ref{inst:0025},\ref{inst:0108}}
\and D.        ~Hatzidimitriou                \orcit{0000-0002-5415-0464}\inst{\ref{inst:0109}}
\and U.        ~Heiter                        \orcit{0000-0001-6825-1066}\inst{\ref{inst:0110}}
\and J.        ~Hern\'{a}ndez                 \inst{\ref{inst:0031}}
\and D.        ~Hestroffer                    \orcit{0000-0003-0472-9459}\inst{\ref{inst:0083}}
\and S.T.      ~Hodgkin                       \inst{\ref{inst:0025}}
\and B.        ~Holl                          \orcit{0000-0001-6220-3266}\inst{\ref{inst:0026},\ref{inst:0064}}
\and K.        ~Jan{\ss}en                    \inst{\ref{inst:0012}}
\and G.        ~Jevardat de Fombelle          \inst{\ref{inst:0026}}
\and S.        ~Jordan                        \orcit{0000-0001-6316-6831}\inst{\ref{inst:0023}}
\and A.        ~Krone-Martins                 \orcit{0000-0002-2308-6623}\inst{\ref{inst:0119},\ref{inst:0120}}
\and A.C.      ~Lanzafame                     \orcit{0000-0002-2697-3607}\inst{\ref{inst:0085},\ref{inst:0122}}
\and W.        ~L\"{ o}ffler                  \inst{\ref{inst:0023}}
\and A.        ~Lorca                         \inst{\ref{inst:0027}}
\and M.        ~Manteiga                      \orcit{0000-0002-7711-5581}\inst{\ref{inst:0125}}
\and O.        ~Marchal                       \inst{\ref{inst:0105}}
\and P.M.      ~Marrese                       \inst{\ref{inst:0127},\ref{inst:0128}}
\and A.        ~Moitinho                      \orcit{0000-0003-0822-5995}\inst{\ref{inst:0119}}
\and A.        ~Mora                          \inst{\ref{inst:0027}}
\and K.        ~Muinonen                      \orcit{0000-0001-8058-2642}\inst{\ref{inst:0131},\ref{inst:0132}}
\and P.        ~Osborne                       \inst{\ref{inst:0025}}
\and E.        ~Pancino                       \orcit{0000-0003-0788-5879}\inst{\ref{inst:0037},\ref{inst:0128}}
\and T.        ~Pauwels                       \inst{\ref{inst:0055}}
\and A.        ~Recio-Blanco                  \inst{\ref{inst:0024}}
\and P.J.      ~Richards                      \inst{\ref{inst:0138}}
\and M.        ~Riello                        \orcit{0000-0002-3134-0935}\inst{\ref{inst:0025}}
\and L.        ~Rimoldini                     \orcit{0000-0002-0306-585X}\inst{\ref{inst:0064}}
\and A.C.      ~Robin                         \orcit{0000-0001-8654-9499}\inst{\ref{inst:0141}}
\and T.        ~Roegiers                      \inst{\ref{inst:0142}}
\and J.        ~Rybizki                       \orcit{0000-0002-0993-6089}\inst{\ref{inst:0042}}
\and C.        ~Siopis                        \inst{\ref{inst:0035}}
\and M.        ~Smith                         \inst{\ref{inst:0044}}
\and A.        ~Sozzetti                      \orcit{0000-0002-7504-365X}\inst{\ref{inst:0045}}
\and A.        ~Ulla                          \inst{\ref{inst:0147}}
\and E.        ~Utrilla                       \inst{\ref{inst:0027}}
\and M.        ~van Leeuwen                   \inst{\ref{inst:0025}}
\and W.        ~van Reeven                    \inst{\ref{inst:0027}}
\and U.        ~Abbas                         \orcit{0000-0002-5076-766X}\inst{\ref{inst:0045}}
\and A.        ~Abreu Aramburu                \inst{\ref{inst:0096}}
\and S.        ~Accart                        \inst{\ref{inst:0153}}
\and C.        ~Aerts                         \orcit{0000-0003-1822-7126}\inst{\ref{inst:0154},\ref{inst:0155},\ref{inst:0042}}
\and J.J.      ~Aguado                        \inst{\ref{inst:0004}}
\and M.        ~Ajaj                          \inst{\ref{inst:0022}}
\and G.        ~Altavilla                     \orcit{0000-0002-9934-1352}\inst{\ref{inst:0127},\ref{inst:0128}}
\and M.A.      ~\'{A}lvarez                   \orcit{0000-0002-6786-2620}\inst{\ref{inst:0161}}
\and J.        ~\'{A}lvarez Cid-Fuentes       \orcit{0000-0001-7153-4649}\inst{\ref{inst:0162}}
\and J.        ~Alves                         \orcit{0000-0002-4355-0921}\inst{\ref{inst:0163}}
\and R.I.      ~Anderson                      \orcit{0000-0001-8089-4419}\inst{\ref{inst:0164}}
\and E.        ~Anglada Varela                \orcit{0000-0001-7563-0689}\inst{\ref{inst:0096}}
\and T.        ~Antoja                        \orcit{0000-0003-2595-5148}\inst{\ref{inst:0001}}
\and M.        ~Audard                        \orcit{0000-0003-4721-034X}\inst{\ref{inst:0064}}
\and D.        ~Baines                        \orcit{0000-0002-6923-3756}\inst{\ref{inst:0077}}
\and S.G.      ~Baker                         \orcit{0000-0002-6436-1257}\inst{\ref{inst:0044}}
\and L.        ~Balaguer-N\'{u}\~{n}ez        \orcit{0000-0001-9789-7069}\inst{\ref{inst:0001}}
\and Z.        ~Balog                         \orcit{0000-0003-1748-2926}\inst{\ref{inst:0023},\ref{inst:0042}}
\and C.        ~Barache                       \inst{\ref{inst:0079}}
\and D.        ~Barbato                       \inst{\ref{inst:0026},\ref{inst:0045}}
\and M.        ~Barros                        \orcit{0000-0002-9728-9618}\inst{\ref{inst:0119}}
\and M.A.      ~Barstow                       \orcit{0000-0002-7116-3259}\inst{\ref{inst:0177}}
\and S.        ~Bartolom\'{e}                 \orcit{0000-0002-6290-6030}\inst{\ref{inst:0001}}
\and J.-L.     ~Bassilana                     \inst{\ref{inst:0153}}
\and N.        ~Bauchet                       \inst{\ref{inst:0083}}
\and A.        ~Baudesson-Stella              \inst{\ref{inst:0153}}
\and U.        ~Becciani                      \orcit{0000-0002-4389-8688}\inst{\ref{inst:0085}}
\and M.        ~Bellazzini                    \orcit{0000-0001-8200-810X}\inst{\ref{inst:0003}}
\and M.        ~Bernet                        \inst{\ref{inst:0001}}
\and S.        ~Bertone                       \orcit{0000-0001-9885-8440}\inst{\ref{inst:0185},\ref{inst:0186},\ref{inst:0045}}
\and L.        ~Bianchi                       \inst{\ref{inst:0188}}
\and S.        ~Blanco-Cuaresma               \orcit{0000-0002-1584-0171}\inst{\ref{inst:0189}}
\and T.        ~Boch                          \orcit{0000-0001-5818-2781}\inst{\ref{inst:0105}}
\and A.        ~Bombrun                       \inst{\ref{inst:0191}}
\and D.        ~Bossini                       \orcit{0000-0002-9480-8400}\inst{\ref{inst:0192}}
\and S.        ~Bouquillon                    \inst{\ref{inst:0079}}
\and A.        ~Bragaglia                     \orcit{0000-0002-0338-7883}\inst{\ref{inst:0003}}
\and L.        ~Bramante                      \inst{\ref{inst:0061}}
\and E.        ~Breedt                        \orcit{0000-0001-6180-3438}\inst{\ref{inst:0025}}
\and A.        ~Bressan                       \orcit{0000-0002-7922-8440}\inst{\ref{inst:0197}}
\and N.        ~Brouillet                     \inst{\ref{inst:0039}}
\and B.        ~Bucciarelli                   \orcit{0000-0002-5303-0268}\inst{\ref{inst:0045}}
\and A.        ~Burlacu                       \inst{\ref{inst:0200}}
\and D.        ~Busonero                      \orcit{0000-0002-3903-7076}\inst{\ref{inst:0045}}
\and A.G.      ~Butkevich                     \inst{\ref{inst:0045}}
\and R.        ~Buzzi                         \orcit{0000-0001-9389-5701}\inst{\ref{inst:0045}}
\and E.        ~Caffau                        \orcit{0000-0001-6011-6134}\inst{\ref{inst:0022}}
\and R.        ~Cancelliere                   \orcit{0000-0002-9120-3799}\inst{\ref{inst:0205}}
\and H.        ~C\'{a}novas                   \orcit{0000-0001-7668-8022}\inst{\ref{inst:0027}}
\and T.        ~Cantat-Gaudin                 \orcit{0000-0001-8726-2588}\inst{\ref{inst:0001}}
\and R.        ~Carballo                      \inst{\ref{inst:0208}}
\and T.        ~Carlucci                      \inst{\ref{inst:0079}}
\and M.I       ~Carnerero                     \orcit{0000-0001-5843-5515}\inst{\ref{inst:0045}}
\and J.M.      ~Carrasco                      \orcit{0000-0002-3029-5853}\inst{\ref{inst:0001}}
\and L.        ~Casamiquela                   \orcit{0000-0001-5238-8674}\inst{\ref{inst:0039}}
\and M.        ~Castellani                    \orcit{0000-0002-7650-7428}\inst{\ref{inst:0127}}
\and P.        ~Castro Sampol                 \inst{\ref{inst:0001}}
\and L.        ~Chaoul                        \inst{\ref{inst:0034}}
\and P.        ~Charlot                       \inst{\ref{inst:0039}}
\and A.        ~Chiavassa                     \orcit{0000-0003-3891-7554}\inst{\ref{inst:0024}}
\and G.        ~Comoretto                     \inst{\ref{inst:0218}}
\and W.J.      ~Cooper                        \orcit{0000-0003-3501-8967}\inst{\ref{inst:0219},\ref{inst:0045}}
\and T.        ~Cornez                        \inst{\ref{inst:0153}}
\and S.        ~Cowell                        \inst{\ref{inst:0025}}
\and F.        ~Crifo                         \inst{\ref{inst:0022}}
\and M.        ~Crosta                        \orcit{0000-0003-4369-3786}\inst{\ref{inst:0045}}
\and C.        ~Crowley                       \inst{\ref{inst:0191}}
\and C.        ~Dafonte                       \orcit{0000-0003-4693-7555}\inst{\ref{inst:0161}}
\and A.        ~Dapergolas                    \inst{\ref{inst:0081}}
\and M.        ~David                         \orcit{0000-0002-4172-3112}\inst{\ref{inst:0228}}
\and P.        ~David                         \inst{\ref{inst:0083}}
\and P.        ~de Laverny                    \inst{\ref{inst:0024}}
\and F.        ~De Luise                      \orcit{0000-0002-6570-8208}\inst{\ref{inst:0231}}
\and R.        ~De March                      \orcit{0000-0003-0567-842X}\inst{\ref{inst:0061}}
\and J.        ~De Ridder                     \orcit{0000-0001-6726-2863}\inst{\ref{inst:0154}}
\and R.        ~de Souza                      \inst{\ref{inst:0234}}
\and P.        ~de Teodoro                    \inst{\ref{inst:0031}}
\and A.        ~de Torres                     \inst{\ref{inst:0191}}
\and E.F.      ~del Peloso                    \inst{\ref{inst:0023}}
\and E.        ~del Pozo                      \inst{\ref{inst:0027}}
\and A.        ~Delgado                       \inst{\ref{inst:0025}}
\and J.-B.     ~Delisle                       \orcit{0000-0001-5844-9888}\inst{\ref{inst:0026}}
\and P.        ~Di Matteo                     \inst{\ref{inst:0022}}
\and S.        ~Diakite                       \inst{\ref{inst:0242}}
\and C.        ~Diener                        \inst{\ref{inst:0025}}
\and E.        ~Distefano                     \orcit{0000-0002-2448-2513}\inst{\ref{inst:0085}}
\and C.        ~Dolding                       \inst{\ref{inst:0044}}
\and D.        ~Eappachen                     \inst{\ref{inst:0246},\ref{inst:0155}}
%\and B.        ~Edvardsson                    \inst{\ref{inst:0248}}
\and H.        ~Enke                          \orcit{0000-0002-2366-8316}\inst{\ref{inst:0012}}
\and P.        ~Esquej                        \orcit{0000-0001-8195-628X}\inst{\ref{inst:0250}}
\and C.        ~Fabre                         \inst{\ref{inst:0251}}
\and M.        ~Fabrizio                      \orcit{0000-0001-5829-111X}\inst{\ref{inst:0127},\ref{inst:0128}}
\and S.        ~Faigler                       \inst{\ref{inst:0254}}
\and G.        ~Fedorets                      \inst{\ref{inst:0131},\ref{inst:0256}}
\and P.        ~Fernique                      \orcit{0000-0002-3304-2923}\inst{\ref{inst:0105},\ref{inst:0258}}
\and A.        ~Fienga                        \orcit{0000-0002-4755-7637}\inst{\ref{inst:0259},\ref{inst:0083}}
\and F.        ~Figueras                      \orcit{0000-0002-3393-0007}\inst{\ref{inst:0001}}
\and C.        ~Fouron                        \inst{\ref{inst:0200}}
\and F.        ~Fragkoudi                     \inst{\ref{inst:0263}}
\and E.        ~Fraile                        \inst{\ref{inst:0250}}
\and F.        ~Franke                        \inst{\ref{inst:0265}}
\and M.        ~Gai                           \orcit{0000-0001-9008-134X}\inst{\ref{inst:0045}}
\and D.        ~Garabato                      \orcit{0000-0002-7133-6623}\inst{\ref{inst:0161}}
\and A.        ~Garcia-Gutierrez              \inst{\ref{inst:0001}}
\and M.        ~Garc\'{i}a-Torres             \orcit{0000-0002-6867-7080}\inst{\ref{inst:0269}}
\and P.        ~Gavras                        \orcit{0000-0002-4383-4836}\inst{\ref{inst:0250}}
\and E.        ~Gerlach                       \orcit{0000-0002-9533-2168}\inst{\ref{inst:0030}}
\and R.        ~Geyer                         \orcit{0000-0001-6967-8707}\inst{\ref{inst:0030}}
\and P.        ~Giacobbe                      \inst{\ref{inst:0045}}
\and G.        ~Gilmore                       \orcit{0000-0003-4632-0213}\inst{\ref{inst:0025}}
\and S.        ~Girona                        \orcit{0000-0002-1975-1918}\inst{\ref{inst:0162}}
\and G.        ~Giuffrida                     \inst{\ref{inst:0127}}
\and A.        ~Gomez                         \orcit{0000-0002-3796-3690}\inst{\ref{inst:0161}}
\and I.        ~Gonzalez-Santamaria           \orcit{0000-0002-8537-9384}\inst{\ref{inst:0161}}
\and J.J.      ~Gonz\'{a}lez-Vidal            \inst{\ref{inst:0001}}
\and M.        ~Granvik                       \orcit{0000-0002-5624-1888}\inst{\ref{inst:0131},\ref{inst:0281}}
\and R.        ~Guti\'{e}rrez-S\'{a}nchez     \inst{\ref{inst:0077}}
\and L.P.      ~Guy                           \orcit{0000-0003-0800-8755}\inst{\ref{inst:0064},\ref{inst:0218}}
\and M.        ~Hauser                        \inst{\ref{inst:0042},\ref{inst:0286}}
\and M.        ~Haywood                       \orcit{0000-0003-0434-0400}\inst{\ref{inst:0022}}
\and S.L.      ~Hidalgo                       \orcit{0000-0002-0002-9298}\inst{\ref{inst:0288},\ref{inst:0289}}
\and T.        ~Hilger                        \orcit{0000-0003-1646-0063}\inst{\ref{inst:0030}}
\and N.        ~H\l{}adczuk                   \inst{\ref{inst:0031}}
\and D.        ~Hobbs                         \orcit{0000-0002-2696-1366}\inst{\ref{inst:0005}}
\and G.        ~Holland                       \inst{\ref{inst:0025}}
\and H.E.      ~Huckle                        \inst{\ref{inst:0044}}
\and G.        ~Jasniewicz                    \inst{\ref{inst:0295}}
\and P.G.      ~Jonker                        \orcit{0000-0001-5679-0695}\inst{\ref{inst:0155},\ref{inst:0246}}
\and J.        ~Juaristi Campillo             \inst{\ref{inst:0023}}
\and F.        ~Julbe                         \inst{\ref{inst:0001}}
\and L.        ~Karbevska                     \inst{\ref{inst:0026}}
\and P.        ~Kervella                      \orcit{0000-0003-0626-1749}\inst{\ref{inst:0301}}
\and S.        ~Khanna                        \orcit{0000-0002-2604-4277}\inst{\ref{inst:0007}}
\and A.        ~Kochoska                      \orcit{0000-0002-9739-8371}\inst{\ref{inst:0303}}
\and M.        ~Kontizas                      \orcit{0000-0001-7177-0158}\inst{\ref{inst:0109}}
\and G.        ~Kordopatis                    \orcit{0000-0002-9035-3920}\inst{\ref{inst:0024}}
\and A.J.      ~Korn                          \orcit{0000-0002-3881-6756}\inst{\ref{inst:0110}}
\and Z.        ~Kostrzewa-Rutkowska           \inst{\ref{inst:0017},\ref{inst:0246}}
\and K.        ~Kruszy\'{n}ska                \orcit{0000-0002-2729-5369}\inst{\ref{inst:0309}}
\and S.        ~Lambert                       \orcit{0000-0001-6759-5502}\inst{\ref{inst:0079}}
\and A.F.      ~Lanza                         \orcit{0000-0001-5928-7251}\inst{\ref{inst:0085}}
\and Y.        ~Lasne                         \inst{\ref{inst:0153}}
\and J.-F.     ~Le Campion                    \inst{\ref{inst:0313}}
\and Y.        ~Le Fustec                     \inst{\ref{inst:0200}}
\and Y.        ~Lebreton                      \orcit{0000-0002-4834-2144}\inst{\ref{inst:0301},\ref{inst:0316}}
\and T.        ~Lebzelter                     \orcit{0000-0002-0702-7551}\inst{\ref{inst:0163}}
\and S.        ~Leccia                        \orcit{0000-0001-5685-6930}\inst{\ref{inst:0010}}
\and N.        ~Leclerc                       \inst{\ref{inst:0022}}
\and I.        ~Lecoeur-Taibi                 \orcit{0000-0003-0029-8575}\inst{\ref{inst:0064}}
\and S.        ~Liao                          \inst{\ref{inst:0045}}
\and E.        ~Licata                        \orcit{0000-0002-5203-0135}\inst{\ref{inst:0045}}
\and H.E.P.    ~Lindstr{\o}m                  \inst{\ref{inst:0045},\ref{inst:0324}}
\and T.A.      ~Lister                        \orcit{0000-0002-3818-7769}\inst{\ref{inst:0325}}
\and E.        ~Livanou                       \inst{\ref{inst:0109}}
\and A.        ~Lobel                         \inst{\ref{inst:0055}}
\and P.        ~Madrero Pardo                 \inst{\ref{inst:0001}}
\and S.        ~Managau                       \inst{\ref{inst:0153}}
\and R.G.      ~Mann                          \orcit{0000-0002-0194-325X}\inst{\ref{inst:0093}}
\and J.M.      ~Marchant                      \inst{\ref{inst:0331}}
\and M.        ~Marconi                       \orcit{0000-0002-1330-2927}\inst{\ref{inst:0010}}
\and M.M.S.    ~Marcos Santos                 \inst{\ref{inst:0090}}
\and S.        ~Marinoni                      \orcit{0000-0001-7990-6849}\inst{\ref{inst:0127},\ref{inst:0128}}
\and F.        ~Marocco                       \orcit{0000-0001-7519-1700}\inst{\ref{inst:0336},\ref{inst:0337}}
\and D.J.      ~Marshall                      \inst{\ref{inst:0338}}
\and L.        ~Martin Polo                   \inst{\ref{inst:0090}}
\and J.M.      ~Mart\'{i}n-Fleitas            \orcit{0000-0002-8594-569X}\inst{\ref{inst:0027}}
\and A.        ~Masip                         \inst{\ref{inst:0001}}
\and D.        ~Massari                       \orcit{0000-0001-8892-4301}\inst{\ref{inst:0003}}
\and A.        ~Mastrobuono-Battisti          \orcit{0000-0002-2386-9142}\inst{\ref{inst:0005}}
\and T.        ~Mazeh                         \orcit{0000-0002-3569-3391}\inst{\ref{inst:0254}}
\and S.        ~Messina                       \orcit{0000-0002-2851-2468}\inst{\ref{inst:0085}}
\and D.        ~Michalik                      \orcit{0000-0002-7618-6556}\inst{\ref{inst:0019}}
\and N.R.      ~Millar                        \inst{\ref{inst:0025}}
\and A.        ~Mints                         \orcit{0000-0002-8440-1455}\inst{\ref{inst:0012}}
\and D.        ~Molina                        \orcit{0000-0003-4814-0275}\inst{\ref{inst:0001}}
\and R.        ~Molinaro                      \orcit{0000-0003-3055-6002}\inst{\ref{inst:0010}}
\and L.        ~Moln\'{a}r                    \orcit{0000-0002-8159-1599}\inst{\ref{inst:0351},\ref{inst:0352},\ref{inst:0353}}
\and P.        ~Montegriffo                   \inst{\ref{inst:0003}}
\and R.        ~Morbidelli                    \orcit{0000-0001-7627-4946}\inst{\ref{inst:0045}}
\and T.        ~Morel                         \inst{\ref{inst:0091}}
\and D.        ~Morris                        \inst{\ref{inst:0093}}
\and A.F.      ~Mulone                        \inst{\ref{inst:0061}}
\and D.        ~Munoz                         \inst{\ref{inst:0153}}
\and C.P.      ~Murphy                        \inst{\ref{inst:0031}}
\and I.        ~Musella                       \orcit{0000-0001-5909-6615}\inst{\ref{inst:0010}}
\and L.        ~Noval                         \inst{\ref{inst:0153}}
\and C.        ~Ord\'{e}novic                 \inst{\ref{inst:0024}}
\and G.        ~Orr\`{u}                      \inst{\ref{inst:0061}}
\and J.        ~Osinde                        \inst{\ref{inst:0250}}
\and C.        ~Pagani                        \inst{\ref{inst:0177}}
\and I.        ~Pagano                        \orcit{0000-0001-9573-4928}\inst{\ref{inst:0085}}
\and L.        ~Palaversa                     \inst{\ref{inst:0368},\ref{inst:0025}}
\and P.A.      ~Palicio                       \orcit{0000-0002-7432-8709}\inst{\ref{inst:0024}}
\and A.        ~Panahi                        \orcit{0000-0001-5850-4373}\inst{\ref{inst:0254}}
\and M.        ~Pawlak                        \orcit{0000-0002-5632-9433}\inst{\ref{inst:0372},\ref{inst:0309}}
\and X.        ~Pe\~{n}alosa Esteller         \inst{\ref{inst:0001}}
\and A.        ~Penttil\"{ a}                 \orcit{0000-0001-7403-1721}\inst{\ref{inst:0131}}
\and A.M.      ~Piersimoni                    \orcit{0000-0002-8019-3708}\inst{\ref{inst:0231}}
\and F.-X.     ~Pineau                        \orcit{0000-0002-2335-4499}\inst{\ref{inst:0105}}
\and E.        ~Plachy                        \orcit{0000-0002-5481-3352}\inst{\ref{inst:0351},\ref{inst:0352},\ref{inst:0353}}
\and G.        ~Plum                          \inst{\ref{inst:0022}}
\and E.        ~Poggio                        \orcit{0000-0003-3793-8505}\inst{\ref{inst:0045}}
\and E.        ~Poretti                       \orcit{0000-0003-1200-0473}\inst{\ref{inst:0383}}
\and E.        ~Poujoulet                     \inst{\ref{inst:0384}}
\and A.        ~Pr\v{s}a                      \orcit{0000-0002-1913-0281}\inst{\ref{inst:0303}}
\and L.        ~Pulone                        \orcit{0000-0002-5285-998X}\inst{\ref{inst:0127}}
\and E.        ~Racero                        \inst{\ref{inst:0090},\ref{inst:0388}}
\and S.        ~Ragaini                       \inst{\ref{inst:0003}}
\and M.        ~Rainer                        \orcit{0000-0002-8786-2572}\inst{\ref{inst:0037}}
\and C.M.      ~Raiteri                       \orcit{0000-0003-1784-2784}\inst{\ref{inst:0045}}
\and N.        ~Rambaux                       \inst{\ref{inst:0083}}
\and P.        ~Ramos                         \orcit{0000-0002-5080-7027}\inst{\ref{inst:0001}}
\and M.        ~Ramos-Lerate                  \inst{\ref{inst:0394}}
\and P.        ~Re Fiorentin                  \orcit{0000-0002-4995-0475}\inst{\ref{inst:0045}}
\and S.        ~Regibo                        \inst{\ref{inst:0154}}
\and C.        ~Reyl\'{e}                     \inst{\ref{inst:0141}}
\and A.        ~Riva                          \orcit{0000-0002-6928-8589}\inst{\ref{inst:0045}}
\and G.        ~Rixon                         \inst{\ref{inst:0025}}
\and N.        ~Robichon                      \orcit{0000-0003-4545-7517}\inst{\ref{inst:0022}}
\and C.        ~Robin                         \inst{\ref{inst:0153}}
\and M.        ~Roelens                       \orcit{0000-0003-0876-4673}\inst{\ref{inst:0026}}
\and L.        ~Rohrbasser                    \inst{\ref{inst:0064}}
\and N.        ~Rowell                        \inst{\ref{inst:0093}}
\and F.        ~Royer                         \orcit{0000-0002-9374-8645}\inst{\ref{inst:0022}}
\and K.A.      ~Rybicki                       \orcit{0000-0002-9326-9329}\inst{\ref{inst:0309}}
\and G.        ~Sadowski                      \inst{\ref{inst:0035}}
\and A.        ~Sagrist\`{a} Sell\'{e}s       \orcit{0000-0001-6191-2028}\inst{\ref{inst:0023}}
\and J.        ~Sahlmann                      \orcit{0000-0001-9525-3673}\inst{\ref{inst:0250}}
\and J.        ~Salgado                       \orcit{0000-0002-3680-4364}\inst{\ref{inst:0077}}
\and E.        ~Salguero                      \inst{\ref{inst:0096}}
\and N.        ~Samaras                       \orcit{0000-0001-8375-6652}\inst{\ref{inst:0055}}
\and V.        ~Sanchez Gimenez               \inst{\ref{inst:0001}}
\and N.        ~Sanna                         \inst{\ref{inst:0037}}
\and R.        ~Santove\~{n}a                 \orcit{0000-0002-9257-2131}\inst{\ref{inst:0161}}
\and M.        ~Sarasso                       \orcit{0000-0001-5121-0727}\inst{\ref{inst:0045}}
\and M.        ~Schultheis                    \orcit{0000-0002-6590-1657}\inst{\ref{inst:0024}}
\and E.        ~Sciacca                       \orcit{0000-0002-5574-2787}\inst{\ref{inst:0085}}
\and M.        ~Segol                         \inst{\ref{inst:0265}}
\and J.C.      ~Segovia                       \inst{\ref{inst:0090}}
\and D.        ~S\'{e}gransan                 \orcit{0000-0003-2355-8034}\inst{\ref{inst:0026}}
\and D.        ~Semeux                        \inst{\ref{inst:0251}}
\and H.I.      ~Siddiqui                      \orcit{0000-0003-1853-6033}\inst{\ref{inst:0423}}
\and A.        ~Siebert                       \orcit{0000-0001-8059-2840}\inst{\ref{inst:0105},\ref{inst:0258}}
\and L.        ~Siltala                       \orcit{0000-0002-6938-794X}\inst{\ref{inst:0131}}
\and E.        ~Slezak                        \inst{\ref{inst:0024}}
\and R.L.      ~Smart                         \orcit{0000-0002-4424-4766}\inst{\ref{inst:0045}}
\and E.        ~Solano                        \inst{\ref{inst:0429}}
\and F.        ~Solitro                       \inst{\ref{inst:0061}}
\and D.        ~Souami                        \orcit{0000-0003-4058-0815}\inst{\ref{inst:0301},\ref{inst:0432}}
\and J.        ~Souchay                       \inst{\ref{inst:0079}}
\and A.        ~Spagna                        \orcit{0000-0003-1732-2412}\inst{\ref{inst:0045}}
\and F.        ~Spoto                         \orcit{0000-0001-7319-5847}\inst{\ref{inst:0189}}
\and I.A.      ~Steele                        \orcit{0000-0001-8397-5759}\inst{\ref{inst:0331}}
\and H.        ~Steidelm\"{ u}ller            \inst{\ref{inst:0030}}
\and C.A.      ~Stephenson                    \inst{\ref{inst:0077}}
\and M.        ~S\"{ u}veges                  \inst{\ref{inst:0064},\ref{inst:0440},\ref{inst:0042}}
\and L.        ~Szabados                      \orcit{0000-0002-2046-4131}\inst{\ref{inst:0351}}
\and E.        ~Szegedi-Elek                  \orcit{0000-0001-7807-6644}\inst{\ref{inst:0351}}
\and F.        ~Taris                         \inst{\ref{inst:0079}}
\and G.        ~Tauran                        \inst{\ref{inst:0153}}
\and M.B.      ~Taylor                        \orcit{0000-0002-4209-1479}\inst{\ref{inst:0446}}
\and R.        ~Teixeira                      \orcit{0000-0002-6806-6626}\inst{\ref{inst:0234}}
\and W.        ~Thuillot                      \inst{\ref{inst:0083}}
\and N.        ~Tonello                       \orcit{0000-0003-0550-1667}\inst{\ref{inst:0162}}
\and F.        ~Torra                         \orcit{0000-0002-8429-299X}\inst{\ref{inst:0051}}
\and J.        ~Torra$^\dagger$               \inst{\ref{inst:0001}}
\and C.        ~Turon                         \orcit{0000-0003-1236-5157}\inst{\ref{inst:0022}}
\and N.        ~Unger                         \orcit{0000-0003-3993-7127}\inst{\ref{inst:0026}}
\and M.        ~Vaillant                      \inst{\ref{inst:0153}}
\and E.        ~van Dillen                    \inst{\ref{inst:0265}}
\and O.        ~Vanel                         \inst{\ref{inst:0022}}
\and A.        ~Vecchiato                     \orcit{0000-0003-1399-5556}\inst{\ref{inst:0045}}
\and Y.        ~Viala                         \inst{\ref{inst:0022}}
\and D.        ~Vicente                       \inst{\ref{inst:0162}}
\and S.        ~Voutsinas                     \inst{\ref{inst:0093}}
\and M.        ~Weiler                        \inst{\ref{inst:0001}}
\and T.        ~Wevers                        \orcit{0000-0002-4043-9400}\inst{\ref{inst:0025}}
\and \L{}.     ~Wyrzykowski                   \orcit{0000-0002-9658-6151}\inst{\ref{inst:0309}}
\and A.        ~Yoldas                        \inst{\ref{inst:0025}}
\and P.        ~Yvard                         \inst{\ref{inst:0265}}
\and H.        ~Zhao                          \orcit{0000-0003-2645-6869}\inst{\ref{inst:0024}}
\and J.        ~Zorec                         \inst{\ref{inst:0467}}
\and S.        ~Zucker                        \orcit{0000-0003-3173-3138}\inst{\ref{inst:0468}}
\and C.        ~Zurbach                       \inst{\ref{inst:0469}}
\and T.        ~Zwitter                       \orcit{0000-0002-2325-8763}\inst{\ref{inst:0470}}
}
\institute{
     Institut de Ci\`{e}ncies del Cosmos (ICCUB), Universitat  de  Barcelona  (IEEC-UB), Mart\'{i} i  Franqu\`{e}s  1, 08028 Barcelona, Spain\relax                                                                                                                                                              \label{inst:0001}
\and Centro de Astronom\'{i}a - CITEVA, Universidad de Antofagasta, Avenida Angamos 601, Antofagasta 1270300, Chile\relax                                                                                                                                                                                        \label{inst:0002}
\and INAF - Osservatorio di Astrofisica e Scienza dello Spazio di Bologna, via Piero Gobetti 93/3, 40129 Bologna, Italy\relax                                                                                                                                                                                    \label{inst:0003}
\and Dpto. de Inteligencia Artificial, UNED, c/ Juan del Rosal 16, 28040 Madrid, Spain\relax                                                                                                                                                                                                                     \label{inst:0004}
\and Lund Observatory, Department of Astronomy and Theoretical Physics, Lund University, Box 43, 22100 Lund, Sweden\relax                                                                                                                                                                                        \label{inst:0005}
\and Kapteyn Astronomical Institute, University of Groningen, Landleven 12, 9747 AD Groningen, The Netherlands\relax                                                                                                                                                                                             \label{inst:0007}
\and INAF - Osservatorio Astronomico di Capodimonte, Via Moiariello 16, 80131, Napoli, Italy\relax                                                                                                                                                                                                               \label{inst:0010}
\and Leibniz Institute for Astrophysics Potsdam (AIP), An der Sternwarte 16, 14482 Potsdam, Germany\relax                                                                                                                                                                                                        \label{inst:0012}
\and Leiden Observatory, Leiden University, Niels Bohrweg 2, 2333 CA Leiden, The Netherlands\relax                                                                                                                                                                                                               \label{inst:0017}
\and INAF - Osservatorio astronomico di Padova, Vicolo Osservatorio 5, 35122 Padova, Italy\relax                                                                                                                                                                                                                 \label{inst:0018}
\and European Space Agency (ESA), European Space Research and Technology Centre (ESTEC), Keplerlaan 1, 2201AZ, Noordwijk, The Netherlands\relax                                                                                                                                                                  \label{inst:0019}
\and Univ. Grenoble Alpes, CNRS, IPAG, 38000 Grenoble, France\relax                                                                                                                                                                                                                                              \label{inst:0021}
\and GEPI, Observatoire de Paris, Universit\'{e} PSL, CNRS, 5 Place Jules Janssen, 92190 Meudon, France\relax                                                                                                                                                                                                    \label{inst:0022}
\and Astronomisches Rechen-Institut, Zentrum f\"{ u}r Astronomie der Universit\"{ a}t Heidelberg, M\"{ o}nchhofstr. 12-14, 69120 Heidelberg, Germany\relax                                                                                                                                                       \label{inst:0023}
\and Universit\'{e} C\^{o}te d'Azur, Observatoire de la C\^{o}te d'Azur, CNRS, Laboratoire Lagrange, Bd de l'Observatoire, CS 34229, 06304 Nice Cedex 4, France\relax                                                                                                                                            \label{inst:0024}
\and Institute of Astronomy, University of Cambridge, Madingley Road, Cambridge CB3 0HA, United Kingdom\relax                                                                                                                                                                                                    \label{inst:0025}
\and Department of Astronomy, University of Geneva, Chemin des Maillettes 51, 1290 Versoix, Switzerland\relax                                                                                                                                                                                                    \label{inst:0026}
\and Aurora Technology for European Space Agency (ESA), Camino bajo del Castillo, s/n, Urbanizacion Villafranca del Castillo, Villanueva de la Ca\~{n}ada, 28692 Madrid, Spain\relax                                                                                                                             \label{inst:0027}
\and Lohrmann Observatory, Technische Universit\"{ a}t Dresden, Mommsenstra{\ss}e 13, 01062 Dresden, Germany\relax                                                                                                                                                                                               \label{inst:0030}
\and European Space Agency (ESA), European Space Astronomy Centre (ESAC), Camino bajo del Castillo, s/n, Urbanizacion Villafranca del Castillo, Villanueva de la Ca\~{n}ada, 28692 Madrid, Spain\relax                                                                                                           \label{inst:0031}
\and CNES Centre Spatial de Toulouse, 18 avenue Edouard Belin, 31401 Toulouse Cedex 9, France\relax                                                                                                                                                                                                              \label{inst:0034}
\and Institut d'Astronomie et d'Astrophysique, Universit\'{e} Libre de Bruxelles CP 226, Boulevard du Triomphe, 1050 Brussels, Belgium\relax                                                                                                                                                                     \label{inst:0035}
\and F.R.S.-FNRS, Rue d'Egmont 5, 1000 Brussels, Belgium\relax                                                                                                                                                                                                                                                   \label{inst:0036}
\and INAF - Osservatorio Astrofisico di Arcetri, Largo Enrico Fermi 5, 50125 Firenze, Italy\relax                                                                                                                                                                                                                \label{inst:0037}
\and Laboratoire d'astrophysique de Bordeaux, Univ. Bordeaux, CNRS, B18N, all{\'e}e Geoffroy Saint-Hilaire, 33615 Pessac, France\relax                                                                                                                                                                           \label{inst:0039}
\and Max Planck Institute for Astronomy, K\"{ o}nigstuhl 17, 69117 Heidelberg, Germany\relax                                                                                                                                                                                                                     \label{inst:0042}
\and Mullard Space Science Laboratory, University College London, Holmbury St Mary, Dorking, Surrey RH5 6NT, United Kingdom\relax                                                                                                                                                                                \label{inst:0044}
\vfill\break
\and INAF - Osservatorio Astrofisico di Torino, via Osservatorio 20, 10025 Pino Torinese (TO), Italy\relax                                                                                                                                                                                                       \label{inst:0045}
\and University of Turin, Department of Physics, Via Pietro Giuria 1, 10125 Torino, Italy\relax                                                                                                                                                                                                                  \label{inst:0048}
\and DAPCOM for Institut de Ci\`{e}ncies del Cosmos (ICCUB), Universitat  de  Barcelona  (IEEC-UB), Mart\'{i} i  Franqu\`{e}s  1, 08028 Barcelona, Spain\relax                                                                                                                                                   \label{inst:0051}
\and Royal Observatory of Belgium, Ringlaan 3, 1180 Brussels, Belgium\relax                                                                                                                                                                                                                                      \label{inst:0055}
\and ALTEC S.p.a, Corso Marche, 79,10146 Torino, Italy\relax                                                                                                                                                                                                                                                     \label{inst:0061}
\and Department of Astronomy, University of Geneva, Chemin d'Ecogia 16, 1290 Versoix, Switzerland\relax                                                                                                                                                                                                          \label{inst:0064}
\and Sednai S\`{a}rl, Geneva, Switzerland\relax                                                                                                                                                                                                                                                                  \label{inst:0065}
\and Gaia DPAC Project Office, ESAC, Camino bajo del Castillo, s/n, Urbanizacion Villafranca del Castillo, Villanueva de la Ca\~{n}ada, 28692 Madrid, Spain\relax                                                                                                                                                \label{inst:0074}
\and Telespazio Vega UK Ltd for European Space Agency (ESA), Camino bajo del Castillo, s/n, Urbanizacion Villafranca del Castillo, Villanueva de la Ca\~{n}ada, 28692 Madrid, Spain\relax                                                                                                                        \label{inst:0077}
\and SYRTE, Observatoire de Paris, Universit\'{e} PSL, CNRS,  Sorbonne Universit\'{e}, LNE, 61 avenue de l’Observatoire 75014 Paris, France\relax                                                                                                                                                              \label{inst:0079}
\and National Observatory of Athens, I. Metaxa and Vas. Pavlou, Palaia Penteli, 15236 Athens, Greece\relax                                                                                                                                                                                                       \label{inst:0081}
\and IMCCE, Observatoire de Paris, Universit\'{e} PSL, CNRS, Sorbonne Universit{\'e}, Univ. Lille, 77 av. Denfert-Rochereau, 75014 Paris, France\relax                                                                                                                                                           \label{inst:0083}
\and INAF - Osservatorio Astrofisico di Catania, via S. Sofia 78, 95123 Catania, Italy\relax                                                                                                                                                                                                                     \label{inst:0085}
\and Serco Gesti\'{o}n de Negocios for European Space Agency (ESA), Camino bajo del Castillo, s/n, Urbanizacion Villafranca del Castillo, Villanueva de la Ca\~{n}ada, 28692 Madrid, Spain\relax                                                                                                                 \label{inst:0090}
\and Institut d'Astrophysique et de G\'{e}ophysique, Universit\'{e} de Li\`{e}ge, 19c, All\'{e}e du 6 Ao\^{u}t, B-4000 Li\`{e}ge, Belgium\relax                                                                                                                                                                  \label{inst:0091}
\and CRAAG - Centre de Recherche en Astronomie, Astrophysique et G\'{e}ophysique, Route de l'Observatoire Bp 63 Bouzareah 16340 Algiers, Algeria\relax                                                                                                                                                           \label{inst:0092}
\and Institute for Astronomy, University of Edinburgh, Royal Observatory, Blackford Hill, Edinburgh EH9 3HJ, United Kingdom\relax                                                                                                                                                                                \label{inst:0093}
\and ATG Europe for European Space Agency (ESA), Camino bajo del Castillo, s/n, Urbanizacion Villafranca del Castillo, Villanueva de la Ca\~{n}ada, 28692 Madrid, Spain\relax                                                                                                                                    \label{inst:0096}
\and ETSE Telecomunicaci\'{o}n, Universidade de Vigo, Campus Lagoas-Marcosende, 36310 Vigo, Galicia, Spain\relax                                                                                                                                                                                                 \label{inst:0101}
\and Universit\'{e} de Strasbourg, CNRS, Observatoire astronomique de Strasbourg, UMR 7550,  11 rue de l'Universit\'{e}, 67000 Strasbourg, France\relax                                                                                                                                                          \label{inst:0105}
\and Kavli Institute for Cosmology Cambridge, Institute of Astronomy, Madingley Road, Cambridge, CB3 0HA\relax                                                                                                                                                                                                   \label{inst:0108}
\and Department of Astrophysics, Astronomy and Mechanics, National and Kapodistrian University of Athens, Panepistimiopolis, Zografos, 15783 Athens, Greece\relax                                                                                                                                                \label{inst:0109}
\and Observational Astrophysics, Division of Astronomy and Space Physics, Department of Physics and Astronomy, Uppsala University, Box 516, 751 20 Uppsala, Sweden\relax                                                                                                                                         \label{inst:0110}
\and CENTRA, Faculdade de Ci\^{e}ncias, Universidade de Lisboa, Edif. C8, Campo Grande, 1749-016 Lisboa, Portugal\relax                                                                                                                                                                                          \label{inst:0119}
\and Department of Informatics, Donald Bren School of Information and Computer Sciences, University of California, 5019 Donald Bren Hall, 92697-3440 CA Irvine, United States\relax                                                                                                                              \label{inst:0120}
\vfill\break
\and Dipartimento di Fisica e Astronomia ""Ettore Majorana"", Universit\`{a} di Catania, Via S. Sofia 64, 95123 Catania, Italy\relax                                                                                                                                                                             \label{inst:0122}
\and CITIC, Department of Nautical Sciences and Marine Engineering, University of A Coru\~{n}a, Campus de Elvi\~{n}a S/N, 15071, A Coru\~{n}a, Spain\relax                                                                                                                                                       \label{inst:0125}
\and INAF - Osservatorio Astronomico di Roma, Via Frascati 33, 00078 Monte Porzio Catone (Roma), Italy\relax                                                                                                                                                                                                     \label{inst:0127}
\and Space Science Data Center - ASI, Via del Politecnico SNC, 00133 Roma, Italy\relax                                                                                                                                                                                                                           \label{inst:0128}
\and Department of Physics, University of Helsinki, P.O. Box 64, 00014 Helsinki, Finland\relax                                                                                                                                                                                                                   \label{inst:0131}
\and Finnish Geospatial Research Institute FGI, Geodeetinrinne 2, 02430 Masala, Finland\relax                                                                                                                                                                                                                    \label{inst:0132}
\and STFC, Rutherford Appleton Laboratory, Harwell, Didcot, OX11 0QX, United Kingdom\relax                                                                                                                                                                                                                       \label{inst:0138}
\and Institut UTINAM CNRS UMR6213, Universit\'{e} Bourgogne Franche-Comt\'{e}, OSU THETA Franche-Comt\'{e} Bourgogne, Observatoire de Besan\c{c}on, BP1615, 25010 Besan\c{c}on Cedex, France\relax                                                                                                               \label{inst:0141}
\and HE Space Operations BV for European Space Agency (ESA), Keplerlaan 1, 2201AZ, Noordwijk, The Netherlands\relax                                                                                                                                                                                              \label{inst:0142}
\and Applied Physics Department, Universidade de Vigo, 36310 Vigo, Spain\relax                                                                                                                                                                                                                                   \label{inst:0147}
\and Thales Services for CNES Centre Spatial de Toulouse, 18 avenue Edouard Belin, 31401 Toulouse Cedex 9, France\relax                                                                                                                                                                                          \label{inst:0153}
\and Instituut voor Sterrenkunde, KU Leuven, Celestijnenlaan 200D, 3001 Leuven, Belgium\relax                                                                                                                                                                                                                    \label{inst:0154}
\and Department of Astrophysics/IMAPP, Radboud University, P.O.Box 9010, 6500 GL Nijmegen, The Netherlands\relax                                                                                                                                                                                                 \label{inst:0155}
\and CITIC - Department of Computer Science and Information Technologies, University of A Coru\~{n}a, Campus de Elvi\~{n}a S/N, 15071, A Coru\~{n}a, Spain\relax                                                                                                                                                 \label{inst:0161}
\and Barcelona Supercomputing Center (BSC) - Centro Nacional de Supercomputaci\'{o}n, c/ Jordi Girona 29, Ed. Nexus II, 08034 Barcelona, Spain\relax                                                                                                                                                             \label{inst:0162}
\and University of Vienna, Department of Astrophysics, T\"{ u}rkenschanzstra{\ss}e 17, A1180 Vienna, Austria\relax                                                                                                                                                                                               \label{inst:0163}
\and European Southern Observatory, Karl-Schwarzschild-Str. 2, 85748 Garching, Germany\relax                                                                                                                                                                                                                     \label{inst:0164}
\and School of Physics and Astronomy, University of Leicester, University Road, Leicester LE1 7RH, United Kingdom\relax                                                                                                                                                                                          \label{inst:0177}
\and Center for Research and Exploration in Space Science and Technology, University of Maryland Baltimore County, 1000 Hilltop Circle, Baltimore MD, USA\relax                                                                                                                                                  \label{inst:0185}
\and GSFC - Goddard Space Flight Center, Code 698, 8800 Greenbelt Rd, 20771 MD Greenbelt, United States\relax                                                                                                                                                                                                    \label{inst:0186}
\and EURIX S.r.l., Corso Vittorio Emanuele II 61, 10128, Torino, Italy\relax                                                                                                                                                                                                                                     \label{inst:0188}
\and Harvard-Smithsonian Center for Astrophysics, 60 Garden St., MS 15, Cambridge, MA 02138, USA\relax                                                                                                                                                                                                           \label{inst:0189}
\and HE Space Operations BV for European Space Agency (ESA), Camino bajo del Castillo, s/n, Urbanizacion Villafranca del Castillo, Villanueva de la Ca\~{n}ada, 28692 Madrid, Spain\relax                                                                                                                        \label{inst:0191}
\and CAUP - Centro de Astrofisica da Universidade do Porto, Rua das Estrelas, Porto, Portugal\relax                                                                                                                                                                                                              \label{inst:0192}
\and SISSA - Scuola Internazionale Superiore di Studi Avanzati, via Bonomea 265, 34136 Trieste, Italy\relax                                                                                                                                                                                                      \label{inst:0197}
\and Telespazio for CNES Centre Spatial de Toulouse, 18 avenue Edouard Belin, 31401 Toulouse Cedex 9, France\relax                                                                                                                                                                                               \label{inst:0200}
\and University of Turin, Department of Computer Sciences, Corso Svizzera 185, 10149 Torino, Italy\relax                                                                                                                                                                                                         \label{inst:0205}
\vfill\break
\and Dpto. de Matem\'{a}tica Aplicada y Ciencias de la Computaci\'{o}n, Univ. de Cantabria, ETS Ingenieros de Caminos, Canales y Puertos, Avda. de los Castros s/n, 39005 Santander, Spain\relax                                                                                                                 \label{inst:0208}
\and Vera C Rubin Observatory,  950 N. Cherry Avenue, Tucson, AZ 85719, USA\relax                                                                                                                                                                                                                                \label{inst:0218}
\and Centre for Astrophysics Research, University of Hertfordshire, College Lane, AL10 9AB, Hatfield, United Kingdom\relax                                                                                                                                                                                       \label{inst:0219}
\and University of Antwerp, Onderzoeksgroep Toegepaste Wiskunde, Middelheimlaan 1, 2020 Antwerp, Belgium\relax                                                                                                                                                                                                   \label{inst:0228}
\and INAF - Osservatorio Astronomico d'Abruzzo, Via Mentore Maggini, 64100 Teramo, Italy\relax                                                                                                                                                                                                                   \label{inst:0231}
\and Instituto de Astronomia, Geof\`{i}sica e Ci\^{e}ncias Atmosf\'{e}ricas, Universidade de S\~{a}o Paulo, Rua do Mat\~{a}o, 1226, Cidade Universitaria, 05508-900 S\~{a}o Paulo, SP, Brazil\relax                                                                                                              \label{inst:0234}
\and M\'{e}socentre de calcul de Franche-Comt\'{e}, Universit\'{e} de Franche-Comt\'{e}, 16 route de Gray, 25030 Besan\c{c}on Cedex, France\relax                                                                                                                                                                \label{inst:0242}
\and SRON, Netherlands Institute for Space Research, Sorbonnelaan 2, 3584CA, Utrecht, The Netherlands\relax                                                                                                                                                                                                      \label{inst:0246}
%\and Theoretical Astrophysics, Division of Astronomy and Space Physics, Department of Physics and Astronomy, Uppsala University, Box 516, 751 20 Uppsala, Sweden\relax                                                                                                                                           \label{inst:0248}
\and RHEA for European Space Agency (ESA), Camino bajo del Castillo, s/n, Urbanizacion Villafranca del Castillo, Villanueva de la Ca\~{n}ada, 28692 Madrid, Spain\relax                                                                                                                                          \label{inst:0250}
\and ATOS for CNES Centre Spatial de Toulouse, 18 avenue Edouard Belin, 31401 Toulouse Cedex 9, France\relax                                                                                                                                                                                                     \label{inst:0251}
\and School of Physics and Astronomy, Tel Aviv University, Tel Aviv 6997801, Israel\relax                                                                                                                                                                                                                        \label{inst:0254}
\and Astrophysics Research Centre, School of Mathematics and Physics, Queen's University Belfast, Belfast BT7 1NN, UK\relax                                                                                                                                                                                      \label{inst:0256}
\and Centre de Donn\'{e}es Astronomique de Strasbourg, Strasbourg, France\relax                                                                                                                                                                                                                                  \label{inst:0258}
\and Universit\'{e} C\^{o}te d'Azur, Observatoire de la C\^{o}te d'Azur, CNRS, Laboratoire G\'{e}oazur, Bd de l'Observatoire, CS 34229, 06304 Nice Cedex 4, France\relax                                                                                                                                         \label{inst:0259}
\and Max-Planck-Institut f\"{ u}r Astrophysik, Karl-Schwarzschild-Stra{\ss}e 1, 85748 Garching, Germany\relax                                                                                                                                                                                                    \label{inst:0263}
\and APAVE SUDEUROPE SAS for CNES Centre Spatial de Toulouse, 18 avenue Edouard Belin, 31401 Toulouse Cedex 9, France\relax                                                                                                                                                                                      \label{inst:0265}
\and \'{A}rea de Lenguajes y Sistemas Inform\'{a}ticos, Universidad Pablo de Olavide, Ctra. de Utrera, km 1. 41013, Sevilla, Spain\relax                                                                                                                                                                         \label{inst:0269}
\and Onboard Space Systems, Lule\aa{} University of Technology, Box 848, S-981 28 Kiruna, Sweden\relax                                                                                                                                                                                                           \label{inst:0281}
\and TRUMPF Photonic Components GmbH, Lise-Meitner-Stra{\ss}e 13,  89081 Ulm, Germany\relax                                                                                                                                                                                                                      \label{inst:0286}
\and IAC - Instituto de Astrofisica de Canarias, Via L\'{a}ctea s/n, 38200 La Laguna S.C., Tenerife, Spain\relax                                                                                                                                                                                                 \label{inst:0288}
\and Department of Astrophysics, University of La Laguna, Via L\'{a}ctea s/n, 38200 La Laguna S.C., Tenerife, Spain\relax                                                                                                                                                                                        \label{inst:0289}
\and Laboratoire Univers et Particules de Montpellier, CNRS Universit\'{e} Montpellier, Place Eug\`{e}ne Bataillon, CC72, 34095 Montpellier Cedex 05, France\relax                                                                                                                                               \label{inst:0295}
\and LESIA, Observatoire de Paris, Universit\'{e} PSL, CNRS, Sorbonne Universit\'{e}, Universit\'{e} de Paris, 5 Place Jules Janssen, 92190 Meudon, France\relax                                                                                                                                                 \label{inst:0301}
\and Villanova University, Department of Astrophysics and Planetary Science, 800 E Lancaster Avenue, Villanova PA 19085, USA\relax                                                                                                                                                                               \label{inst:0303}
\and Astronomical Observatory, University of Warsaw,  Al. Ujazdowskie 4, 00-478 Warszawa, Poland\relax                                                                                                                                                                                                           \label{inst:0309}
\vfill\break
\and Laboratoire d'astrophysique de Bordeaux, Univ. Bordeaux, CNRS, B18N, all\'{e}e Geoffroy Saint-Hilaire, 33615 Pessac, France\relax                                                                                                                                                                           \label{inst:0313}
\and Universit\'{e} Rennes, CNRS, IPR (Institut de Physique de Rennes) - UMR 6251, 35000 Rennes, France\relax                                                                                                                                                                                                    \label{inst:0316}
\and Niels Bohr Institute, University of Copenhagen, Juliane Maries Vej 30, 2100 Copenhagen {\O}, Denmark\relax                                                                                                                                                                                                  \label{inst:0324}
\and Las Cumbres Observatory, 6740 Cortona Drive Suite 102, Goleta, CA 93117, USA\relax                                                                                                                                                                                                                          \label{inst:0325}
\and Astrophysics Research Institute, Liverpool John Moores University, 146 Brownlow Hill, Liverpool L3 5RF, United Kingdom\relax                                                                                                                                                                                \label{inst:0331}
\and IPAC, Mail Code 100-22, California Institute of Technology, 1200 E. California Blvd., Pasadena, CA 91125, USA\relax                                                                                                                                                                                         \label{inst:0336}
\and Jet Propulsion Laboratory, California Institute of Technology, 4800 Oak Grove Drive, M/S 169-327, Pasadena, CA 91109, USA\relax                                                                                                                                                                             \label{inst:0337}
\and IRAP, Universit\'{e} de Toulouse, CNRS, UPS, CNES, 9 Av. colonel Roche, BP 44346, 31028 Toulouse Cedex 4, France\relax                                                                                                                                                                                      \label{inst:0338}
\and Konkoly Observatory, Research Centre for Astronomy and Earth Sciences, MTA Centre of Excellence, Konkoly Thege Mikl\'{o}s \'{u}t 15-17, 1121 Budapest, Hungary\relax                                                                                                                                        \label{inst:0351}
\and MTA CSFK Lend\"{ u}let Near-Field Cosmology Research Group\relax                                                                                                                                                                                                                                            \label{inst:0352}
\and ELTE E\"{ o}tv\"{ o}s Lor\'{a}nd University, Institute of Physics, 1117, P\'{a}zm\'{a}ny P\'{e}ter s\'{e}t\'{a}ny 1A, Budapest, Hungary\relax                                                                                                                                                               \label{inst:0353}
\and Ru{\dj}er Bo\v{s}kovi\'{c} Institute, Bijeni\v{c}ka cesta 54, 10000 Zagreb, Croatia\relax                                                                                                                                                                                                                   \label{inst:0368}
\and Institute of Theoretical Physics, Faculty of Mathematics and Physics, Charles University in Prague, Czech Republic\relax                                                                                                                                                                                    \label{inst:0372}
\and INAF - Osservatorio Astronomico di Brera, via E. Bianchi 46, 23807 Merate (LC), Italy\relax                                                                                                                                                                                                                 \label{inst:0383}
\and AKKA for CNES Centre Spatial de Toulouse, 18 avenue Edouard Belin, 31401 Toulouse Cedex 9, France\relax                                                                                                                                                                                                     \label{inst:0384}
\and Departmento de F\'{i}sica de la Tierra y Astrof\'{i}sica, Universidad Complutense de Madrid, 28040 Madrid, Spain\relax                                                                                                                                                                                      \label{inst:0388}
\and Vitrociset Belgium for European Space Agency (ESA), Camino bajo del Castillo, s/n, Urbanizacion Villafranca del Castillo, Villanueva de la Ca\~{n}ada, 28692 Madrid, Spain\relax                                                                                                                            \label{inst:0394}
\and Department of Astrophysical Sciences, 4 Ivy Lane, Princeton University, Princeton NJ 08544, USA\relax                                                                                                                                                                                                       \label{inst:0423}
\and Departamento de Astrof\'{i}sica, Centro de Astrobiolog\'{i}a (CSIC-INTA), ESA-ESAC. Camino Bajo del Castillo s/n. 28692 Villanueva de la Ca\~{n}ada, Madrid, Spain\relax                                                                                                                                    \label{inst:0429}
\and naXys, University of Namur, Rempart de la Vierge, 5000 Namur, Belgium\relax                                                                                                                                                                                                                                 \label{inst:0432}
\and EPFL - Ecole Polytechnique f\'{e}d\'{e}rale de Lausanne, Institute of Mathematics, Station 8 EPFL SB MATH SDS, Lausanne, Switzerland\relax                                                                                                                                                                  \label{inst:0440}
\and H H Wills Physics Laboratory, University of Bristol, Tyndall Avenue, Bristol BS8 1TL, United Kingdom\relax                                                                                                                                                                                                  \label{inst:0446}
\and Sorbonne Universit\'{e}, CNRS, UMR7095, Institut d'Astrophysique de Paris, 98bis bd. Arago, 75014 Paris, France\relax                                                                                                                                                                                       \label{inst:0467}
\and Porter School of the Environment and Earth Sciences, Tel Aviv University, Tel Aviv 6997801, Israel\relax                                                                                                                                                                                                    \label{inst:0468}
\and Laboratoire Univers et Particules de Montpellier, Universit\'{e} Montpellier, Place Eug\`{e}ne Bataillon, CC72, 34095 Montpellier Cedex 05, France\relax                                                                                                                                                    \label{inst:0469}
\and Faculty of Mathematics and Physics, University of Ljubljana, Jadranska ulica 19, 1000 Ljubljana, Slovenia\relax                                                                                                                                                                                             \label{inst:0470}
}

\date{Received date /
Accepted date}

%======================================================================================
% Abstract
%======================================================================================
\abstract{
  % context (optional)
  This work is part of the \gaia Data Processing and Analysis Consortium (DPAC) 
        papers published with the \gaia Early Data Release 3 (EDR3). 
        It is one of the demonstration papers aiming to highlight the improvements 
        and quality of the newly published data by applying them to a scientific case.
}{
  % aims        
  We use the \egdr{3} data to study the structure and kinematics of the
        Magellanic Clouds. The large distance to the Clouds is a challenge for the
        \gaia astrometry. The Clouds lie at the very limits of the usability of the 
				\textit{Gaia} data, which makes the Clouds an excellent
        case study for evaluating the quality and properties of the \textit{Gaia} data.
}{
  % methods
  The basis of our work are two samples selected to provide a representation as clean as possible
         of the stars of the Large Magellanic Cloud (LMC) and the Small Magellanic
        Cloud (SMC). The selection used criteria based on position, parallax, and proper motions
        to remove foreground contamination from the Milky Way, and allowed the separation of the 
        stars of both Clouds. From these two samples we defined a series of subsamples
        based on cuts in the colour-magnitude diagram; these subsamples were used to select
        stars in a common evolutionary phase and can also be used as approximate proxies of
        a selection by age.
}{
  % results
        We compared the \gaia Data Release 2 (DR2) and \egdr{3} performances in the study 
        of the Magellanic Clouds and show the clear improvements in precision and 
        accuracy in the new release. We also show that the systematics still present 
        in the data make the determination of the 3D geometry of the LMC a difficult 
        endeavour; this is at the very limit of the usefulness of the \egdr{3} astrometry, 
        but it may become feasible with the use of additional   external data. 
        
        We derive radial and tangential velocity maps and global profiles for the LMC
        for the several subsamples we defined. To our knowledge,
        this is the first time that the two planar components of the ordered and 
  random motions are derived for multiple stellar evolutionary phases in a galactic 
        disc outside the Milky Way, showing the differences between younger and older
        phases. We also analyse the spatial structure and motions in the central region, 
        the bar, and the disc,   providing new insights into features and kinematics. 

        Finally, we show that the \egdr{3} data allows clearly resolving the
        Magellanic Bridge, and we trace the density and velocity flow of
        the stars from the SMC towards the LMC not only globally, but also
        separately for young and evolved populations. This allows us to confirm an evolved population in the Bridge that is slightly shift from the 
        younger population. Additionally, we were able to study the outskirts of both Magellanic
        Clouds, in which we detected some well-known features and indications of new ones.
}
{} % Conclusions (optional, not included)


\titlerunning{Structure and properties of the Magellanic Clouds}
\authorrunning{Gaia Collaboration, Luri et al.}
\maketitle

%======================================================================================
% Introduction
%======================================================================================
\section{Introduction}

This paper takes advantage and highlights the improvements from \gaia Data Release 2 (DR2)
to \gaia Early Data Release 3 (EDR3) in the context of astrometry, photometry, and completeness in the 
Magellanic Cloud sky area. A previous \gdrtwo science-demonstration paper on 
dwarf galaxies \cite{2018A&A...616A..12G} 
only scratched the surface of what \gaia can tell us about these objects; it 
only considered their basic parameters, and barely used the photometry. Here we demonstrate 
how much more \egdr{3} shows us compared to \gdrtwo, thus demonstrating the value added by 
this new data release. A summary of the contents and survey properties of the 
\egdr{3} release can be found in \cite{EDR3-DPACP-130}, and a general description
of the \gaia mission can be found in \cite{DR1-DPACP-18}. Specifically, as described 
in \cite{EDR3-DPACP-130}, we use:
%\LEt{you include four lists that need to be rephrased so they become proper text. In addition to this list here, they are in Sect. 2.2. (2 numbered lists)  and two lists in Sect. 5.1. with bullet points. You need to form proper sentences and a paragraph}  

\begin{itemize}
        \item A reduction of a factor 2 in the proper motion uncertainty.
        \item A new transit cross-match that provides a significant improvement in crowded 
              areas and increases completeness.
        \item 33 months of data significantly reduce the \gaia scanning-law effects observed
              in \gdrtwo when means and medians of parallaxes and proper motions are computed
        \item New photometry, with reduced systematic effects, that is less affected by 
              crowding effects in the centre of the clouds (see \figref{fig:dr2dr3_phot_excess}). 
							This helps us to unveil different stellar populations in the area of the Magellanic Clouds.
\end{itemize}

In \secref{sec:DR2DR3} we provide an analysis of the improvements since \gdrtwo in \egdr{3}.
In \secref{sec:samples} we define the samples we use throughout the paper. We start
by selecting objects in a radius around the centre of each cloud, and then we filter the 
objects using parallax, proper motions, and $G$ magnitude. The result is two clean
samples, one for the Large Magellanic Cloud (LMC) and one for the Small Magellanic 
Cloud (SMC). They constitute the baseline for our work. By selecting objects
based on their position in the $(G,\gbp-\grp)$ diagram, we then further split these
samples into a set of evolutionary phase subsamples that can be used as a proxy
for age selection.

In \secref{sec:DR2DR3} we compare \gdrtwo and \egdr{3}
using the LMC and SMC samples. We compare the parallax and proper motion
fields and show that the systematics and noise are significantly reduced. 
We also show that the photometry has improved by comparing the excess flux. 

In \secref{sec:Spatial_Structure} we use the \egdr{3} astrometry
to resolve the 3D structure of the LMC by modelling it as a disc. We determine its parameters using a Bayesian approach. We show that the \egdr{3} 
level of parallax systematics (essentially the zero-point variations), 
combined with the parallax uncertainties for a distant object such as the LMC, 
place this determination at the very limit of feasibility. We do not reach a satisfactory 
result, but we conclude that it might be possible with \egdr{3} combined with 
external data, and certainly with future releases, in which the  systematics and uncertainties will be reduced. 

In \secref{sec:kinematics} we study the kinematics of the LMC in detail.
We analyse the general kinematic trends and consider the velocity 
profiles across the disc in detail, focusing on the separation of the rotation
velocities as a function of the evolutionary stage.

In \secref{sec:Bridge} we study the outskirts of the two Magellanic Clouds,
and we specifically focus on one of its more prominent features: the
Magellanic Bridge, a structure joining the Magellanic Clouds that formed as a result of 
tidal forces that stripped gas and stars from the SMC towards the LMC. We show
that using \egdr{3} data, the Bridge becomes apparent without the need
of sophisticated statistical treatment, and we can determine its
velocity field and study it for different stellar populations.

In \secref{sec:Spiral_structure} we study the structure and kinematics
of the spiral arms of the LMC using samples of different evolutionary phases, 
so that we can compare its outline as it becomes visible through different types of objects. 
We also study the streaming motions in the arms and produce radial 
velocity profiles for the different evolutionary phases.
In the appendices we finally compile a variety of additional material
based on \egdr{3} data.

%======================================================================================
% Section: samples
%======================================================================================
\section{Sample selection}\label{sec:samples}

We describe here the samples that we used in this paper. The selection 
was made in three steps that we describe below.
First, we applied a spatial selection (radius around a predefined centre) to 
generate two base samples (LMC and SMC) in order to select objects in the 
general direction of the two clouds. Second, for each one of these samples, we 
introduced an additional selection to retain objects whose proper motions are compatible with the mean motion of each cloud. This second selection ensured 
that most of the contamination from foreground (Milky Way) objects was 
removed. Finally, we defined a set of eight subsets for each cloud based on 
the position in the colour-magnitude diagram (CMD) with the aim to produce groups of 
objects in similar evolutionary phases as a proxy of ages (see the discussion in 
\secref{sec:selection_evophase}). We did not apply the correction to $G$ magnitudes for sources with 6p solutions that was suggested in 
Section 7.2 of \cite{EDR3-DPACP-130}. The correction is small enough (around
0.01 mag) to not have relevant effects for the methods applied in this paper, 
and we verified that it only very marginally affects the composition of 
our samples (0.04\% or less of the sample size).

%%%%%%%%%%%%%%%%%%%%%%%%
\subsection{Spatial selection \label{sec:spatial_selection}}

\subsubsection{LMC}

The base sample for the LMC was obtained using a selection with a 20$^\circ$ radius 
around a centre defined as 
$(\alpha,\delta)=(81.28^\circ,-69.78^\circ)$ \cite{vanderMarel2001} and a 
limiting $G$ magnitude of $20.5$. This selection can be reproduced using the following ADQL query in the Gaia archive:
\vspace{0.25cm}

\noindent {\tt \small
                SELECT * FROM user\_edr3int4.gaia\_source as g   \newline
                WHERE 1=CONTAINS(POINT('ICRS',g.ra,g.dec),       \newline
                ~\hspace{1.5cm} CIRCLE('ICRS',81.28,-69.78,20)) \newline
								AND g.parallax IS NOT NULL \newline
                AND g.phot\_g\_mean\_mag < 20.5 
}
\vspace{0.25cm}

The resulting sample contains $27,231,400$ objects. The large selection 
radius causes the selection to include part of the SMC, as is shown in 
\figref{fig:clean_SkyPlots}. The purpose of such a large selection area was 
to ensure the inclusion of the outer parts of the LMC and the regions
where the LMC-SMC bridge is located. 

\subsubsection{SMC}

The base sample for the SMC was obtained using a selection with an 11$^\circ$ radius 
around a centre defined as
$(\alpha,\delta)=(12.80^\circ, -73.15^\circ)$ \cite{Cioni2000} and a limiting 
$G$ magnitude of $20.5$. This selection can be reproduced using the following 
ADQL query in the Gaia archive:
\vspace{0.25cm}

\noindent {\tt \small
                SELECT * FROM user\_edr3int4.gaia\_source as g   \newline
                WHERE 1=CONTAINS(POINT('ICRS',g.ra,g.dec),       \newline
                ~\hspace{1.5cm} CIRCLE('ICRS',12.80,-73.15,11)) \newline
								AND g.parallax IS NOT NULL \newline
                AND g.phot\_g\_mean\_mag < 20.5
}
\vspace{0.25cm}

The resulting sample contains $4,709,622$ objects.

%%%%%%%%%%%%%%%%%%%%%%%%
\subsection{Proper motion selection \label{sec:pm_selection}}

Starting from the base samples described above, we followed the procedure 
described in \citet{2018A&A...616A..12G}
to remove foreground (Milky Way) contamination of objects based on proper 
motion selection. For the proper motions
to be relatively easy to interpret in terms of internal velocities, we 
defined an orthographic projection, 
$\{\alpha,\delta,\mu_{\alpha^*},\mu_\delta\} \rightarrow \{x,y,\mu_x,\mu_y\}$ 
(see Eqn. 2 from \cite{2018A&A...616A..12G} and also \secref{sec:DR2DR3}). 
To determine the proper motions of the LMC and SMC and build the 
filters that lead to the clean samples of both clouds, we then used the following procedure.
First, we computed a robust estimate of the  proper motions of the clouds by:

\begin{enumerate}

   \item We retained objects with $\sqrt{x^2+y^2}<\sin{r_{\text{sel}}}$, where $r_{\text{sel}}$ is $5$ 
               deg for the LMC and $1.5$ deg for the SMC.
                                
   \item We minimised the foreground contamination by selecting stars with $\varpi/\sigma_{\varpi} < 5$. 
         This parallax cut excludes solutions that are not compatible with being distant enough to be part of 
         the LMC or SMC, and therefore possible foreground contamination from Milky Way stars. This filter 
				 was kept for the final clean samples, as described below. 
                                
   \item We also introduced a magnitude limit $G < 19$. This limit aims to remove
         the less precise astrometry from the estimation of proper motions, and was relaxed to build
         the final  clean samples, as described below.
                                                                
   \item We then computed median values for $\mu_x$ and $\mu_y$ with the above selection 
         $(\mu_{x,\mathrm{med}},\mu_{y,\mathrm{med}})$. These values are our reference 
				 for the typical LMC and SMC proper motions in the orthographic plane.
         Using these values, we determined the covariance matrix of the proper motion distribution 
         ($\Sigma_{\mu_x,\mu_y}$).
                                
   \item We retained only stars with proper motions within $\vec\mu'^T\Sigma^{-1}\vec\mu' < 9.21$, 
         where $\vec{\mu'} =(\mu_x-\mu_{x,\mathrm{med}}, \mu_y-\mu_{y,\mathrm{med}})$. 
         This corresponds to a $99\%$ confidence region. For simplicity, we did not take 
         the covariance matrix of individual stars into account. The aim was simply to remove clear foreground
         objects, and we considered the given formulation just an approximation, but sufficient for this purpose.

    \item We determined the median parallax of this sample, $\varpi_{\mathrm{med}}$, and for each star in our 
          full sample, we determined the proper motion 
					conditional on $\varpi_{\mathrm{med}}$ being the true parallax of the star, taking the relevant 
					uncertainties $\sigma$ and correlations $\rho$ into account. for example, $\hat\mu_{\alpha*}= \mu_{\alpha*} - 
          (\varpi-\overline\varpi)\rho_{\mu_\alpha*\varpi}\sigma_{\mu_\alpha*}/\sigma_\varpi$. 

    \item We computed new $\mu_x,\mu_y$ from $\hat\mu_{\alpha*}, \hat\mu_\delta$. We used these to 
          repeat steps 1-4 to derive a final estimate of 
          $\mu_{x,\mathrm{med}},\mu_{y,\mathrm{med}}$ , and $\Sigma_{\mu_x,\mu_y}$.

\end{enumerate}
\vspace{0.25cm}

\noindent Using these results, we applied the following two conditions to the base samples defined in the previous section:

\begin{enumerate}
                                
    \item We retained only stars with proper motions within $\vec\mu'^T\,\Sigma_{\mu_x,\mu_y}^{-1}\,\vec\mu' < 9.21$.
                                        
    \item As before, we selected only stars with $\varpi/\sigma_{\varpi} < 5$ to minimise any remaining foreground 
                      contamination, but now we set a fainter magnitude limit, $G < 20.5$.
                                        
\end{enumerate}

\begin{figure*}[h]
   \begin{subfigure}[b]{0.5\textwidth} 
      \begin{center}
      \includegraphics[width=\columnwidth]{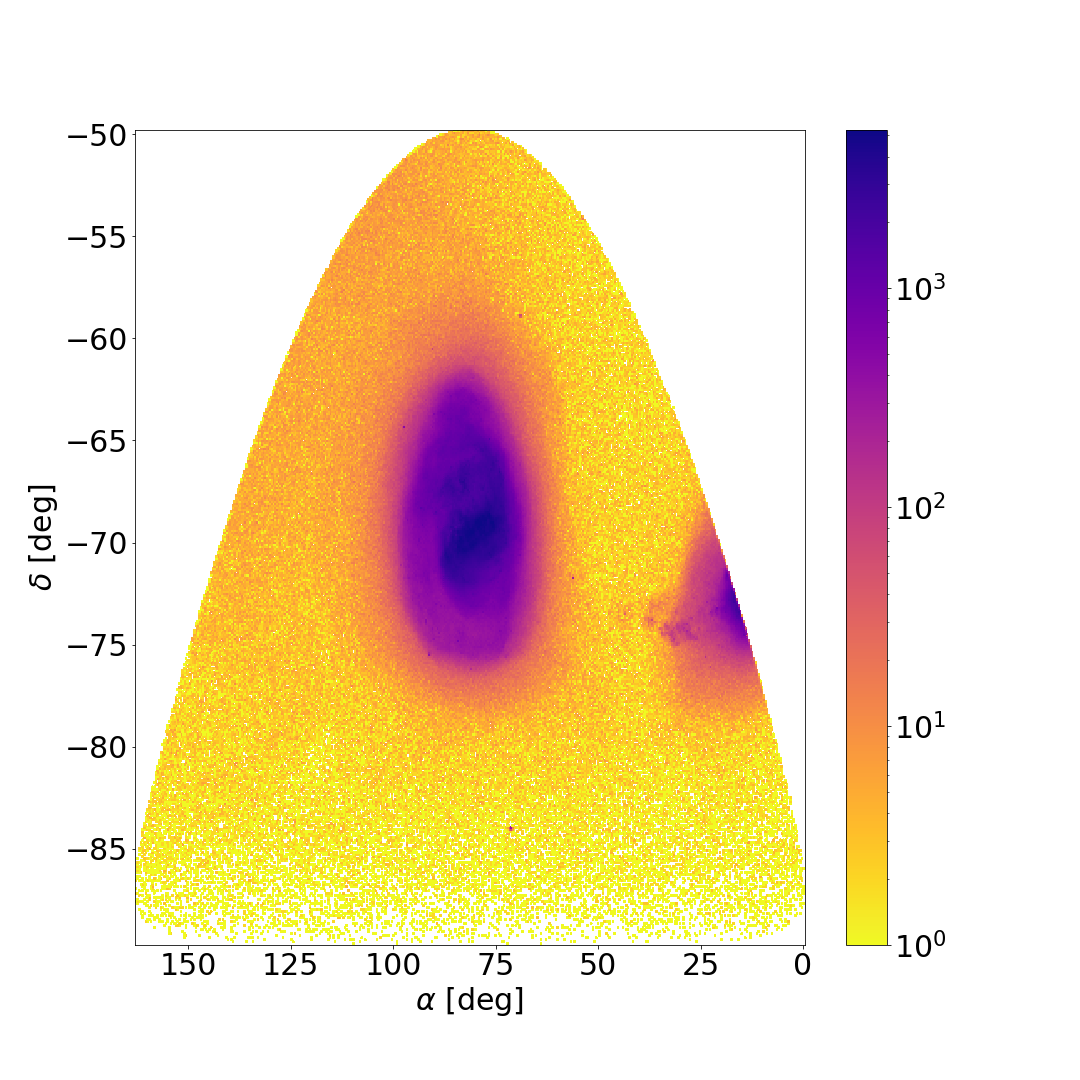}
      \end{center}
         \end{subfigure}
   \begin{subfigure}[b]{0.5\textwidth} 
      \begin{center}
      \includegraphics[width=\columnwidth]{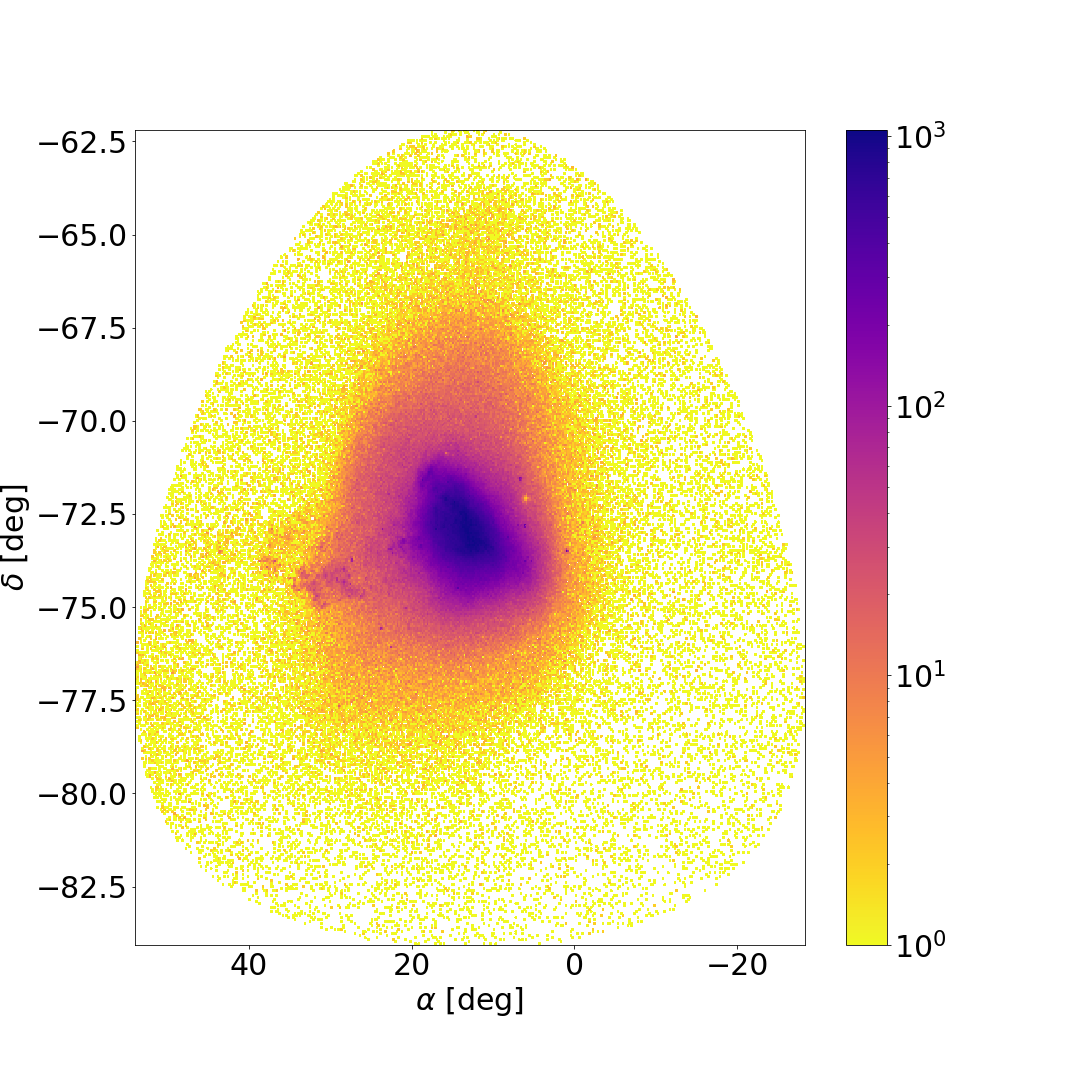}
      \end{center}
         \end{subfigure}
   \begin{subfigure}[b]{0.5\textwidth} 
      \begin{center}
      \includegraphics[width=\columnwidth]{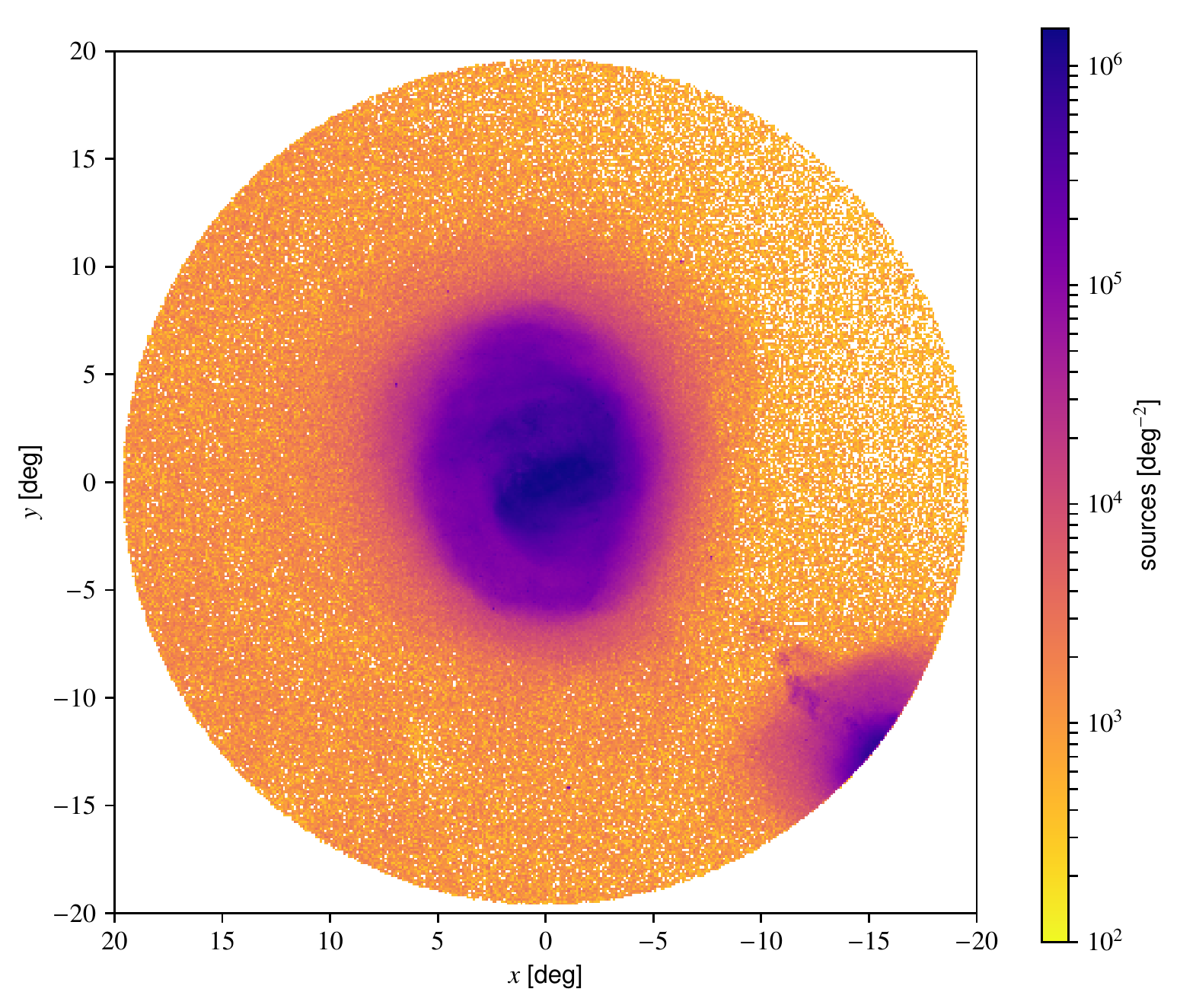}
      \end{center}
         \end{subfigure}
   \begin{subfigure}[b]{0.5\textwidth} 
      \begin{center}
      \includegraphics[width=\columnwidth]{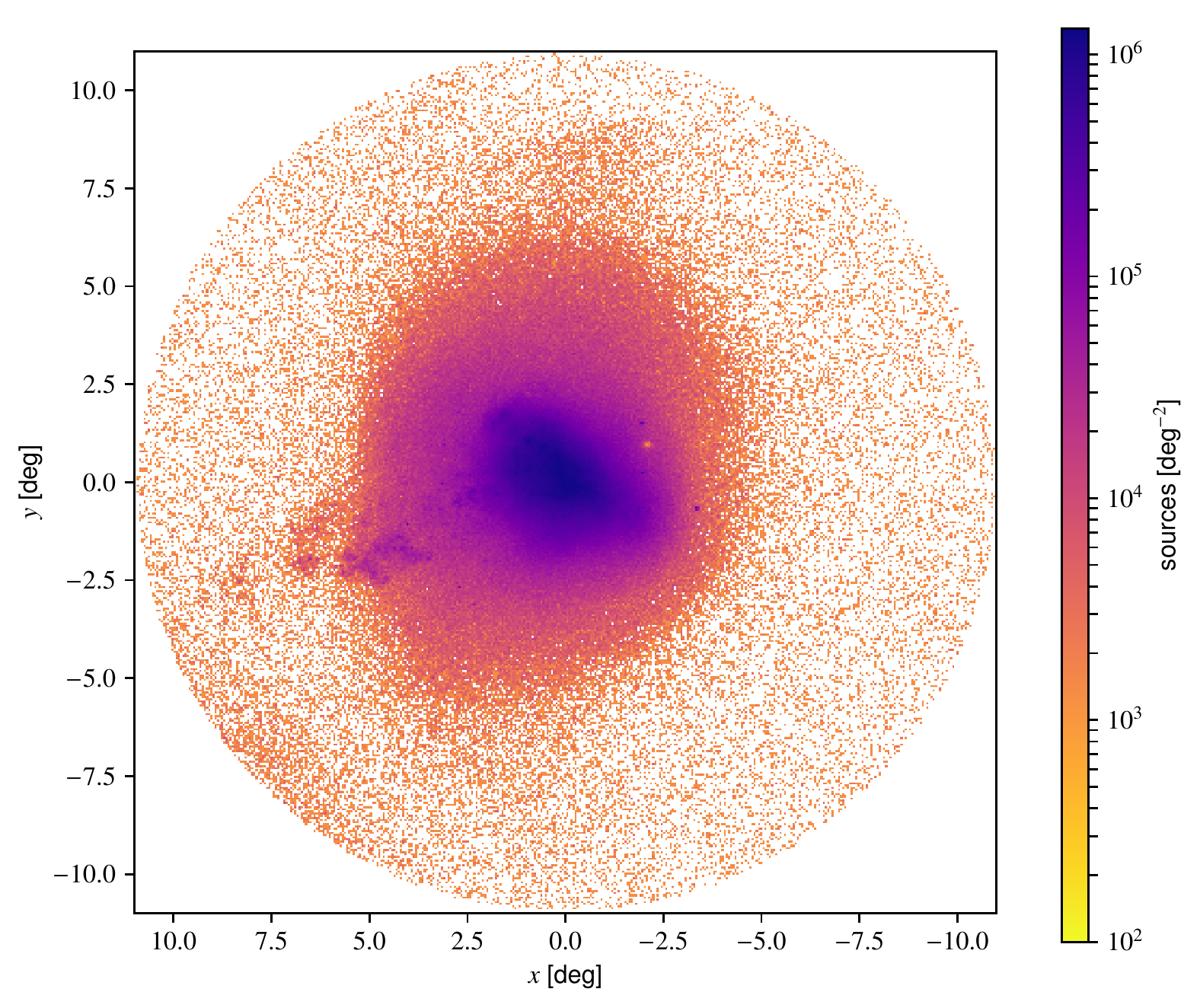}
      \end{center}
         \end{subfigure}
   \caption{Sky density plots for the LMC (left) and SMC (right) clean samples (after spatial and proper motion selection).
                  Top row: Plots in equatorial coordinates. Bottom row: Orthographic projection (as used in \secref{sec:DR2DR3}) }
         \label{fig:clean_SkyPlots}
\end{figure*}

The resulting clean sample for the LMC contains a total of $11,156,431$ objects, and the sample for the SMC 
contains $1,728,303$ objects; their distribution in the sky is depicted in \figref{fig:clean_SkyPlots} and the 
mean astrometry is presented in \tabref{tab:Clouds_astrometry_stats}. The mean parallaxes of both objects are negative, 
while the expected values would be 
$\varpi_{LMC} \simeq \frac{1}{49.5 kpc} = 0.0202\text{mas}$ \citep{Pietrzynski2019}
and
$\varpi_{SMC} \simeq \frac{1}{62.8 kpc} = 0.0159\text{mas}$ \citep{2000A&A...359..601C}.
This is due to the zero-point offset in the \gaia parallaxes that was discussed in \cite{EDR3-DPACP-132};
using the values in this paper, the (rough) estimates of the LMC ($-0.0242\text{mas}$) and  
SMC ($-0.0185\text{mas}$) zero-points are in line with a global value of $-0.020\text{mas}$, as 
discussed in Sec. 4.2 of \cite{EDR3-DPACP-132}.

\vspace{0.25cm}

\begin{table*}[h]
        \centering
                \begin{tabular}{|l|r|r|r|r|r|r|}
                   \hline
         & \multicolumn{1}{c|}{$\overline{\varpi}$}     & 
                                   \multicolumn{1}{c|}{$\sigma_\varpi$} & 
                                   \multicolumn{1}{c|}{$\overline{\mu_\alpha*}$} & 
                                         \multicolumn{1}{c|}{$\sigma_{\mu_\alpha*}$} & 
                                         \multicolumn{1}{c|}{$\overline{\mu_\delta}$}  & 
                                         \multicolumn{1}{c|}{$\sigma_{\mu_\delta}$} \\
                   \hline
                          {\bf LMC} & {\bf -0.0040} & {\bf 0.3346} & {\bf 1.7608} & {\bf 0.4472} & {\bf 0.3038} & {\bf 0.6375} \\
                            Young 1 & -0.0049 & 0.0729 & 1.7005 & 0.2700 &  0.2073 & 0.4733 \\                       Young 2 &  0.0058 & 0.1154 & 1.7376 & 0.3260 &  0.2083 & 0.5067 \\                       Young 3 & -0.0095 & 0.4245 & 1.7491 & 0.4814 &  0.2859 & 0.6586 \\                            RGB     & -0.0010 & 0.3239 & 1.7690 & 0.4372 &  0.3255 & 0.6344 \\                       AGB     & -0.0164 & 0.0414 & 1.8387 & 0.2686 &  0.3217 & 0.4486 \\                       RRL     & -0.0046 & 0.3201 & 1.7698 & 0.4818 &  0.2947 & 0.6742 \\                       BL      &  0.0047 & 0.1341 & 1.7103 & 0.3996 &  0.2852 & 0.6260 \\                       RC      & -0.0050 & 0.2314 & 1.7719 & 0.4167 &  0.3093 & 0.6113 \\              \hline
                                {\bf SMC} & {\bf -0.0026} & {\bf 0.3273} & {\bf 0.7321} & {\bf 0.3728} & {\bf -1.2256} & {\bf 0.2992} \\
                            Young 1 & -0.0099 & 0.0995 & 0.7754 & 0.2495 & -1.2560 & 0.1195 \\                       Young 2 &  0.0036 & 0.1585 & 0.7708 & 0.2981 & -1.2555 & 0.1951 \\                       Young 3 & -0.0012 & 0.4382 & 0.7721 & 0.4224 & -1.2336 & 0.3472 \\                            RGB     & -0.0034 & 0.3244 & 0.7106 & 0.3593 & -1.2183 & 0.2883 \\                       AGB     & -0.0145 & 0.0545 & 0.7267 & 0.2247 & -1.2432 & 0.1222 \\                       RRL     & -0.0028 & 0.4196 & 0.7372 & 0.4368 & -1.2214 & 0.3637 \\                       BL      & -0.0080 & 0.1401 & 0.7647 & 0.2907 & -1.2416 & 0.2070 \\                       RC      & -0.0050 & 0.2576 & 0.7130 & 0.3572 & -1.2196 & 0.2890 \\              \hline
                \end{tabular}
        \caption{Mean astrometry of the LMC and SMC clean (after spatial and proper motion selection) samples 
                 and the evolutionary phase subsamples extracted from them.
                 Parallax is in mas and proper motions in $\text{mas}\,\text{yr}^{-1}$.
                                         As discussed in the text, the negative mean parallaxes arise because 
                                         zero-point parallax corrections were not applied.}
        \label{tab:Clouds_astrometry_stats}
\end{table*}

%%%%%%%%%%%%%%%%%%%%%%%%
\subsection{Evolutionary phase subsamples \label{sec:selection_evophase}}

The two samples obtained following the procedure outlined in the two previous sections constitute our 
basic selection of objects for the LMC and SMC, our clean samples for the stars of the clouds. 
These were used for analysis of the LMC and SMC as a whole. A selection of basic statistics and 
maps using these samples ispresented in Appendix \ref{sec:appendix}.

Several cases required a definition of subsamples that were adequate for the study of different substructures 
of the clouds (disc, halo, etc.), however. Ideally, we would like to select these subsamples by age,
but this would require either generating our own age estimates or a cross-match with external 
catalogues, which is beyond the scope of a \egdr{3} demonstration paper such as this. Instead, we used 
a different approach, using a selection of samples based on the CMD of the clouds.
We defined cut-outs in the shape of polygonal regions in the $(G,G_{BP}-G_{RP})$ diagram to select 
the following target evolutionary phases:

\begin{description}

  \item[Young 1:] very young main sequence (ages $< 50$ Myr)
  \item[Young 2:] young main sequence ($~50 < $ age $< ~400$ Myr)
  \item[Young 3:] intermediate-age main-sequence population (mixed ages reaching up $1-2$ Gyr)
  \item[RGB:] red giant branch
  \item[AGB:] asymptotic giant branch (including long-period variables)
  \item[RRL:] RR-Lyrae region of the diagram
  \item[BL:] blue loop (including classical Cepheids)
  \item[RC:] red clump

\end{description}

The defined areas are shown in \figref{fig:EvoPhases_CMD}. There 
are unassigned areas in the CMD diagrams: this is on purpose because these unassigned areas are too mixed, affected by blended stars, or too contaminated 
by foreground (Milky Way) stars. The areas are exclusive, that is, they do not overlap.

\begin{figure}[h]
   \begin{center}
     \includegraphics[width=\columnwidth]{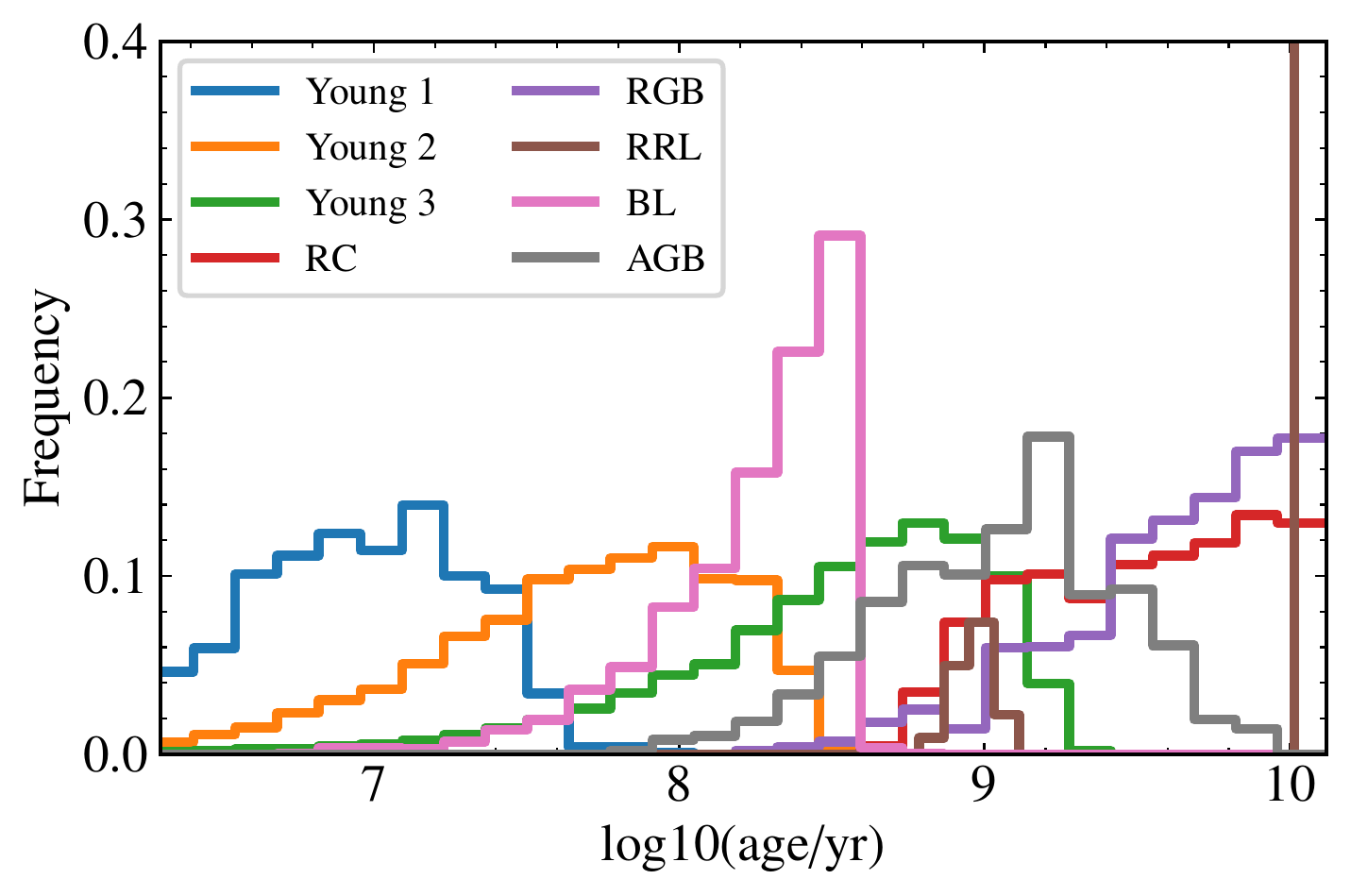}
   \end{center}
   \caption{Estimated age distribution of the selected evolutionary phase. 
                  Based on a simulation using a constant star formation rate, the age-metallicity 
                                                relation by Harris \& Zaritsky (2009), and PARSEC1.2 models}
         \label{fig:Subsample_Ages}
\end{figure}

This rather raw selection is not even corrected for reddening, but to some 
extent, it can be used as an age-selected proxy. Based on a simulation using 
a constant star formation rate, the age-metallicity relation by Harris \& Zaritsky (2009),  
and PARSEC1.2 models, the estimated age distribution of the 
resulting subsamples is shown in \figref{fig:Subsample_Ages}. The 
figure shows that the resulting subsamples indeed have different age 
distributions that suffice for the purposes of this demonstration paper. 
For the sake of brevity, we refer to these 
subsamples as ``evolutionary phases''.

\subsubsection{LMC evolutionary phases}

The polygons in the CMD diagram defining the LMC subsamples are as follows, 
and they are represented in \figref{fig:EvoPhases_CMD} (left panel):

\begin{description}

  \item[Young 1:]     [0.18, 16.0],
                     [-0.3, 10.0],
                     [-1.0, 10.0],
                     [-1.0, 16.0],
                     [0.18, 16.0]
                                                                                
  \item[Young 2:     [-1.0, 16.0],
                     [0.18, 16.0],
                     [0.34, 18.0],
                     [-1.0, 18.0],
                     [-1.0, 16.0]
                                                                                
  \item[Young 3:]    [-0.40, 20.5],
                     [-0.6, 19.0],
                     [-0.6, 18.0],
                     [0.34, 18.0],
                     [0.40, 18.9],
                     [0.45, 19.5],
                     [0.70, 20.5],
                     [-0.40, 20.5]
                                                                                
  %\item[Cont:]       [-1.0, 20.5],
                     %[-0.4, 20.5],
                     %[-0.6, 19.0],                    
                     %[-0.6, 18.0],                    
                     %[-1.0, 18.0],
                     %[-1.0, 20.5]
                                                                                
  \item[RGB:]      [0.80, 20.5],
                   [0.90, 19.5],
                   [1.60, 19.8],
                   [1.60, 19.0],
                   [1.05, 18.41],
                   [1.30, 16.56], 
                   [1.60, 15.3],
                   [2.40, 15.97],
                   [1.95, 17.75],
                   [1.85, 19.0],
                   [2.00, 20.5],
                   [0.80, 20.5]

  \item[AGB:]      [1.6, 15.3],
                   [1.92, 13.9],
                   [3.5, 15.0],
                   [3.5, 16.9],
                   [1.6, 15.3]  
                                                                                        
  \item[RRL:]      [0.45, 19.5],
                   [0.40, 18.9],
                   [0.90, 18.9],
                   [0.90, 19.5],
                   [0.45, 19.5]
                                                                        
  \item[BL:]       [0.90, 18.25],
                   [0.1, 15.00],
                   [-0.30, 10.0],
                   [2.85, 10.0],
                   [1.30, 16.56], 
                   [1.05, 18.41],
                   [0.90, 18.25] 
                                                                        
  \item[RC:]       [0.90, 19.5],
                   [0.90, 18.25],
                   [1.60, 19.0],
                   [1.60, 19.8],
                   [0.90, 19.5]
                                                                                        
\end{description}

The number of objects per subsample is listed in \tabref{tab:LMC_EvoPhases_Count}. The sky
distribution of the stars in the samples is shown in \figref{fig:LMC_SkyPlots}.

\begin{table}[h]
        \centering
                \begin{tabular}{|l|r|}
                   \hline
       {\bf Total objects LMC}  & {\bf 11,156,431} \\
                   \hline
       Young 1           &   23,869 \\
       Young 2           &   233,216 \\
       Young 3           &   3,514,579 \\
       %Cont              &   11,621 \\
       RGB               &   2,642,458 \\
       AGB               &   34,076 \\                  
       RRL               &   221,100 \\
       BL                &   261,929 \\
       RC                &   3,730,351 \\
                   \hline
                \end{tabular}
        \caption{Object counts of LMC evolutionary phases}
        \label{tab:LMC_EvoPhases_Count}
\end{table}

\subsubsection{SMC evolutionary phases}

\begin{figure*}[h]
   \begin{subfigure}[b]{0.5\textwidth} 
      \begin{center}
      \includegraphics[width=\columnwidth]{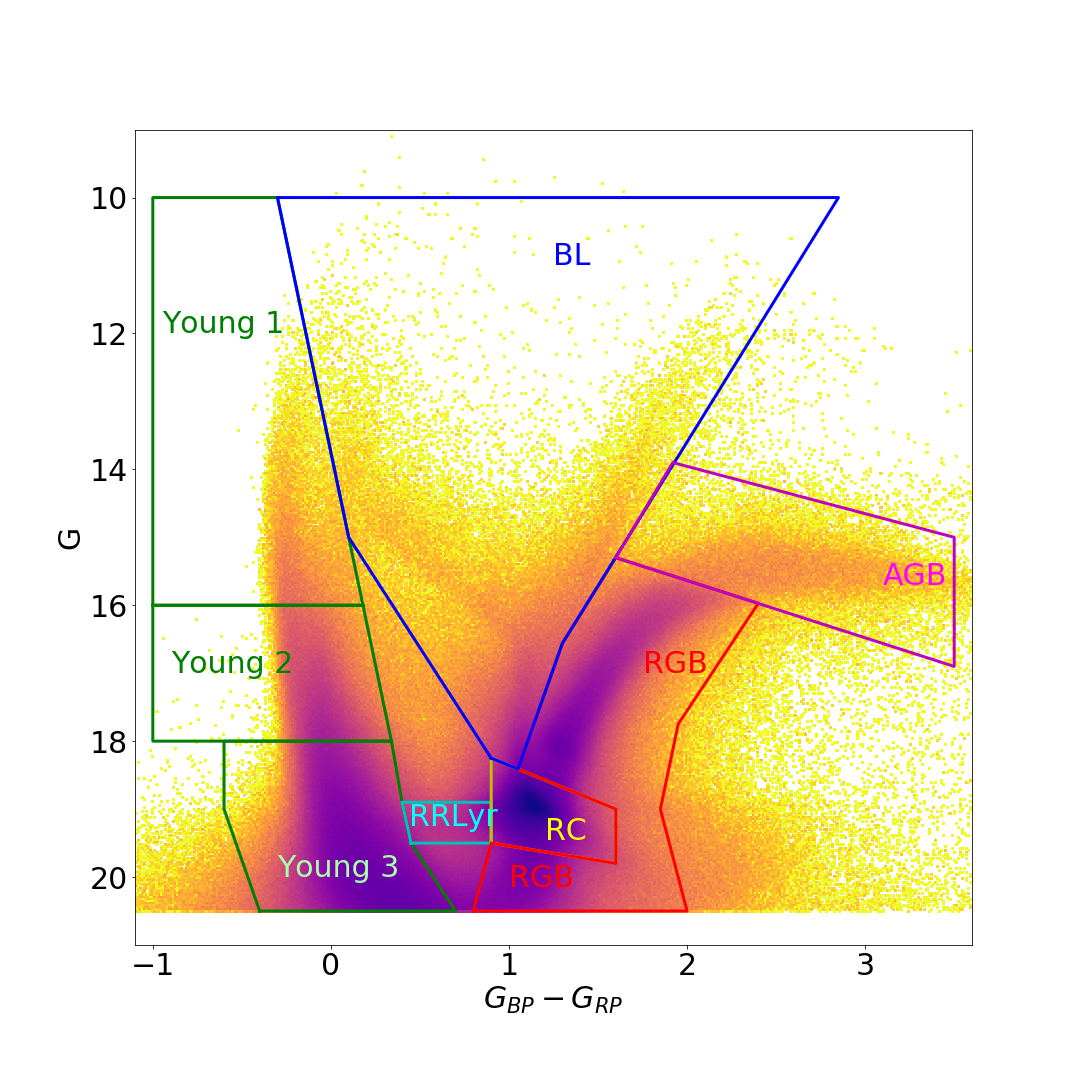}
      \end{center}
         \end{subfigure}
   \begin{subfigure}[b]{0.5\textwidth} 
      \begin{center}
      \includegraphics[width=\columnwidth]{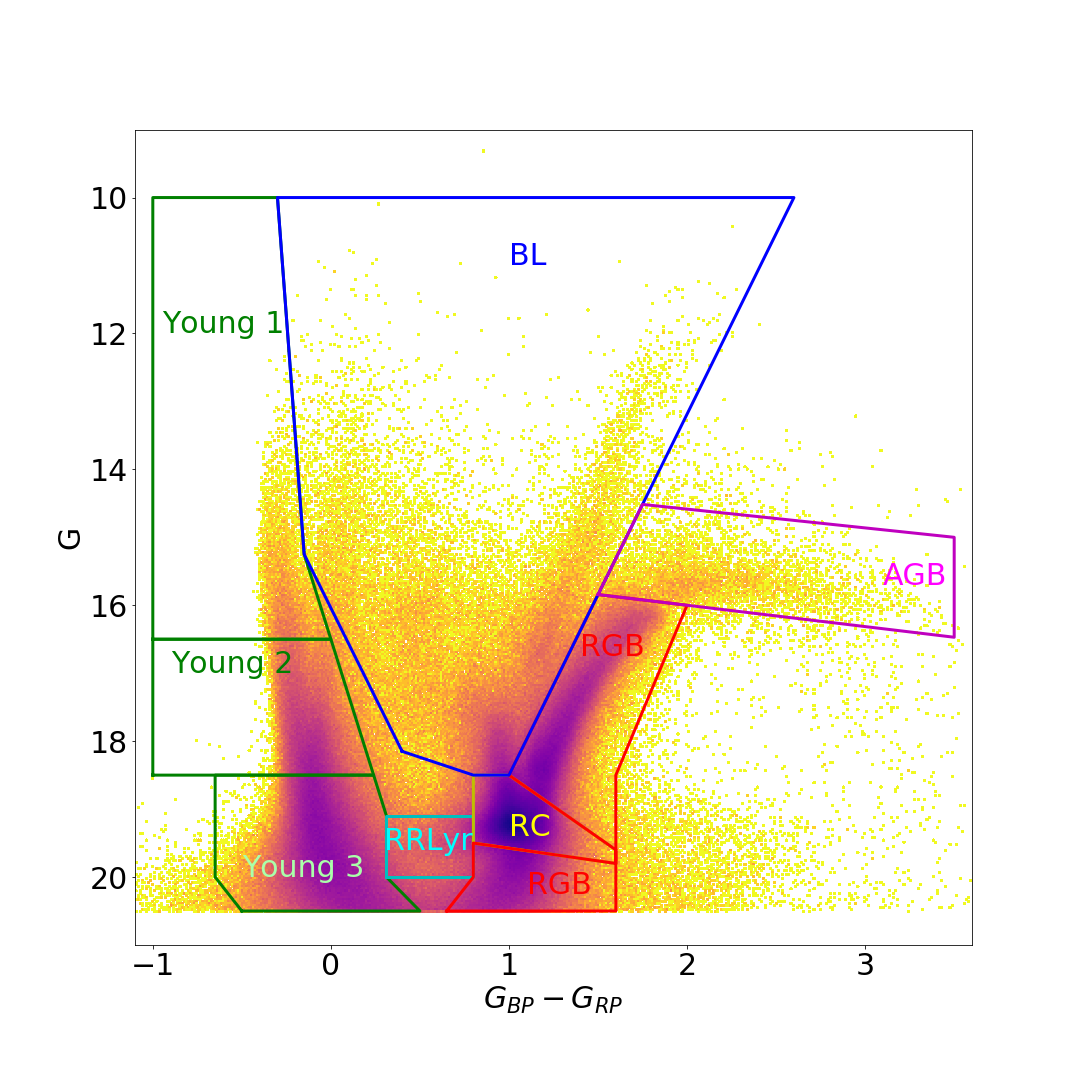}
      \end{center}
         \end{subfigure}
   \caption{Areas (as defined by the polygons given in the text) of the CMD for the LMC (left) and SMC (right) evolutionary phases. The colours are
                                                not corrected for reddening for the selection.}
         \label{fig:EvoPhases_CMD}
\end{figure*}

The polygons in the CMD diagram defining the SMC subsamples are as follows, 
and they are represented in \figref{fig:EvoPhases_CMD} (right panel):

\begin{description}

  \item[Young 1:]    [-1.00, 16.50],
                     [-1.00, 10.00],
                     [-0.30, 10.00],
                     [-0.15, 15.25],
                     [ 0.00, 16.50],
                     [-1.00, 16.50]
                                                                                
  \item[Young 2:]    [-1.00, 18.50],
                     [-1.00, 16.50],
                     [ 0.00, 16.50],
                     [ 0.24, 18.50],
                     [-1.00, 18.50]
                                                                                
  \item[Young 3:]    [-0.50, 20.50],
                     [-0.65, 20.00],
                     [-0.65, 18.50],
                     [ 0.24, 18.50],
                     [ 0.312, 19.10],
                     [ 0.312, 20.00],
                     [ 0.50, 20.50],
                     [-0.50, 20.50]

  %\item[Cont:]       [-1.00, 20.50],
                     %[-1.00, 18.50],
                     %[-0.65, 18.50],
                     %[-0.65, 20.00],
                     %[-0.50, 20.50],
                     %[-1.00, 20.50]
                                                                                
  \item[RGB:]      [0.65, 20.50],
                   [0.80, 20.00],
                   [0.80, 19.50],
                   [1.60, 19.80],
                   [1.60, 19.60],
                   [1.00, 18.50],
                   [1.50, 15.843],
                   [2.00, 16.00],
                   [1.60, 18.50],
                   [1.60, 20.50],
                   [0.65, 20.50]
                                                                        
  \item[AGB:]
                  [1.50, 15.843],
                  [1.75, 14.516],
                  [3.50, 15.00],
                  [3.50, 16.471],
                  [1.50, 15.843]
                                                                                
  \item[RRL:]      [ 0.312, 20.00],
                   [ 0.312, 19.10],
                   [ 0.80 , 19.10],
                   [ 0.80 , 20.00],
                   [ 0.312, 20.00]
                                                                        
  \item[BL:]       [0.40, 18.15],
                   [-0.15, 15.25],
                   [-0.3, 10.00],
                   [2.60, 10.00],
                   [1.00, 18.50],
                   [0.80, 18.50],
                   [0.40, 18.15]

  \item[RC:]       [0.80, 19.50],
                   [0.80, 18.50],
                   [1.00, 18.50],
                   [1.60, 19.60],
                   [1.60, 19.80],
                   [0.80, 19.50]

\end{description}

The number of objects per subsample is listed in \tabref{tab:SMC_EvoPhases_Count}. The sky
distribution of the stars in the samples is shown in \figref{fig:SMC_SkyPlots}.

\begin{table}[h]
        \centering
                \begin{tabular}{|l|r|}
                   \hline
       {\bf Total objects SMC}  & {\bf 1,728,303} \\
                   \hline
       Young 1            &   7,166 \\
       Young 2            &   83,417 \\
       Young 3            &   379,234 \\
       %Cont               &   1,175 \\
       RGB                &   448,948 \\
       AGB                &   5,887 \\  
       RRL                &   40,421 \\
       BL                 &   86,212 \\
       RC                 &   634,569 \\
                   \hline
                \end{tabular}
        \caption{Object counts of SMC evolutionary phases}
        \label{tab:SMC_EvoPhases_Count}
\end{table}

%======================================================================================
% Section: DR2-DR3
%======================================================================================
\section{Comparison with DR2 results} \label{sec:DR2DR3}

In this section we show the improvement in astrometry and photometry of sources in 
the Magellanic clouds in \egdr{3} 
compared to \gdrtwo. The selection of sources from \gdrtwo for the 
comparison was made in the same way as for our 
main clean samples (as described in \secref{sec:samples}).

One of the scientific demonstration papers released with \gdrtwo, \cite{2018A&A...616A..12G} 
studied the LMC and SMC, in addition to the kinematics of 
globular clusters and dwarf galaxies around the Milky Way. Following this
study, and to ensure that the quoted (and plotted) proper motions are 
relatively easy to interpret in terms of internal velocities, it is 
particularly helpful to define an orthographic projection 
of the usual celestial coordinates and proper motions,

\begin{equation}\label{eq:xy}
        \begin{aligned}
                x & = \cos\delta\sin(\alpha-\alpha_C)\\
                y & = \sin\delta\cos\delta_C-\cos\delta\sin\delta_C\cos(\alpha-\alpha_C) 
        \end{aligned}
,\end{equation}  

\noindent where $\alpha_C$ and $\delta_C$ are the reference centres of the respective clouds (see \secref{sec:spatial_selection}). 

The corresponding proper motions $\vec{\mu_{xy}} = (\mu_x,\mu_y)$ and 
uncertainties in the form of a covariance matrix $\tens {C_{\mu_{xy}}}$ can 
be found from $\vec{\mu_{\alpha*\delta}} = (\mu_{\alpha*},\mu_\delta)$, and 
their uncertainty covariance matrix $\tens{C_{\mu_{\alpha*\delta}}}$  by the 
conversions

\begin{equation}\label{eq:muxy}
        \begin{aligned}
                \vec{\mu_{xy}} = & \tens{M}\,\vec{\mu^T_{\alpha*,\delta}} \\
                \tens {C_{\mu_{xy}}} = & \tens{M}\,\tens{C_{\mu_{\alpha*,\delta}}}\, \tens{M}^T
        \end{aligned}
,\end{equation}

where 

\begin{equation}
        \tens{M} = \begin{bmatrix} 
                        \cos(\alpha-\alpha_C) & - \sin\delta\sin(\alpha-\alpha_C) \\
                        \sin\delta_C\sin(\alpha-\alpha_C)
                        &\cos\delta\cos\delta_C+\sin\delta\sin\delta_C\cos(\alpha-\alpha_C)
        \end{bmatrix}
.\end{equation}

We note that at ($\alpha_C,\delta_C$), we have $\mu_x =  \mu_{\alpha*}$, $\mu_y = 
\mu_\delta$. We use these coordinates throughout.

\begin{figure*}[h]
   \includegraphics[width=\textwidth]{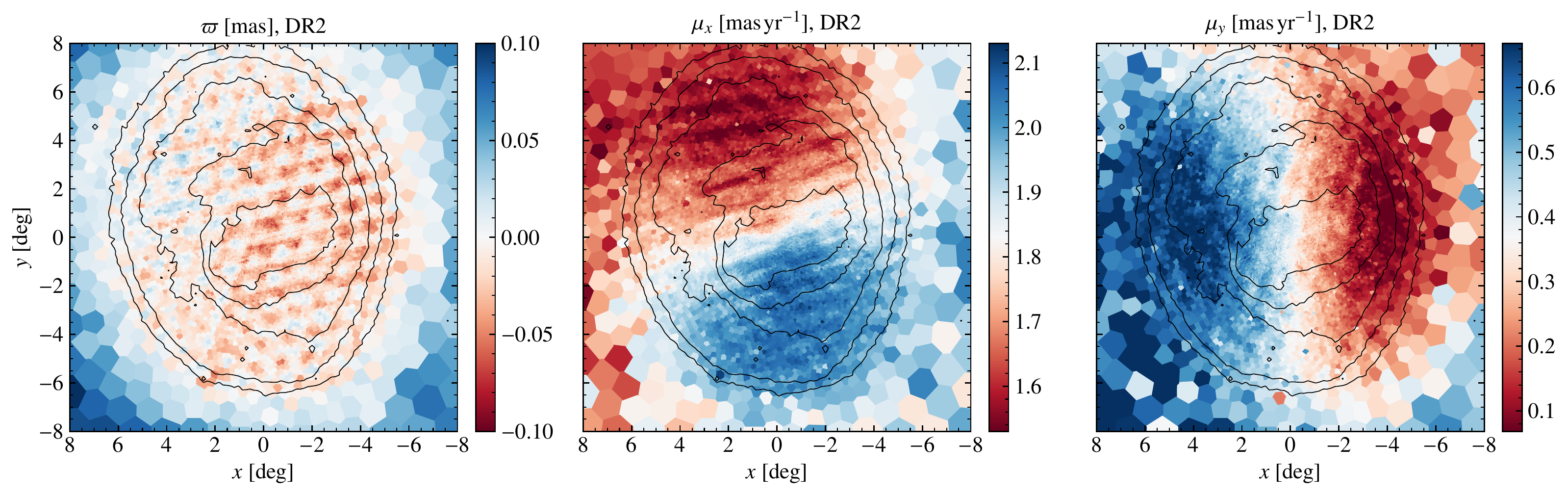}
     \includegraphics[width=\textwidth]{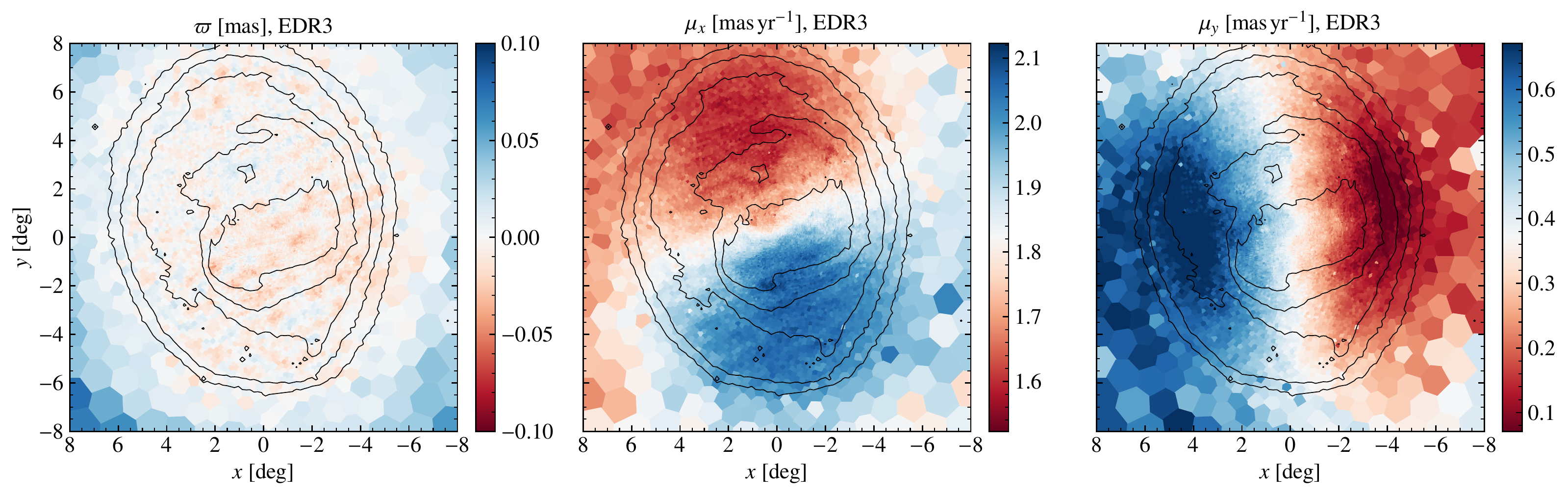}
                \caption{\label{fig:ParallaxLMC} Comparison of the parallaxes (left) and proper motions in the $x$ and $y$ directions (middle and right, respectively) of LMC sources in \gdrtwo 
                         (upper panels) and \egdr{3} (lower panels).}
\end{figure*}

\begin{figure*}[h]
   \includegraphics[width=\textwidth]{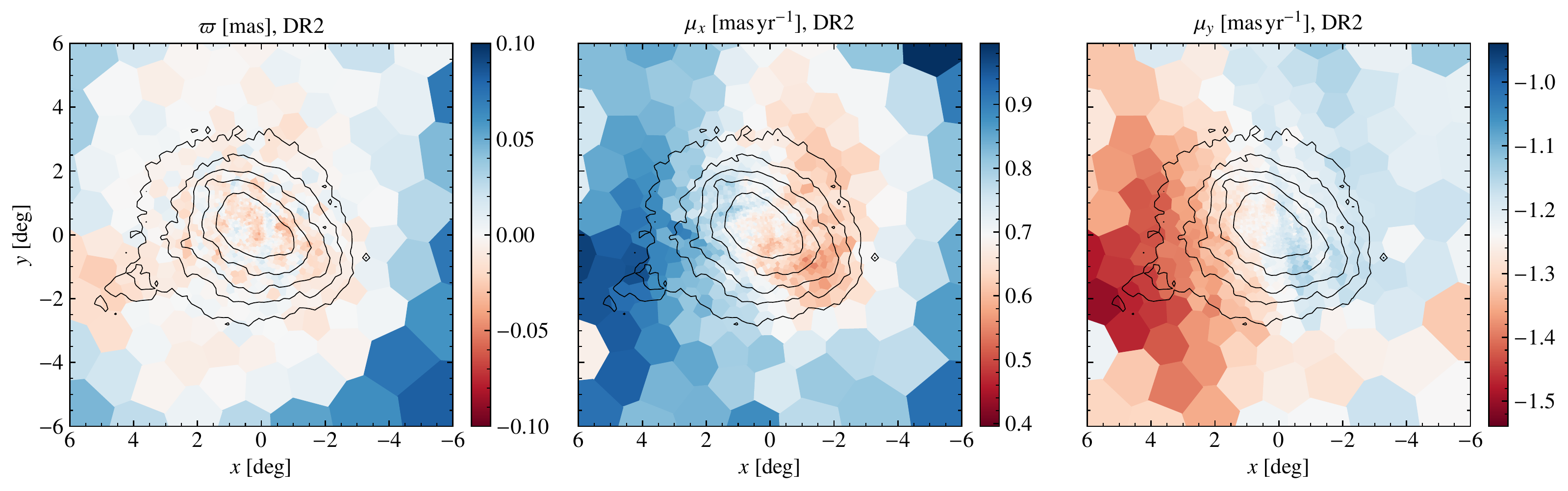}
     \includegraphics[width=\textwidth]{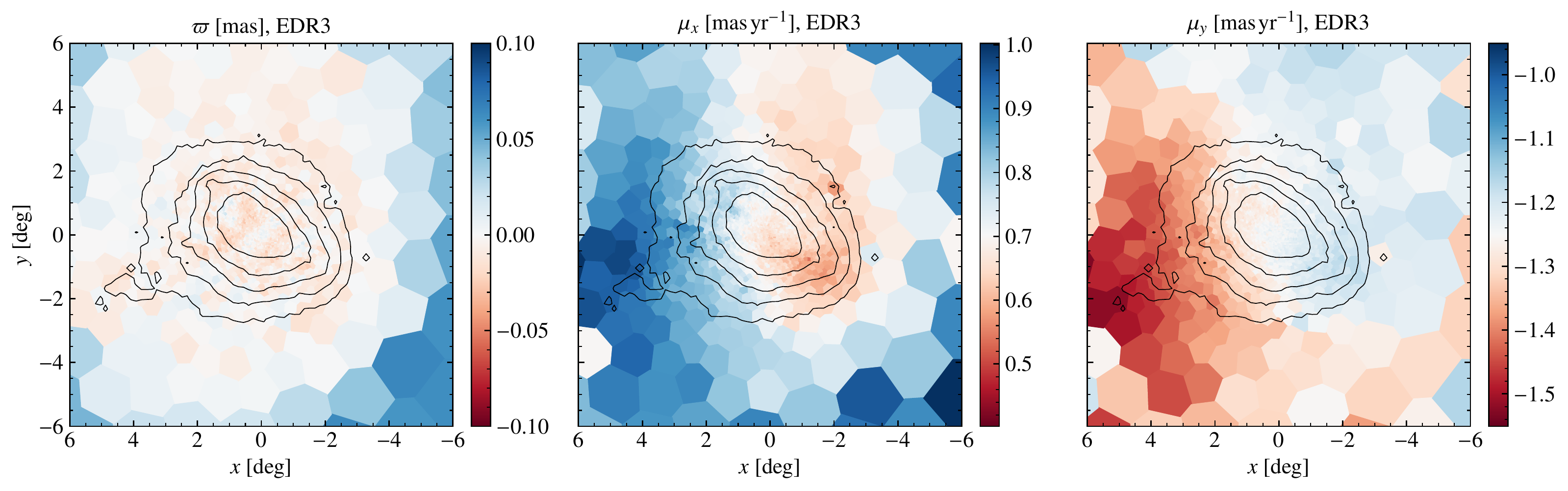}
    \caption{\label{fig:ParallaxSMC} 
    % Comparison of the parallaxes between \gdrtwo (left) and \egdr{3} (right) for sources in the SMC.
    Same as in Fig.~\ref{fig:ParallaxLMC}, but for the SMC.}
\end{figure*}

\begin{figure}[h]
   \includegraphics[width=\columnwidth]{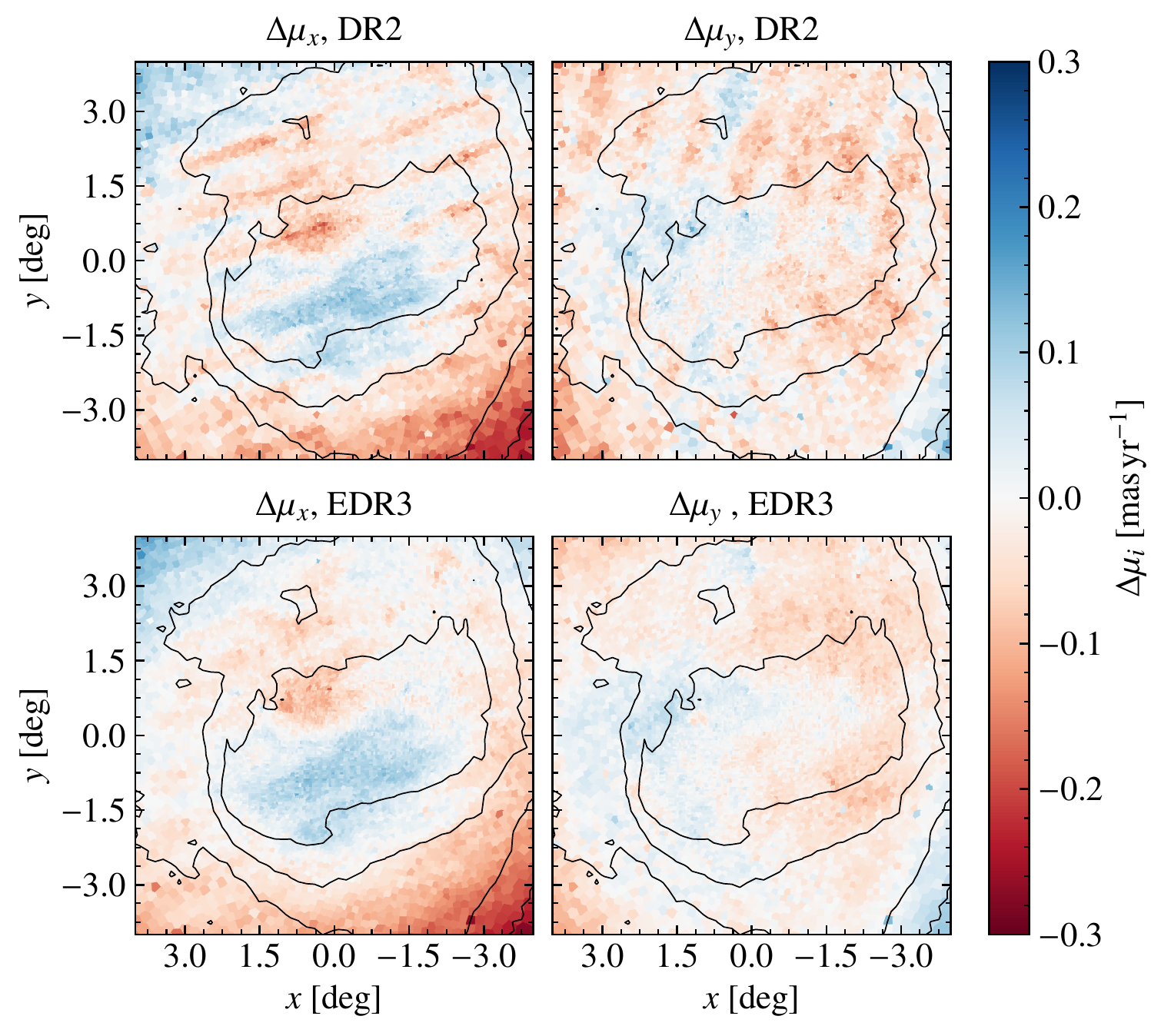}
     \caption{\label{fig:PMresidLMC} Comparison of the residual proper motion fields of the LMC 
                         after a first-order approximation of the field was subtracted to emphasise 
                                                 the systematic errors in \gdrtwo (upper panels) and \egdr{3} (lower panels).}
\end{figure}

\begin{figure}[h]
   \includegraphics[width=\columnwidth]{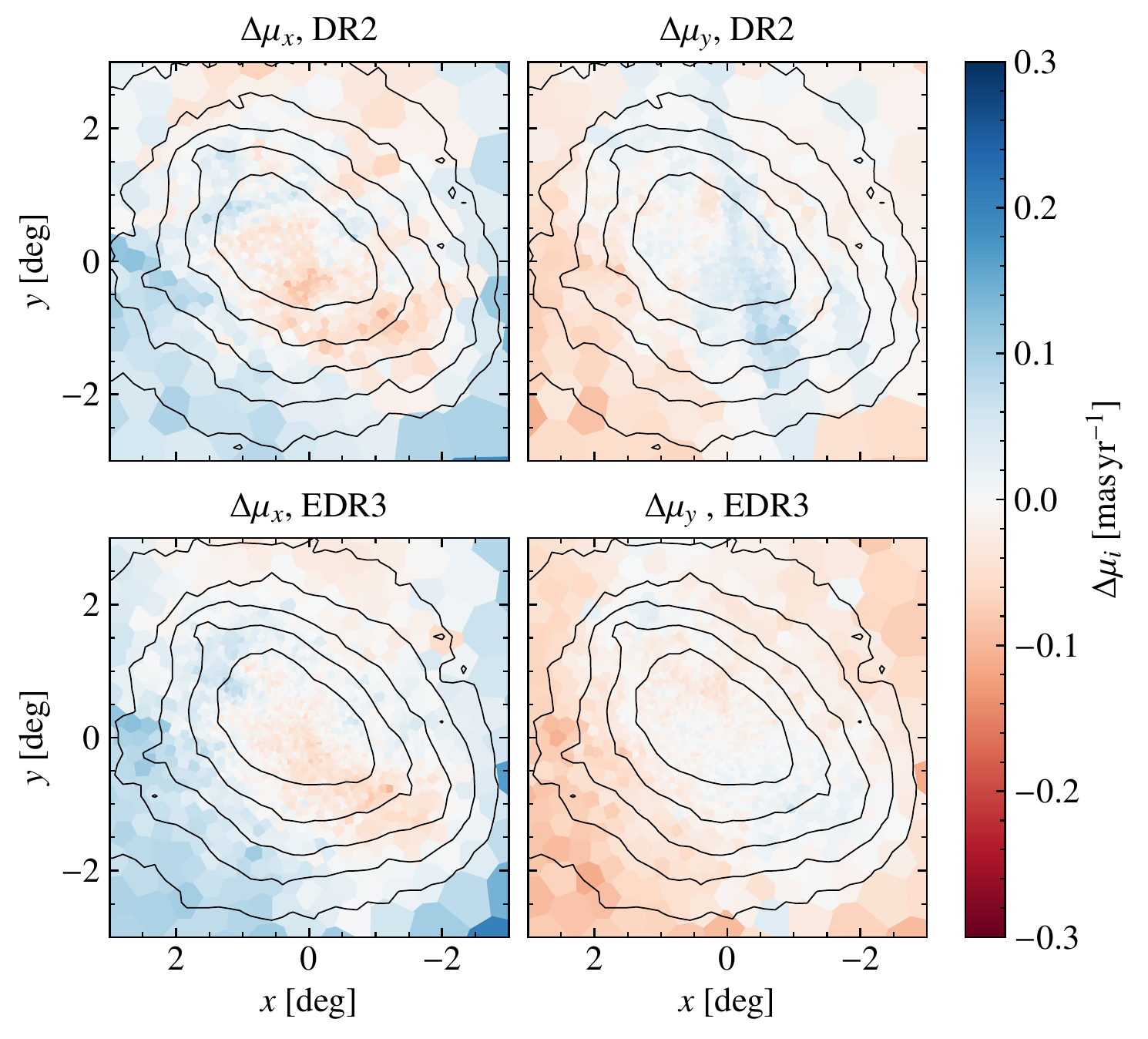}
     \caption{\label{fig:PMresidSMC} Same as in Fig.~\ref{fig:PMresidLMC}, but for the SMC.}
\end{figure}

In Figures~\ref{fig:ParallaxLMC}~to~\ref{fig:PMresidSMC} we show the parallax 
and proper motion fields of the area around each of the cloud centres, as shown in 
the filtered \gdrtwo and \egdr{3} data. 
We use a Voronoi binning scheme \citep{Cappellari2003}, which produces bins 
with approximately 1000 stars each. The bins are therefore irregularly shaped and
become large far from the centre of the clouds. 
Each bin is coloured according to the error-weighted mean of the indicated quantity.
In each case, the dark lines are density contours. 

These figures show that the \egdr{3} data are a clear 
improvement to \gdrtwo data: the sawtooth variation that was seen in 
parallax and proper motion is significantly reduced. The outer bins of both 
the LMC and SMC still show a net positive parallax, which indicates that for these 
bins, foreground contamination that passes our proper motion and parallax 
filter makes a small but non-negligible contribution.

In Figures~\ref{fig:PMresidLMC}~and~\ref{fig:PMresidSMC} we show the proper motions that  remain when we subtracted a linear gradient from each, so we show in each case

\begin{equation}
        \Delta \mu_i = \mu_i - \left( \mu_{i,0} + \left.\frac{\partial\mu_{i}}{\partial x}\right|_0 x + 
        \left.\frac{\partial\mu_{i}}{\partial y}\right|_0 y\right)
,\end{equation}

\noindent where the central values, $\mu_{i,0}$, and partial derivatives ${\partial\mu_{
i}}/{\partial x}$ and ${\partial\mu_{i}}/{\partial y}$ were evaluated as a 
linear fit to the values within a radius of $3^\circ$ around the centre. The 
values found using \egdr{3} are shown in 
\tabref{tab:pm_linear}. This allows us to show the sawtooth 
pattern in proper motions more clearly. The patterns are again significantly reduced in 
\egdr{3}. The faint indications of a streaming motion along the bar that were pointed out 
in \gdrtwo stand out much more clearly in \egdr{3}, and we investigate them 
further in Section~\ref{sec:Spiral_structure}.

\begin{table}[t]
        \centering
                \begin{tabular}{ccccccc}
                   \hline
        & $\mu_x$ & $\mu_y$ & $\frac{\partial\mu_x}{\partial x}$&  $\frac{\partial\mu_x}{\partial y}$
        &  $\frac{\partial\mu_y}{\partial x}$& $\frac{\partial\mu_y}{\partial y}$ \\
                   \hline
       LMC           &  1.871 & 0.391 & -1.561 & -4.136 & 4.481 & -0.217 \\
       SMC & 0.686 & -1.237 & 1.899 & 0.288 & -1.488 & 0.213 \\
                   \hline
                \end{tabular}
        \caption{Linear fit to the proper motions in the $x,y$ directions using \egdr{3}. Proper motions 
                 are in $\mathrm{mas}\,\mathrm{yr}^{-1}$, and $x,y$ positions in radians.}
        \label{tab:pm_linear}
\end{table}

As explained in \citet[eq.  3]{2018A&A...616A..12G}, we can use the simple 
linear gradients to estimate the inclination, orientation and angular 
velocity of the disc under the assumptions that this angular velocity $\omega
$ is constant, which is  valid for a linearly rising rotation curve,  and that the 
average motion is purely azimuthal in a flat disc. We define the inclination $i$ 
to be the angle between the line-of-sight direction to the cloud centre and 
the rotation axis of the disc, and 
orientation $\Omega$ is the position angle of the receding node, measured from $
\vec{y}$ towards $\vec{x}$, that is,\ from north towards east. Here and elsewhere, we 
assume that the distances to the LMC and SMC are 
$D_{LMC} = 49.5\,{\rm kpc}$ \citep{Pietrzynski2019} and 
 $D_{SMC} = 62.8\,{\rm kpc}$ \citep{2000A&A...359..601C}, respectively.

The line-of-sight velocity of the disc can either be derived from these 
gradients, or (as we do here) assumed given the known line-of-sight velocity 
of the LMC \citep[$262.2\pm3.4\, \rm{km}\,\rm{s}^{-1}$]{vanderMarel2002} or 
SMC \citep[$145.6\pm0.6\,\rm{km}\,\rm{s}^{-1}$]{2006AJ....131.2514H}. The 
values we find for $i$, $\Omega,$ and $\omega$
are $34.538^\circ, 298.121^\circ, 4.732\;\mathrm{mas}\,\mathrm{
yr}^{-1}$ and $ 78.763^\circ , 8.955^\circ , 0.854\;\mathrm{mas}\,\mathrm{yr}^
{-1}$ for the LMC and SMC, respectively. This is broadly consistent with the values 
found for \gdrtwo. The LMC values are consistent with those found by the more 
detailed investigation in \secref{sec:kinematics}.

\begin{figure*}[h]
   \includegraphics[width=0.5\textwidth]{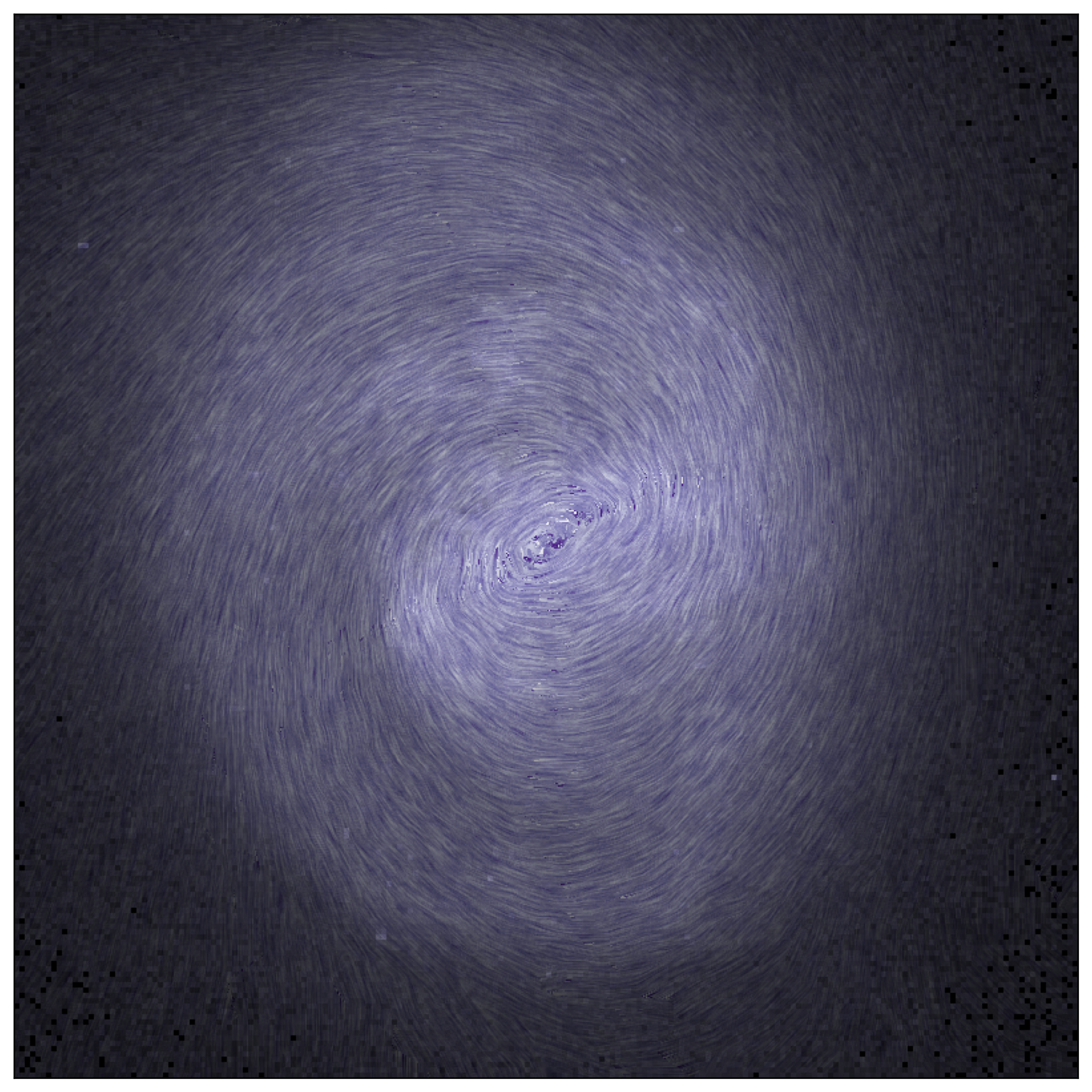}
     \includegraphics[width=0.5\textwidth]{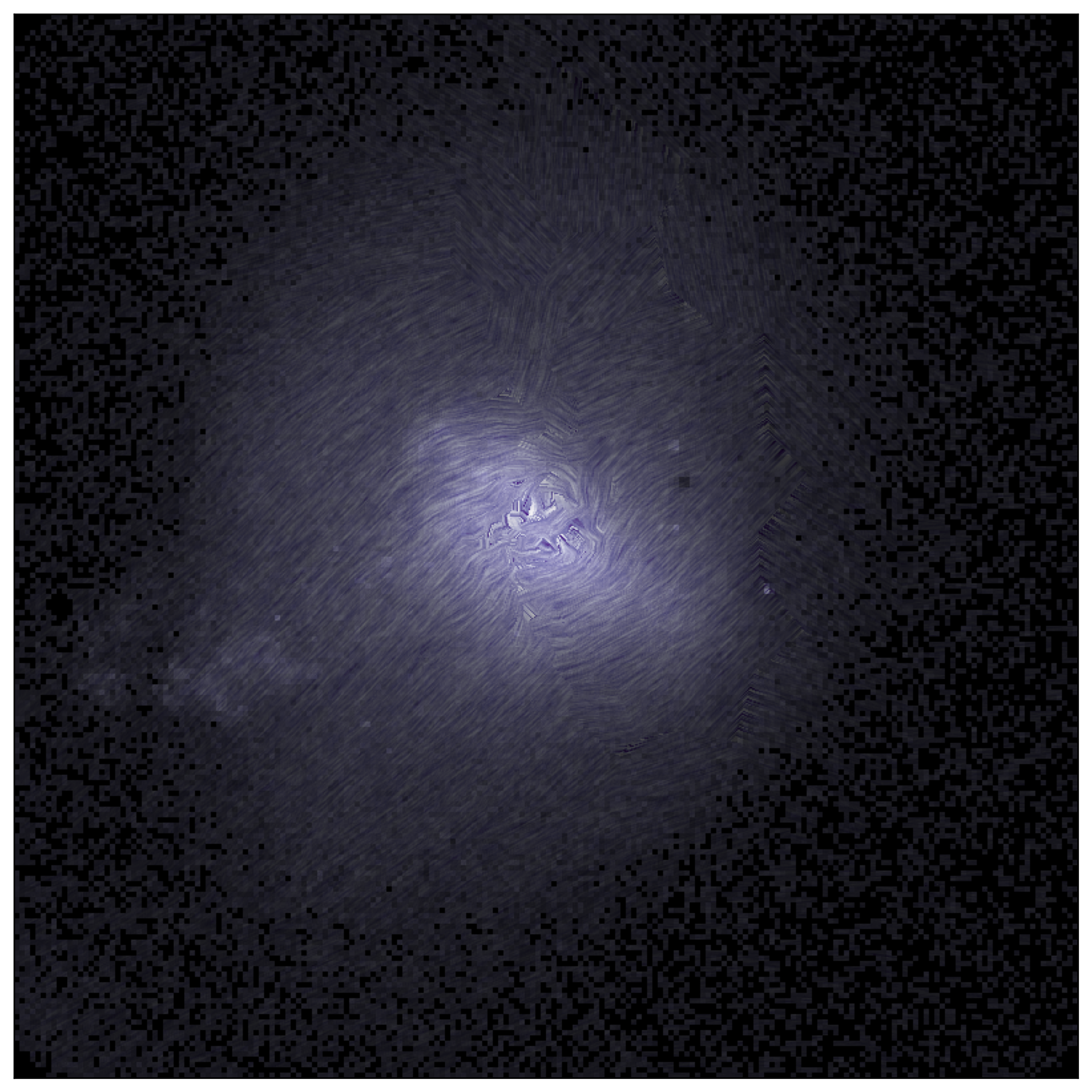}
   \caption{\label{fig:LIC} Illustration of the proper motion field of the LMC (left) 
          and SMC (right) using line-integral convolution. We set the alpha parameter (opacity) of the 
                coloured lines according to the density, with the densest regions being the most opaque.}
\end{figure*}

In \figref{fig:LIC} we use the technique of line-integral convolution 
\citep{LIC} to better illustrate the proper motion field of the  
Magellanic Clouds. The direction of the lines illustrates the vector field of 
the proper motions, while their brightness illustrates the density (more precisely, we set the 
alpha parameter in \textsc{matplotlib} to be proportional to the $1/4$ power of 
the star count). The 
ordered rotation of the LMC is very clear from this image, while the SMC is 
more jumbled.

\begin{figure*}[h]
                \includegraphics[width=0.5\textwidth]{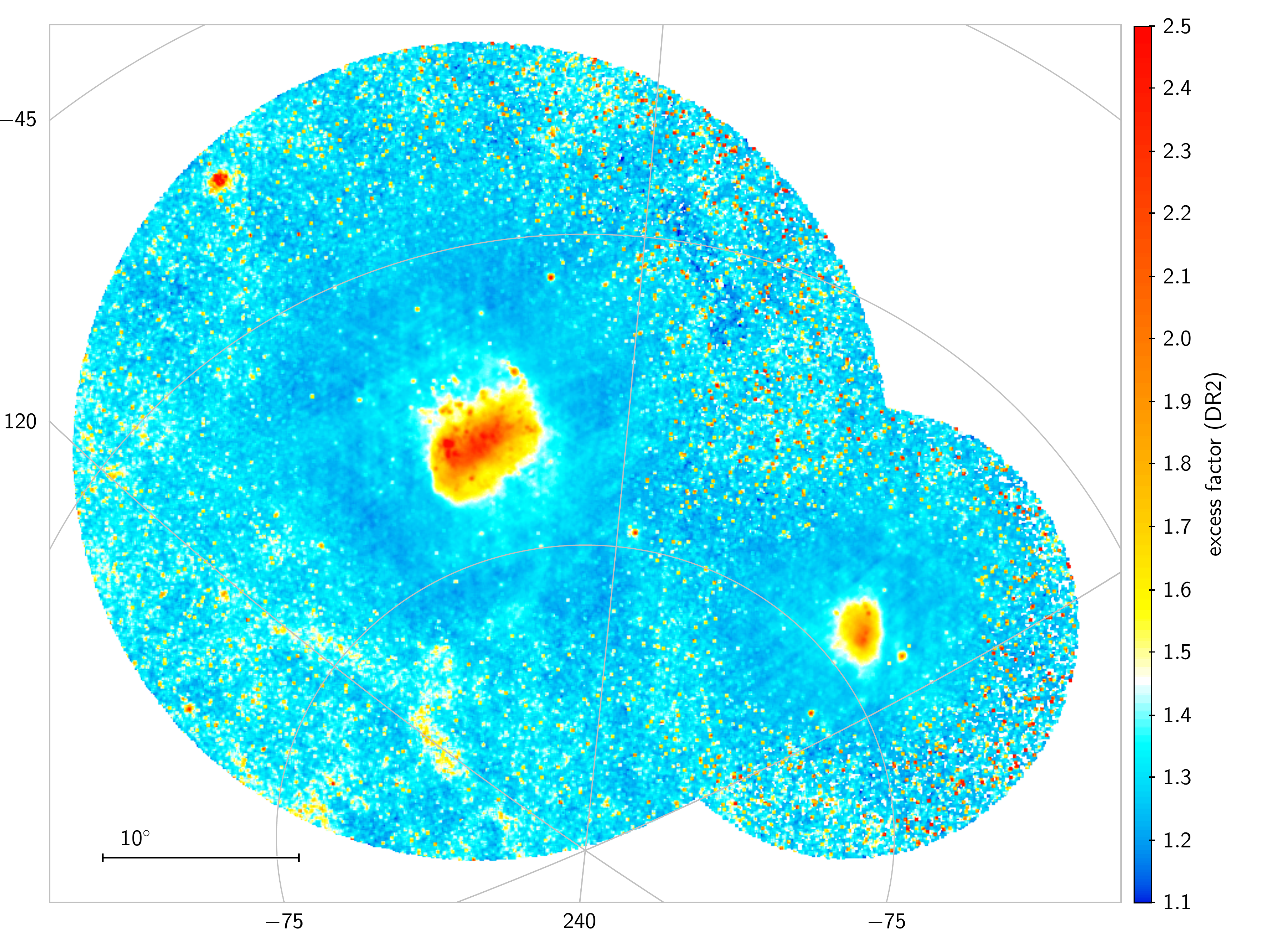}
                \includegraphics[width=0.5\textwidth]{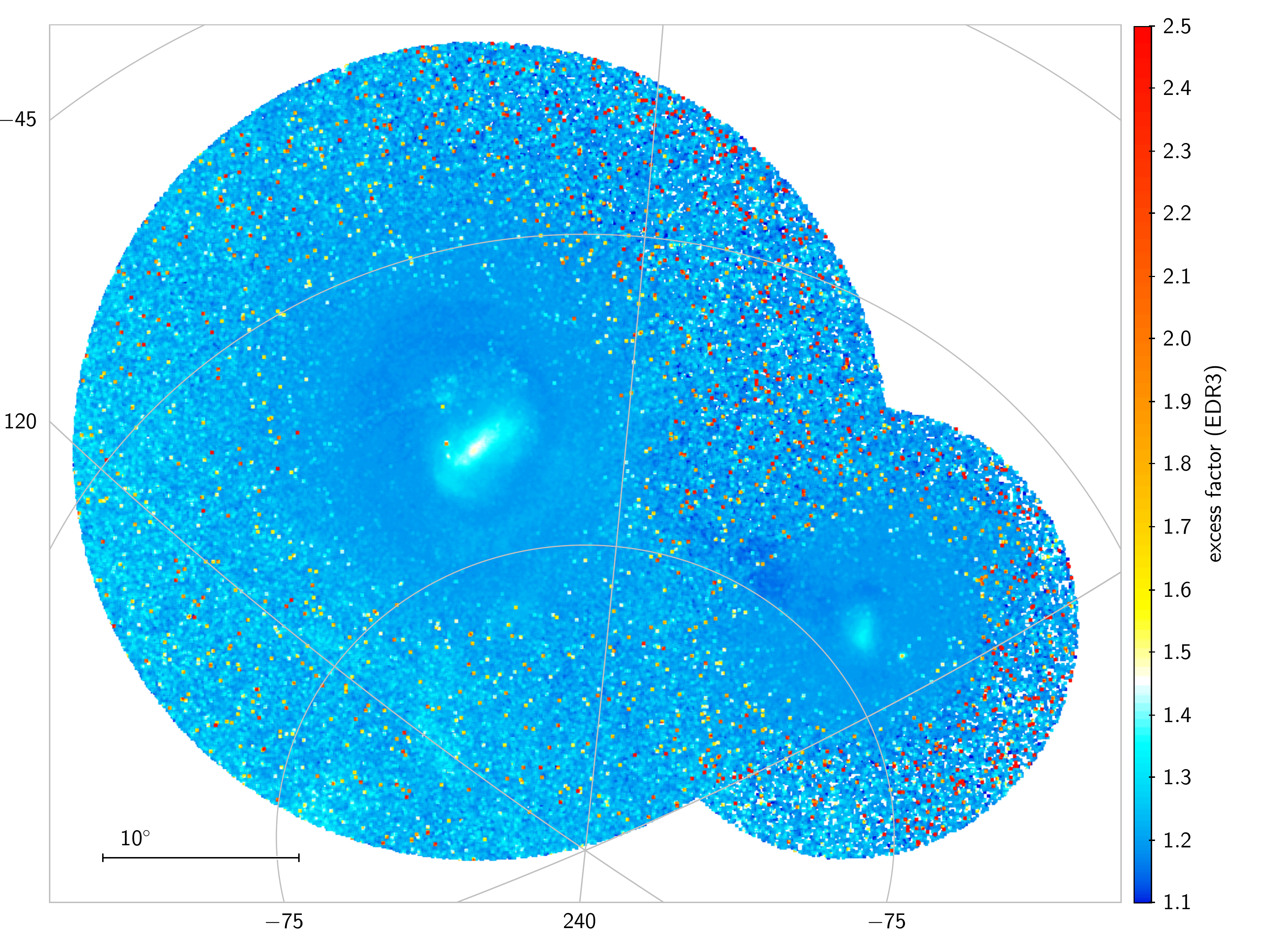}           \caption{\label{fig:dr2dr3_phot_excess} 
                                Photometric excess factor, i.e.\ the sum of fluxes in the \gbp\ and \grp\
                                bands over the \gmag\ flux. {\em Left:} for \gdrtwo, and {\em right:} for \egdr{3}.
                }
\end{figure*}

Finally, to complete this section, we compare the quality of the photometry
in the LMC and SMC areas. Extracting \gbp\ and \grp\ photometry from prism spectra is challenging in the
dense, central parts of the Magellanic Clouds. A simple diagnostic for the consistency
of the photometry for a source is the photometric excess factor (included in the archive), which is defined as the
ratio of the flux of the prism spectra (\gbp\ and \grp) and the \gmag\
flux. Because the two spectra overlap slightly and have a higher response in
the red, this ratio typically lies in the range 1.1--1.4 for isolated point
sources, with higher values for the redder sources. 
\figref{fig:Appendix_bprp} shows that the centres of the 
clouds are not very red, and \figref{fig:dr2dr3_phot_excess}
shows that the mean excess factor increases in these centres,
but with abnormally high values in \gdrtwo (left panel) and typical 
values in \egdr{3} (right panel). 
In \egdr{3}\ the background estimation has changed significantly as compared 
to \gdrtwo \citep{EDR3-DPACP-117}, while crowding is still left uncorrected for. 
We conclude that the photometry in \gdrtwo\ was strongly affected by background 
issues in the central areas, and that this problem has greatly diminished in 
\egdr{3}, where traces of crowding are still visible. 
The \gmag\ flux has only changed slightly between the two
releases, that is,\ by a few hundredths of a magnitude, while \gbp\ and \grp\ have
been revised by a few tenths of a magnitude. It is therefore a fair assumption
that the improved excess factor is driven by the improvement of \gbp\ and \grp\ 
photometry in \egdr{3}.

%======================================================================================
% Section: spatial structure
%======================================================================================
\section{Spatial structure of the Large Magellanic Cloud 
\label{sec:Spatial_Structure}}

In this section we summarise our attempts to infer the spatial distribution of 
sources in the LMC using a simplified model without separating the various 
stellar populations that constitute the galaxy. This is an oversimplification 
\citep[see e.g.][for a recent summary of the complexity of the problem when the 
different populations are taken into account]{ElYoussoufi2019}, aimed only at 
exemplifying the use of the \gaia astrometry for this type of studies.

Despite the significant improvement of the \egdr{3} astrometry with respect to 
\gdrtwo, systematic problems remain, as described in \cite{EDR3-DPACP-128} and 
exemplified in the spatial distribution of median parallaxes shown in 
\figref{fig:ParallaxLMC}. In order to infer the parameters of the 
LMC spatial distribution, we therefore modelled the observed parallaxes as affected 
by a zero-point offset.

We assumed, for the sake of illustrating the magnitude of these zero-point 
offsets, that the sources selected as candidate members of the LMC have a mean 
parallax of 0.02 mas, corresponding to a distance of 50 kpc from the Sun \citep{Pietrzynski2019}. The central 90\% interval around the median 
(binned) \egdr{3} parallaxes  
shown in \figref{fig:ParallaxLMC} extends from -0.075 to 0.05 mas. We can therefore estimate the range of zero-point offsets as (-0.095,0.03). This
means that the zero-point offsets are of the same order of magnitude,
but larger than the expected value of the mean parallax of the
LMC. Variations in parallaxes around the mean value due to the
spatial distribution of the LMC sources (e.g. due to its depth or
inclination angle) are expected to be much smaller. 
In addition, these systematics occur in combination with the usual random 
uncertainties associated with the individual measurements that propagate to 
yield the catalogue parallax uncertainty of each source. In the case of the data 
set used here, these parallax uncertainties have a median value of
0.17\,mas. Estimating the zero-point offsets therefore is a critical 
element of 
the modelling effort we describe in this section and plays a central role in the 
inference of the parameters of the spatial structure of the LMC.

Unfortunately, we did not succeed in our aim 
of inferring geometric properties of the LMC from the \egdr{3} astrometric 
measurements. We tried several degrees of model complexity and two approaches 
to the inference problem: Markov chain Monte Carlo posterior sampling (MCMC) 
\citep{RobertCasella2013},  and approximate Bayesian computation (ABC) 
\citep{Beaumont2002, Marjoram2003}, always in the context of the Bayesian 
approach to inference. In the MCMC posterior sampling we used the parallaxes of 
the individual LMC sources to compute the full likelihood, while in the ABC 
approach, we binned the data in a certain number of constant-size right 
ascension and declination bins and employed a distance metric to compare simulations and observations in order to avoid computing the full 
likelihood.

Both approaches used the same probabilistic generative model for the  
distribution of  the \egdr{3} parallax measurements. This model assumes 
that the LMC sources  are  spatially distributed as an elliptic double -exponential disc (similarly as in Eq. (1) of \cite{Mancini-2004}, but with the 
vertical distances from the disc mid-plane modelled by a central Laplace prior) and generates as many (proper to the disc) location coordinates as 
there are sources in the \egdr{3} sample. The model applies a number 
of geometrical transformations (see e.g. \citealt{Weinberg-Nikolaev-2001}) to 
generate a set of true parallaxes that are unaffected by the measurement 
uncertainties and/or zero-point offsets. 
Our generative model has nine global parameters: the disc scale length
$R_{0}$, the disc scale height $h_{0}$, the disc ellipticity parameter
$\epsilon$, the disc minor-axis position angle $\theta_{\mathsf{ma}}$ , and the
LMC line of nodes position angle $\theta_{\mathsf{LON}}$ (both angles measured
with respect to the west direction), the inclination angle $i$ of the LMC plane
with respect to the plane of the sky, and the spherical coordinates
$\left(\alpha_{0},\delta_{0},D_{LMC}\right)$ of the centre of the LMC disc.

To simulate observed parallaxes, we took 
the \egdr{3} parallax uncertainties (the variance error component) and
the parallax zero-point offset patterns (the systematic error
component) observed in the \egdr{3} data into account. We modelled the latter as part of the 
inference process by means of a linear combination of Gaussian radial basis
functions (RBFs) using the observed patterns and a canonical distance to the 
LMC as initial guess. Finally, each parallax measurement was simulated using
a Gaussian distribution centred at the sum of the true simulated
parallax and the offset generated using the RBF model.

In addition to modelling the parallax zero-point offsets using the RBF 
parametrisation as part of the inference process, we also tried to correct 
individual source parallaxes using an early version of the fit proposed in 
\cite{EDR3-DPACP-128} as a function of the apparent magnitude and colour. 
Unfortunately, the correction is not useful for our purposes. The mentioned 
correction \citep[from][]{EDR3-DPACP-128} is obtained by a combination of 
information from quasars, physical stellar pairs, and LMC sources. However, it 
is not able to reproduce the local variations of the parallax zero-point in the 
LMC field because its only dependence on positions is of the form of the sinus 
of the ecliptic latitude, which is almost constant in the LMC area. 
Additionally, the correction assumes that all the LMC stars are at the same 
distance embedding its internal 3D structure, which is what we aimed to 
determine.

In what follows we describe our attempt of using the probabilistic 
generative model to perform the parameter inference using the MCMC algorithm.
We attempted to evaluate the full likelihood for several of the populations 
defined in  \secref{sec:samples}. The inference process was based on a 
hierarchical Bayesian model and an MCMC no-U-turn posterior sampler (NUTS)
\citep{Hoffman2014}. In this approach the true parallax of individual LMC 
sources was used to compute the likelihood. This implies the 
inclusion of one additional parameter per source (its true distance).
The computational demands were so high that we were forced to distribute the 
likelihood computations in a TensorFlow \citep{Abadi2016} Probability 
\citep{Dillon2017} framework in the Mare Nostrum supercomputer at the
Barcelona Supercomputing Centre. Unfortunately, the maturity level 
of the TensorFlow libraries involved was not sufficient and we did not 
achieve the required performance accelerations. Then,  our main problem 
was that we were unable to scale our models to 
the size of the \egdr{3} sample  using the MCMC NUTS algorithm.

Because of the scalability issues found when using the MCMC, we decided to try with 
a sequential Monte Carlo approximate Bayesian computation algorithm (SMC-ABC), 
which is further described in \cite{Jennings2017} and section 5 of 
\cite{Mor2018}. The theoretical basis for these algorithms can be found in 
\cite{Marin2011} , \cite{Beaumont2008}, and \cite{Sisson2010}. 
        
The choice of the summary statistics is  crucial for the performance of the 
SMC-ABC algorithm. For the purposes of the present work, we defined the 
summary statistics as the median parallax of the stars in the LMC sample, 
distributed in a grid of 50$\times$50 bins in right ascension (from 
$50\deg$ to $120\deg$) and declination ($-50\deg$ to $-80\deg$). The stellar 
sample 
used for this inference was the combination of the following subsamples: 
\textit{Young 1}, \textit{Young 2}, \textit{Young 3}, \textit{RGB}, \textit{AGB}, 
\textit{RRL}, \textit{BL}, and \textit{RC}.
        
With the SMC-ACB technique we attempted to infer up to seven parameters of the 
structure of the LMC: the distance to the centre $D_{LMC}$, the inclination 
angle $i$, the position angle of the line of nodes $\theta_{\mathsf{LON}}$, 
the position angle of the disc minor axis $\theta_{\mathsf{ma}}$, the 
ellipticity factor $\epsilon,$ and the position in the sky of the LMC centre 
$\left(\alpha_{0},\delta_{0}\right)$. To infer  these 
structural parameters, we chose Gaussian priors centred on the standard values found in the literature; the prior in distance is the most 
restrictive. Furthermore, we  simultaneously inferred the model parameters 
of  the parallax zero-point variations (i.e. the coefficients of the RBF linear model described above)  using 50 basis functions. Additionally,  
we fixed the scale height  and the radial scale length of the disc at $1.6$ 
and $0.35$ kpc, respectively. 
        
From the SMC-ABC attempt, our conclusion is that the local parallax zero -point of the LMC in  \egdr{3} distorts most of the signal of the 3D structure 
of the LMC (in the astrometry), and that there is not enough information in our 
summary statistics to simultaneously infer the local parallax zero-point variations and the 3D structure of the LMC. However, it may be possible if the 
former is constrained with additional external restrictions and/or finding an 
optimal way to add the information from the density distribution of the stars 
in the LMC area.

%======================================================================================
% Section: Kinematics
%======================================================================================
\section{Kinematics of the Large Magellanic Cloud}
\label{sec:kinematics}

In this section we use the \egdr{3} data to study the kinematics of the 
Magellanic Clouds. The analysis is focused on the LMC because it has a clear disc
structure that can be meaningfully modelled and understood; the SMC has a more 
complex, irregular structure that would require a more extensive and deep analysis, 
which is beyond the scope of this demonstration paper. 

In the \secref{sec:kinematics_method} we describe the method and tools we used
in our analysis, and in \secref{sec:kinlmc} we present an analysis of the general
kinematic trends and a detailed look at the velocity profiles in the disc,
focusing on the segregation of the rotation velocities as a function of 
the evolutionary stage.

%%%%%%%%%%%%%%%%%%%
\subsection{Method and tools \label{sec:kinematics_method}}

\cite{2018A&A...616A..12G} presented formulae relating the in-plane velocities of 
stars to their observed proper motions under the assumption that the stars all move in a flat disc\footnote{See the erratum, 
\cite{2020A&A...642C...1G}, for 
corrections required for some of the formulae given in Appendix B of \cite{
2018A&A...616A..12G}}. 
Here we summarise the key results and equations.

\noindent Defining: 

\begin{itemize}
  \item $a = \tan i \cos\Omega$
        \item $b = -\tan i\sin\Omega$
        \item $(l_x,l_y) = (\sin\Omega,\cos\Omega)$
        \item $(m_x, m_y, m_z) = (-\cos i\cos\Omega,\cos i \sin\Omega,\sin i)$  
\end{itemize}

\noindent \cite{2018A&A...616A..12G} show that 
Cartesian coordinates can be defined in the plane of the disc $\xi,\eta,$ where

\begin{equation}\label{equ:xi_eta}
        \begin{aligned}
                \xi &= \frac{l_xx+l_yy}{z+ax+by}\\[3pt]
                \eta &= \frac{(m_x-am_z)x+(m_y-bm_z)y}{z+ax+by} 
        \end{aligned}
,\end{equation}

\noindent and derive simultaneous equations relating the velocities $\dot{\xi},
\dot{\eta}$ to $\mu_x,\mu_y$ for a given disc inclination, orientation, and 
bulk velocity of the galaxy can  derived. The bulk velocity of the galaxy is expressed as
$(\mu_{x,0},\mu_{y,0},\mu_{z,0})$, where $\mu_{x,0}$ and $\mu_{y,0}$ are the 
associated proper motions in the $x$ and $y$ directions at the centre of the 
disc, and $\mu_{z,0} = v_{z,0}/D_{LMC}$, the associated line-of-sight velocity, 
expressed on the same scale as the proper motions by dividing by $D_{LMC}$. We derive

\begin{equation}\label{equ:xi_eta_vel}
        \begin{aligned}
                (l_x-&x(l_xx+l_yy))\dot{\xi} + (m_x-x(m_xx+m_yy+m_zz))\dot{\eta} \\
                &=-\mu_{x,0}+x(\mu_{x,0}x+\mu_{y,0}y+\mu_{z,0}z) + \frac{\mu_x}{ax+by+\sqrt{1-x^2-y^2}}\\[2pt]
                (l_y-&y(l_xx+l_yy))\dot{\xi} + (m_y-y(m_xx+m_yy+m_zz))\dot{\eta} \\
                &=-v_y+y(\mu_{x,0}x+\mu_{y,0}y+\mu_{z,0}z) +  \frac{\mu_y}{ax+by+\sqrt{1-x^2-y^2}} .
        \end{aligned} 
\end{equation}

Furthermore, we can gain much more physical insight by converting these Cartesian 
coordinates $\xi,\eta,\dot\xi, and\dot\eta$ into polar coordinates $R,\phi,v_R, \text{and }v_
\phi$.

Our strategy in this paper therefore was to fit the proper motion of the 
filtered LMC population as a flat rotating disc with average $v_R=0 $ and 
$v_\phi = v_\phi(R)$. Our model has ten parameters, some of which can be kept 
fixed (based on the other knowledge of the Magellanic Clouds):

\begin{itemize}
        \item Rotational centre of the disc on sky, parametrised as $(\alpha_0,\delta_0)$
        
        \item Bulk velocity in the $x$ direction, which we parametrise as $\mu_{x,0}$, 
              the associated proper motion at the centre of the disc.
                                
        \item Bulk velocity in the $y$ direction, which we parametrise as $\mu_{y,0}$, 
              the associated proper motion at the centre of the disc.
                                
        \item Bulk velocity in the $z$ direction, which we parametrise as 
              $\mu_{z,0} = v_{z,0}/D_{LMC}$. 
                                
        \item Inclination, $i$.
        
        \item Orientation, $\Omega$.
        
        \item Three parameters ($v_0$, $r_0$ and $\alpha$) are used to describe the rotation curve,
                                \[
                                v_{\phi,M}(R) = v_0\left(1+\left({r_0\over R}\right)^\alpha\right)^{-1/\alpha}
                                \]
\end{itemize}

To analyse the data, we considered bins of $0.08^\circ$ by $0.08^\circ$ in $x, y$
in the range $-8^\circ<x<8^\circ$, $-8^\circ<y<8^\circ$. For each bin with 
centre $x_i,y_i$, we derived a maximum likelihood estimate of the typical proper motion, that is, for the $i$th bin, $\vec{\mu_i} = (\mu_{x,i},\mu_{y,i}$), and dispersion matrix

\begin{equation}
        \tens \Sigma_i = \begin{bmatrix} \sigma_{x,i}^2 & \rho_{\sigma xy,i} \sigma_{x,i}\sigma_{y,i} \\ 
         \rho_{\sigma xy,i} \sigma_{x,i}\sigma_{y,i} &  \sigma_{y,i}^2   \end{bmatrix}
\end{equation}

\noindent by maximising

\begin{equation}
        \begin{aligned}
                \mathcal{L}_i = \prod_{j=1}^{N_i} & \frac{1}{2\pi\sqrt{|\tens \Sigma+\tens C_{\mu_{xy},j}|}} \\ 
                                &\;\times\exp\left(-\frac 1 2 (\vec\mu_j-\vec{\mu_i})^T(\tens \Sigma_i+\tens C_{\mu_{xy},j})^{-1}(\vec\mu_j-\vec{\mu_i})\right)
        \end{aligned}
        \label{eq:likelihoodpixel}
,\end{equation}

\noindent where the product is over all $N_i$ sources in our sample in the $i$th 
bin,  $\vec\mu_j$ is the quoted proper motion of the source $(\mu_{x,j},\mu_
{y,j})$, and $\tens {C_{\mu_{xy},j}} $ is the covariance matrix associated 
with the uncertainties as derived in \secrefalt{sec:DR2DR3}.

We estimated the uncertainty of $\vec{\mu_i}$ by bootstrap resampling within 
each pixel. This gave us an estimate of the error covariance matrix in 
proper motion for the bin, $\tens {C_{\mu_{xy},i}} $. As a simple way of taking systematic errors in proper motion into account, we added a systematic 
uncertainty of 0.01 $\mathrm{mas}\,\mathrm{yr}^{-1}$ for each component of 
proper motion, isotropically. This is smaller than the statistical 
uncertainty in most bins outside the inner $\sim$$3^\circ$. We chose this 
value because it is of the same order as the spatially dependent systematic errors 
found by \cite{EDR3-DPACP-128}. 
Binning the data allowed us to make this correction 
for systematic uncertainty and reduced the computational difficulty of fitting 
the model.

The parameters $\mu_{x,0}, \mu_{y,0}, \mu_{z,0} , i,$ and $ \Omega$ give a 
conversion between the $(x_i,y_i,\mu_{x,i},\mu_{y,i})$ values for each pixel 
and the corresponding positions and velocities in the frame of the LMC, $(R_i,
\phi_i,v_{R,i},\text{and }v_{\phi,i})$ thorugh \equref{equ:xi_eta_vel}. We also converted 
the corresponding uncertainty matrix in proper motion into one for $v_{R,i},v_
{\phi,i}$ (for these values of  $\mu_{x,0}, \mu_{y,0}, \mu_{z,0} , i,$ and $ 
\Omega$), which we refer to as $\tens {C_{(v_R,v_\phi),i}} $. The statistic 
that we then calculate is chi-square-like,

\begin{equation}
        \chi^2 = \sum_{i=1}^{N_{bins}} \vec{(\Delta v_i)}^T \tens {C_{(v_R,v_\phi),i}}
                                                \vec{(\Delta v_i)}
\end{equation}

\noindent with $\vec{(\Delta v_i)} = \left(v_{R,i},v_{\phi,i}-v_{\phi,M}(R_i)\right)$.

We note that the statistical uncertainties on the 
values we quote are very small. They are 
$\sim$$0.2\,\mu\mathrm{as}\,\mathrm{yr}^{-1}$ on $\mu_{x,0}$ and $\mu_{y,0}$, 
and less than $0.5$ \%\ on the derived quantities such as $\mu_{z,0}$ or $i$. 
We emphasise therefore that systematic errors, particularly 
those due to our simple model, are the dominant uncertainty. 
The difference between values in table~\ref{tab:KinematicStructuralParameters} 
can be seen as a gauge of these systematic errors.

Our main analysis takes the centre of rotation $(\alpha_0,\delta_0)$ as fixed 
at the photometric centre $\alpha_C,\delta_C$ (\secref{sec:samples}), and $\mu
_{z,0}=1.115\mathrm{mas}\,\mathrm{yr}^{-1}$ taking the value from 
spectroscopy (\secref{sec:DR2DR3}). The parameters of this model, found by 
minimising $\chi^2$, are given in \tabref{tab:KinematicStructuralParameters} 
(along with those from the other models we considered).

\begin{table*}[t]
        \centering
                \begin{tabular}{ccccccccccc}
                                \hline
                                Model & $\alpha_0$ & $\delta_0$ & $\mu_{x,0}$ & $\mu_{y,0}$ & $\mu_{z,0}$ 
                                                        & $i$ & $\Omega$ & $v_0$ & $r_0$ & $\alpha_{RC}$ \\
                                                        & (deg) & (deg) & ($\mathrm{mas}\,\mathrm{yr}^{-1}$) & ($\mathrm{mas}\,\mathrm{yr}^{-1}$) &
                                                        ($\mathrm{mas}\,\mathrm{yr}^{-1}$) & (deg) & (deg) & (\kms) & (kpc) & \\
                                \hline
                                Main& $[81.28]$ & $[-69.78]$ & $1.858$ & $0.385$ & $[1.115]$ & $34.08$ & $309.92$ & $75.9$ & $2.94$ & $5.306$ \\
                                $\mu_{z,0}$ free& $[81.28]$ & $[-69.78]$ & $1.858$ & $0.385$ & $1.179$ & $34.95$ & $310.93$ & $76.5$ & $2.96$ & $5.237$ \\
                                Centre free& $81.07$ & $-69.41$ & $1.847$ & $0.371$ & $[1.115]$ & $33.28$ & $310.97$ & $74.2$ & $2.89$ & $6.160$ \\
                                Centre free, $r_{\rm min} = 1^\circ$& $81.14$ & $-69.42$ & $1.847$ & $0.374$ & $[1.115]$ & $33.21$ & $311.26$ & $74.0$ & $2.96$ & $7.110$ \\                           Centre free, $r_{\rm min} = 2^\circ$& $81.34$ & $-69.48$ & $1.845$ & $0.383$ & $[1.115]$ & $33.24$ & $312.74$ & $73.5$ & $3.21$ & $13.529$ \\
                                Centre free, $r_{\rm min} = 3^\circ$& $81.59$ & $-69.55$ & $1.844$ & $0.394$ & $[1.115]$ & $33.31$ & $313.35$ & $72.1$ & $0.20$ & $4.901$ \\           \hline
                \end{tabular}
                \caption{Parameters of the kinematic model fit to our data. Values in square brackets are 
                         held fixed for that model.}
                \label{tab:KinematicStructuralParameters}
\end{table*}

We also considered the case where we did not fix $\mu_{z,0}$, but left it as a 
free parameter. We find a value of $1.179\mathrm{mas}\,\mathrm{yr}^{-1}$, 
corresponding to a line-of-sight velocity of $288$\kms, which is a difference of 
about 7 \%\ from the value known from spectroscopy. The difference in 
inclination and orientation is around $1.5^\circ$ in each case, and the bulk 
motion in $x$ and $y$ is almost unchanged. The ability of measuring $\mu_{z,0}$ 
from the proper motions alone comes from the perspective contraction of the 
LMC as it moves away from the Sun, but we cannot expect this model-dependent 
result to provide a more accurate measure than from a spectroscopic study.

Finally, we considered the question of the rotational centre of the LMC. The 
easiest way to do this within our analysis is to allow the centre of the $x,y$
coordinate system to shift, and then recalculate the binned values $x_i,y_i,
\mu_{x,i},\text{and }\mu_{y,i}$ and uncertainties in the new coordinate system (in 
practice, we converted the binned values into equatorial 
coordinates, and then converted into the new coordinate system, rather than 
rebinning each time).
The rotational centre of the LMC has been a matter of debate, most notably 
with the photometric centre  and the centre of rotation for the H\textsc{i} 
gas  lying at different positions. The photometric centre was found to be ($81
^\circ.28,-69^\circ.78$) by \citet{vanderMarel2001}, who also found that the 
centre of the outer isopleths (corrected for viewing angle) was at $(82^\circ.
25, -69^\circ.50)$. The kinematic centre of the H\textsc{i} gas disc has been 
found to be $(79^\circ.40,-69^\circ.03)$ by \cite{Kim1998} or $(78^\circ.13,-
69^\circ.00)$ by \cite{Luks1992}. 
Using Hubble Space Telescope (HST) proper motions in the LMC, 
\cite{vanderMarel2014} found a rotational centre 
($78^\circ.76\pm0^\circ.52, -69^\circ.19\pm0.25^\circ$) that lies close 
to the centre of rotation for the H\textsc{i} gas, but pointed out that this was inconsistent with the 
rotational centre derived from studies of the line-of-sight velocity distribution in 
carbon stars \citep[e.g. 
$81^\circ.91 \pm 0^\circ.98, -69^\circ.87 \pm0^\circ.41$]{vanderMarel2002}. 
More recently, \cite{2020MNRAS.492..782W} used \gdrtwo proper motions, 
along with SkyMapper photometry \citep{2018PASA...35...10W} to find dynamic 
centres for carbon stars, RGB stars, and young stars -- $(80^\circ.90 \pm 0^\circ.29, -68^\circ.74 \pm 0^\circ.12)$,
$(81^\circ.23 \pm 0^\circ.04, -69^\circ.00 \pm 0^\circ.02),$ 
and $(80^\circ.98 \pm 0^\circ.07, -69^\circ.69 \pm 0^\circ.02),$ respectively.
% Better as a table?

We derive a centre of $(81^\circ.01,-69^\circ.38)$ when this was left as a 
free parameter, which is somewhat closer to the photometric centre than to 
the  H\textsc{i} centre. 
The inner regions of the galaxy do not provide much information in the 
proper motion field to find the centre because to first order, a linearly 
rising rotation curve produces a linearly varying proper motion field, so that the 
position of the centre is degenerate with the bulk velocity. The centre of 
the LMC does, however, have a significant non-circular motion, which is not 
captured by our model, and large statistical weight in our calculations 
(because of the high density of stars). We therefore investigated whether cutting data from the inner few degrees of the LMC changed our results. 
We did this by cutting data from our analysis with $x^2+y^2<r^2_{\rm min}$ 
(taking $x$ and $y$ from our original coordinate system, so that the data were the 
same for any centre considered), and re-deriving the parameters. The results 
are again listed in \tabref{tab:KinematicStructuralParameters}. The rotational 
centre moves slightly closer to the photometric centre as we cut larger areas 
from the centre of our dataset, suggesting that this result is robust against 
some of the incompleteness of our kinematic model. We tested whether 
changing the centre of our cut region affects the results (e.g. 
cutting data centred on the rotational centre of the H\textsc{i} gas instead), 
and the differences are very small.

\begin{figure}[h]
   \includegraphics[width=\columnwidth]{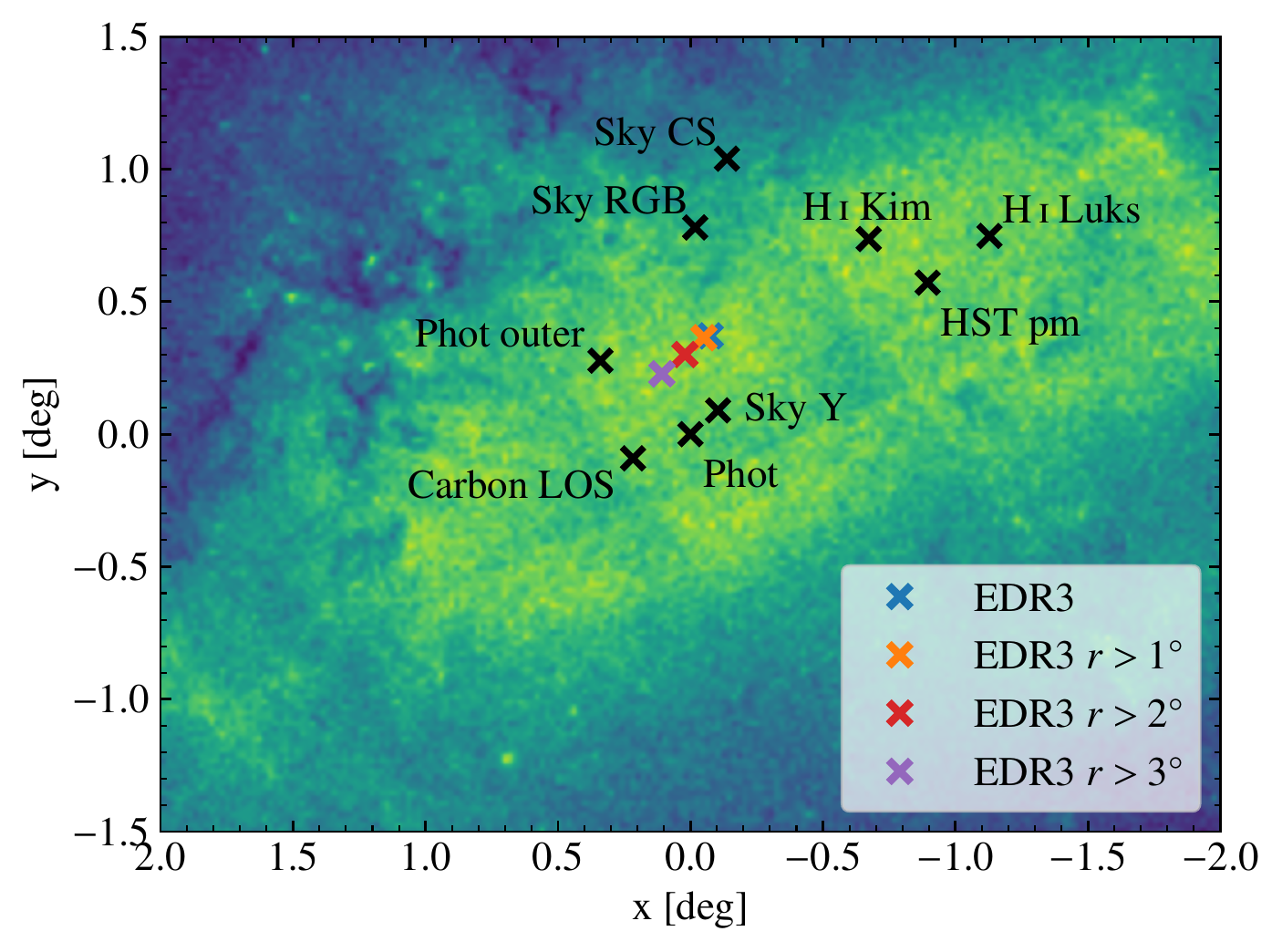}
        \caption{\label{fig:WhereIsTheCentre} Positions of the centre of the LMC as 
                 found by different studies (as described in the text, 
                 SkyMapper estimates are referred to as `Sky' in the figure) superimposed on an image coloured 
                 according to the total density in star counts in the inner few degrees of the LMC. The 
                                         centres found in this study lie closer to the photometric centre 
                                         than to the centre of H\,\textsc{i} rotation.}
\end{figure}

\begin{figure*}[h]
     \includegraphics[width=\textwidth]{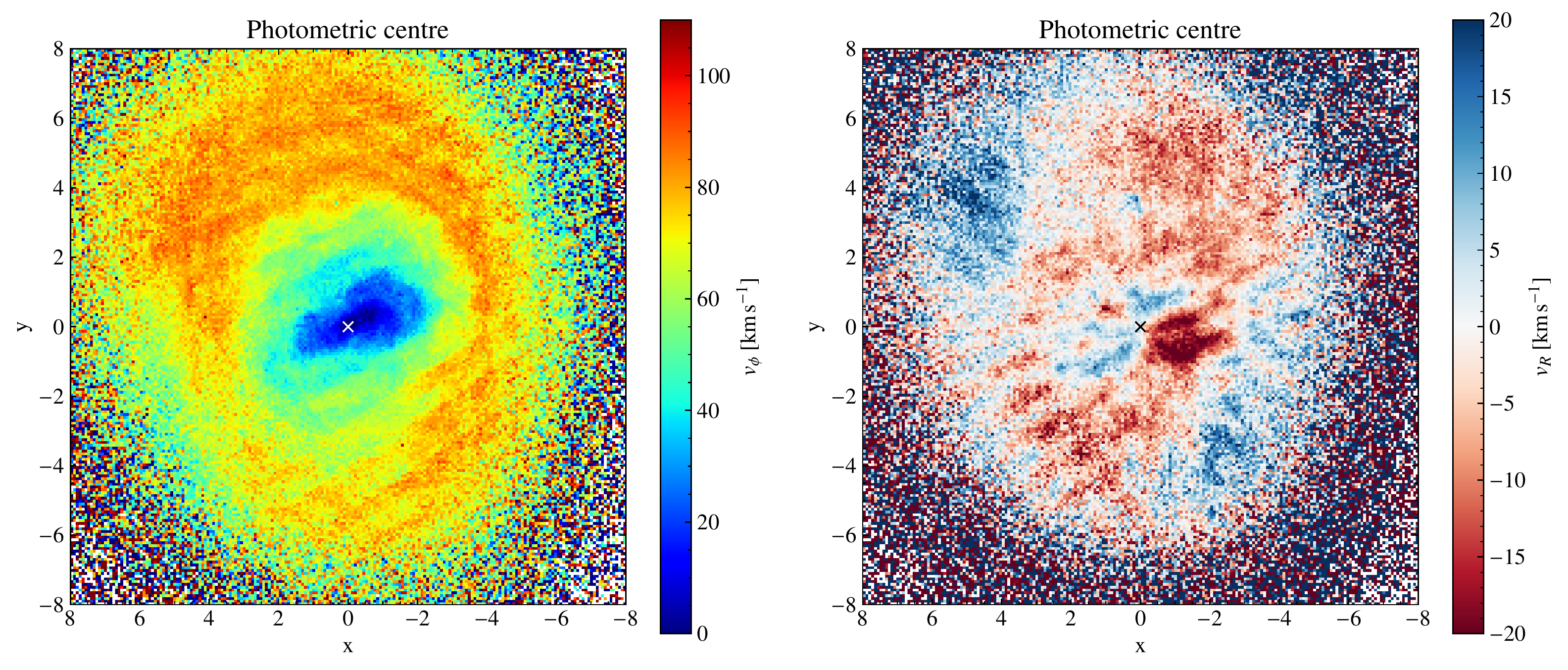}
     \includegraphics[width=\textwidth]{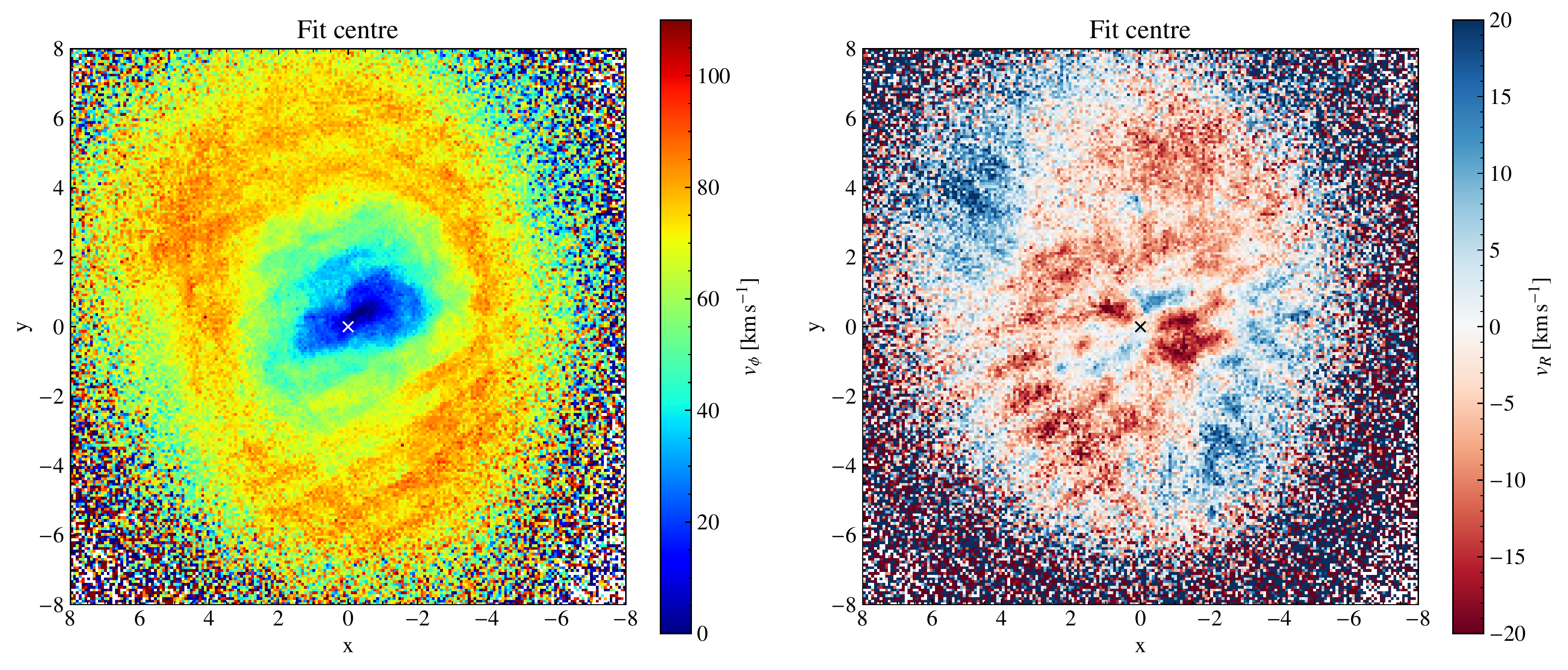}
     \caption{\label{fig:vRvphiTwoCentres} Plots of $v_\phi$ and $v_R$ for our main 
                          model (top) and for our model with the centre left as a free parameter (bottom).}
  \end{figure*}

In \figrefalt{fig:WhereIsTheCentre} we show the different proposed centres of 
rotation on a stellar density map of the centre of the LMC. The centres 
derived from \egdr3 are closer to those from photometric studies than to those from the rotation of H\,\textsc{i} gas or from proper motions measured 
by the HST.
The change in centre also naturally produces a change in derived bulk 
velocity, inclination, and orientation of the disc. The bulk velocity 
changes by $\sim$$0.02\,\mathrm{mas}\,\mathrm{yr}^{-1}$, which at the 
distance of the LMC corresponds to a velocity difference of $5$\kms. The inclination and orientation only change by about $1^\circ$.
We show plots of $v_\phi$ and $v_R$ for our main model, and our model with 
the centre left as a free parameter (considering all data), in 
\figref{fig:vRvphiTwoCentres}. As expected, the differences are relatively 
minor, although the outer parts the north-south asymmetry 
of the $v_\phi$ field is clearly reduced when the centre is left as a free parameter. 
The strong east-west asymmetry in $v_R$ near the centre is also reduced 
(but because the minimum in $v_\phi$ also appears to be offset from the centre, we are cautious about giving too much weight to this fact).
  
%%%%%%%%%%%%%%%%%%%
\subsection{Kinematics analysis}
\label{sec:kinlmc}

After we robustly constrained the main parameters with the simple 
rotation model, we built maps of the azimuthal and radial velocities and 
velocity dispersions for each of the stellar evolutionary phases of the 
LMC, as well as for a sample combining all phases. This latter sample 
is referred to as the combined sample in this section and in Sect.~\ref{sec:Spiral_structure}. 
These maps were thus derived at fixed 
and constant parameters with radius, as given by the main model of
\secref{sec:kinematics_method} ($i=34\degr$, $\Omega=310\degr$, $\alpha_0=81.
28\degr$, $\delta_0=-69.78\degr$, 
$\mu_{x,0}=1.858\, \rm mas\, yr^{-1}$, $\mu_{y,0}=0.385\, \rm mas\, yr^{-1}$
, and  $\mu_{z,0}=1.104\, \rm mas\, yr^{-1}$). 

The angular resolution of the maps can be chosen to be as high as possible. In practice, the maps were made of $400 \times 400$ squared pixels of 0.04\degr\ 
size, which is sufficiently resolved for the simple analysis of the kinematics proposed here,  
and it avoids more significant statistical noise inherent to higher 
resolutions. At the assumed distance to the LMC, it corresponds to a linear scale of 35 pc, which 
is equivalent to that of observations made at 0.7\arcsec\  resolution (i.e. the typical 
seeing at e.g. the ESO Very Large Telescope) of galaxies located 
at the periphery of the local 10 Mpc volume.  At this resolution, the maximum 
number of stars per pixel is 
599, 288, 265, 239, 136, 105, 54, 52, and 13 for the combined, \textit{Young 3}, 
\textit{RC}, \textit{RGB}, \textit{Young 2}, \textit{BL}, \textit{RRL}, \textit{Young 1},  
and  \textit{AGB} samples, respectively. Despite the low surface density of the 
\textit{AGB}, we were able to infer useful quantities, and we found that on average, 
\textit{AGB} star kinematics compare well with other evolutionary phases. The maximum 
likelihood of \equref{eq:likelihoodpixel} then yields the tangential and radial components of the velocity and velocity dispersion
for each pixel. 

\subsubsection{General trends}

 Appendix~\ref{sec:appendixvelolmc}  presents 
the \vt, \vr, \st\ , and \sr\  maps for the eight stellar LMC subsamples, as well as those of the combined sample. These maps are the first of their kind
ever obtained for an extragalactic disc, and the first maps that cover the integrality of the stellar kinematics for a galactic disc. 
To keep the description short in view of such a large quantity of kinematic 
data for a single galaxy, we present here 
example maps for two  evolutionary phases only. We selected an evolved phase (\textit{RC} stars) and a less evolved phase (\textit{Young 2} stars), which are both assumed to trace the kinematics of older and younger stellar ages.

\begin{figure*}[h]
                \includegraphics[width=0.95\textwidth]{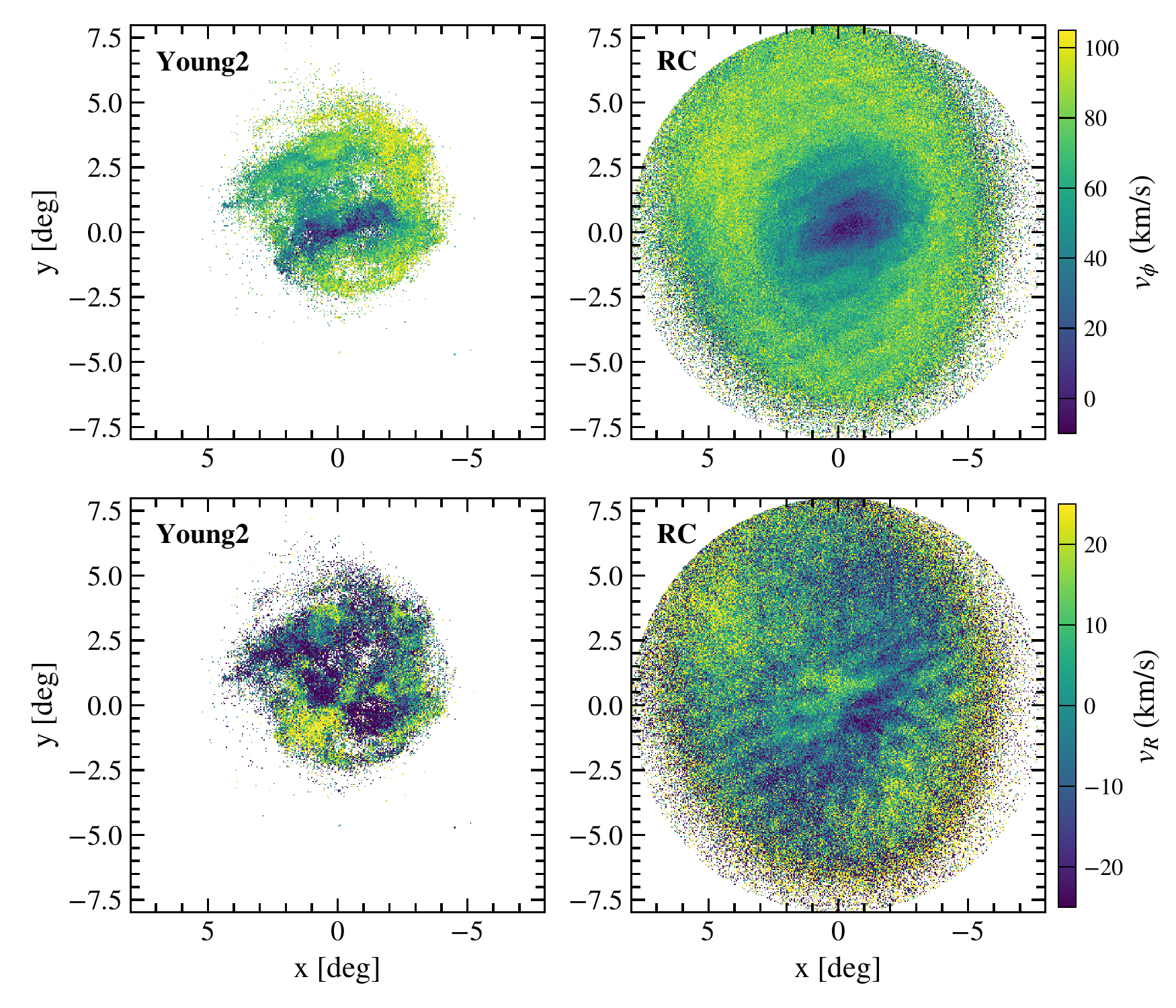}
                \caption{\label{fig:lmcexamplevelo} Example of velocity maps for the LMC. The left and right 
                         columns show young and evolved evolutionary phases (\textit{Young 2} and \textit{RC}, respectively,
                                                 see \secref{sec:selection_evophase}). 
                                                 The upper and bottom rows show \vt\ and \vr, respectively. The linear velocity scales 
                                                 shown by colour bars are the same for the two stellar evolutionary phases and were 
                                                 chosen to show the structure inside the velocity fields more clearly.}
\end{figure*}

\begin{figure*}[h]
                \includegraphics[width=0.95\textwidth]{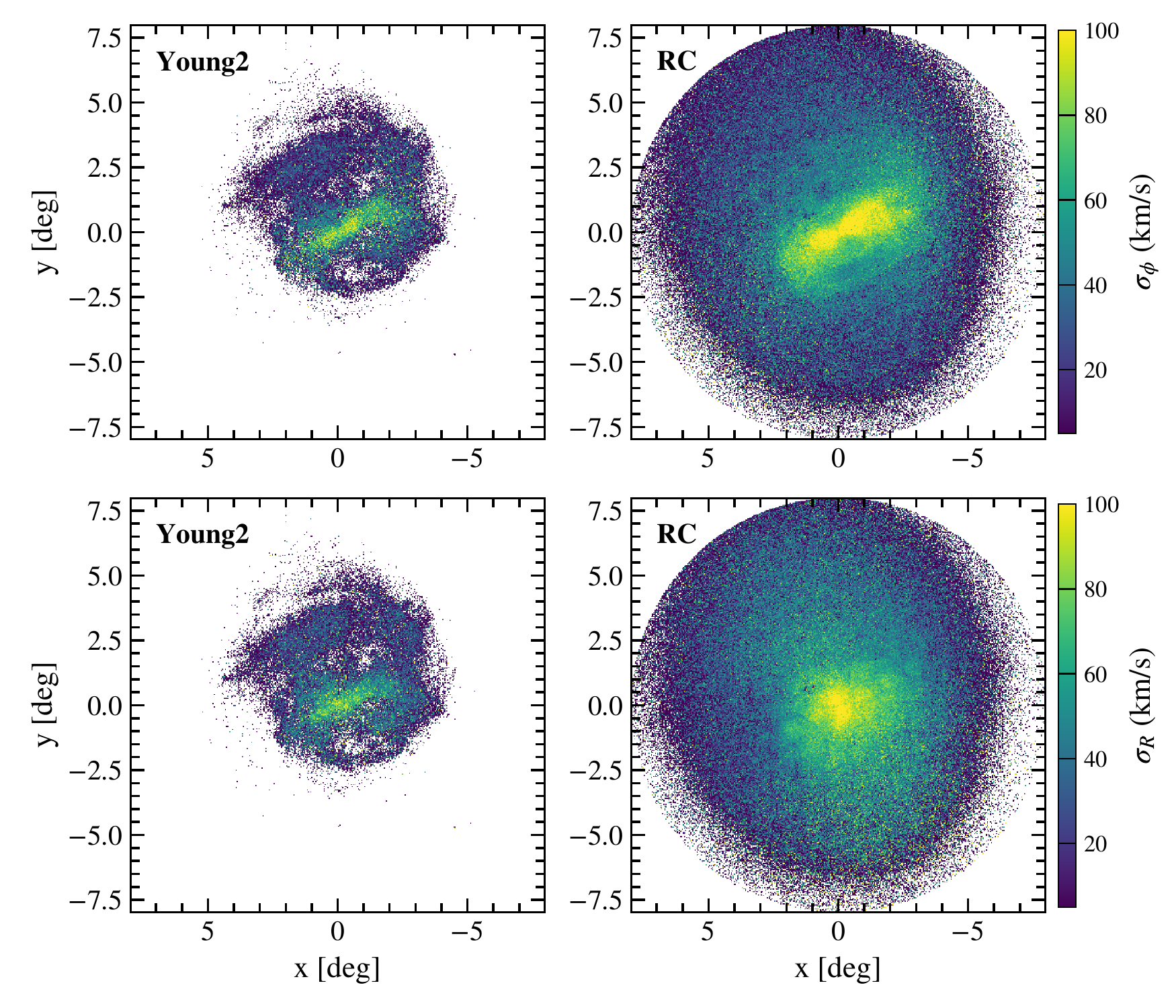}
                \caption{\label{fig:lmcexamplesigvelo} Same as in Fig.~\ref{fig:lmcexamplevelo}, but for the velocity dispersion.}
\end{figure*}

The \vt\ and \vr\ maps of the two phases are shown in Fig.~\ref{fig:lmcexamplevelo} and their 
corresponding velocity dispersion maps in Fig.~\ref{fig:lmcexamplesigvelo}. 
They all exhibit the noisy sawtooth patterns visible in the \gaia proper 
motion fields (\secref{sec:DR2DR3}), 
as well as variations occurring at larger angular scales that may likely reflect the perturbed kinematics in the spiral arms and the bar 
(see also \secref{sec:Spiral_structure}).

Such maps present the diversity and similarity in kinematics of the 
various stellar evolutionary phases. For instance, the younger phase 
presents higher tangential motions than the older phase (e.g. 45 versus 27 
\kms\ at $R=1$ kpc, or 88 versus 77 \kms\ at $R=4$ kpc, on average), which is 
a beautiful signature of the asymmetric drift, while both of them 
present lower velocities in a region that is apparently aligned with the stellar bar, 
with tens of pixels sometimes at negative values (e.g. down to $\sim -25$ 
\kms). It needs to be investigated further whether these negative values reveal counter-rotation in the bar or 
artificial features resulting from 
incorrect assumptions in this perturbed region of the disc, that is, that the 
stars only orbit in the $z=0$ mid-plane and with
$v_z =0$. 

The radial velocity map shows similar trends, with stronger motions for
the young phase than for the \textit{RC} sample.
Overall, the radial motion is mostly negative for $R\lesssim 5$ kpc, 
indicating inwards bulk motions towards the centre of the LMC, although this 
picture strongly depends on the location in the disc. Alternating 
negative and positive velocity patterns as a function of the azimuthal 
position, apparently centred on the assumed photocentre at $x=y=0$, are  
indeed visible in the bar and spiral arms. Similarly to the rotation velocity, 
the velocity streaming of $v_R$ appears to be weaker for the older stars than for the less evolved stars.

The radial and tangential velocity dispersion maps are also rich in 
information. Globally, the radial dispersion dominates the tangential dispersion in both samples, and the difference between the components increases with radius. 
There is an extended pattern of large random motions aligned with the 
bar in both kinematic tracers, but also a dominant feature 
in \sr\ that is perpendicular to the bar (only for the \textit{RC} sample). In this 
inner region of the bar, \sr\ is also 
observed to be larger where \st\ is lower for \textit{RC} stars, which indicates a variation in the velocity 
anisotropy as a function of the azimuthal position in the bar region. 
As for the comparison of the samples, random motions of the \textit{RC} sample 
are always larger than those for the young stars (e.g. 105 versus 80 \kms\ 
in the innermost pixels, or 45 versus 20 \kms\ at $R=3$ kpc, on average), as expected for more evolved stars 
that lie in a thicker disc component  than younger stars. 

\subsubsection{Velocity profiles}

The 36 velocity profiles\footnote{The velocity profiles are only 
available in electronic form at the Centre de Donn\'ees Astronomiques de 
Strasbourg via anonymous ftp to cdsarc.u-strasbg.fr (130.79.128.5)
or via \url{http://cdsweb.u-strasbg.fr/cgi-bin/qcat?J/A+A/}}
are shown in Appendix~\ref{sec:appendixvelolmc}.
The profiles are the median values of all pixels from the maps located in 
radial bins of 200 pc width. This angular sampling suffices to identify 
variations of slope and amplitude in curves in the evolutionary phases.
Radial bins with fewer than 5 pixels were discarded. The 
associated errors were derived from bootstrap resamplings of the velocity distributions and velocity dispersion at a given radial bin, 
at the 16th and 84th percentiles.  

\begin{figure*}[h]
                \includegraphics[width=\columnwidth]{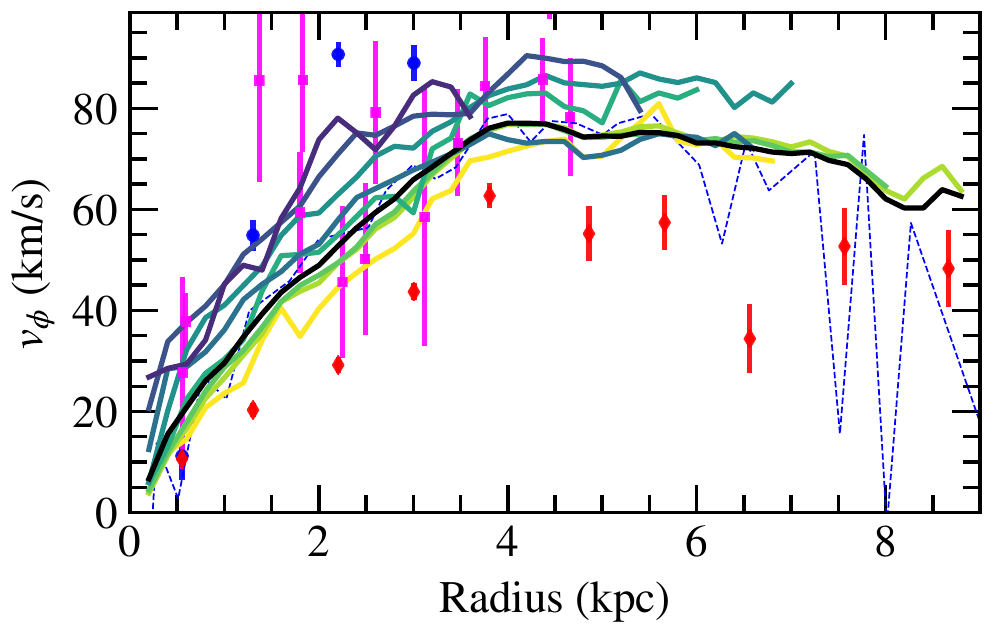}
    \includegraphics[width=\columnwidth]{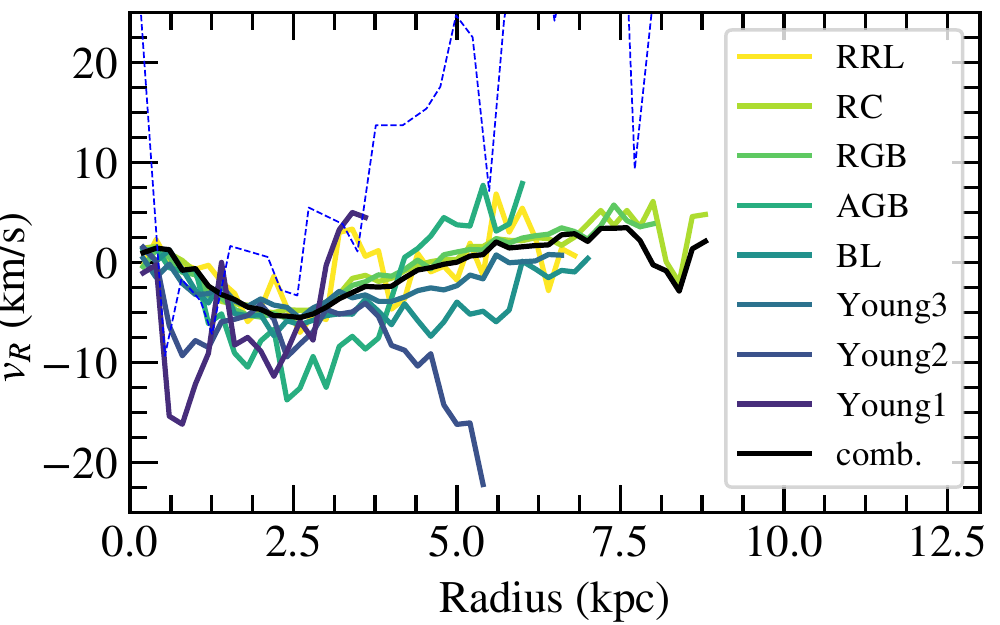}
                \caption{\label{fig:lmccompvphi} Stellar velocity curves of the LMC. The left and right 
                                                        panels show the rotation curves and radial motions, respectively.
                                                        Coloured lines are for the eight evolutionary phases and the combined sample, as given by the legend.
                                                        In the left panel, blue circles, red 
                                                        diamonds and magenta squares are the rotation velocities for the samples of younger 
                                                        and older stars from line-of-sight spectroscopic and HST astrometric 
                                                        measurements published in \citet{vanderMarel2014}. The dashed blue lines are the tangential 
                                                        and radial velocity profiles derived by \cite{2020MNRAS.492..782W} from \gdrtwo\ astrometry.}

\end{figure*}

Figure~\ref{fig:lmccompvphi} summarises the segregation of \vt\ as a function of the evolutionary stage. 
The more evolved the stellar population, the shallower the rotation curve at low radius, 
and the lower the amplitude; this is an expected result from the asymmetric 
drift. Taking \textit{Young 1} as a reference sample with the highest values, we find that on average, the amplitude of the rotation curve of \textit{Young 2}
stars is smaller by 0.6 \kms\  (thus similar  within the errors), and that of the 
\textit{BL, Young 3, AGB, RGB, RC, \textup{and} RRL} samples  by 6, 10, 13, 17, 18, and 22 \kms\ , respectively. The amplitude of  
 the combined sample lags by 15 \kms, as it is indeed dominated by the more numerous evolved stars. 
The \textit{BL} curve is always above the \textit{Young 3} curve, and 
the \textit{AGB} curve is above the \textit{Young 3} curve as well, but only 
beyond $R \sim 3$ kpc. 
Younger phases tend to have flatter rotation curves than more evolved 
stars. Finally, the curves of younger stars show wiggles, which are likely caused
by the perturbed kinematics from the bar and spiral perturbations.  
The effects from the sawtooth pattern in the proper motion fields are averaged when 
the curves are derived, and should contribute little to the observed wiggles. 

Figure~\ref{fig:lmccompvphi} compares our rotation curves with a  \vt\ profile 
of carbon stars, as obtained by \cite{2020MNRAS.492..782W} from modelling the 
\gdrtwo\ astrometry. The curves of the more evolved stars from our samples agree well with their curve for $R < 6$ kpc. Beyond this radius, the 
scatter is large in the kinematics of the carbon stars, and the curves disagree. 
The difference is likely caused by more significant noise in \gdrtwo\ astrometry 
than in \egdr{3}.  

Comparisons with stellar rotation curves derived from line-of-sight 
velocities and HST astrometry as published in \citet{vanderMarel2014} are also shown 
in Fig.~\ref{fig:lmccompvphi}. The HST rotation curve of mixed stellar 
populations shown as magenta squared symbols
agrees well with the \gaia curves within the quoted errors and 
scatter, but it has three outliers (one above 80 \kms\ is not shown).  
The rotation velocity of old stars shown as red diamonds is systematically 
lower than that of our curves, while those of the young stars shown as blue 
circles are in fair agreement with the kinematics 
of the less evolved population, despite the discrepant point at $R=2.2$ kpc. 
The  large difference with the line-of-sight velocities of the older 
stars is not understood because the orientation parameters quoted in 
\citet{vanderMarel2014} do not differ strongly from those adopted here. 

The \vr\ profiles (right panel of Fig.~\ref{fig:lmccompvphi})  mainly 
show dips with minima located at $R=2.5-3$ kpc, near the end of the bar, 
except for the least evolved stars. The \textit{Young 1} and 
\textit{Young 2} samples indeed exhibit stronger average inwards motion 
at lower radius (down to $v_R\sim -15$ \kms, $R \sim 1.5$ kpc). The 
radial motion of \textit{Young 2} stars also strongly decreases 
beyond $R=3$ kpc. Figure~\ref{fig:lmccompvphi} also shows large 
discrepancies between the curves of the more evolved stars with 
the \vr\ profile of carbon stars derived by \cite{2020MNRAS.492..782W}. 
Most of their radial velocities are $>10$ \kms, and show radial motions 
that significantly increase as a function of radius.

Appendix~\ref{sec:appendixvelolmc} also shows the variation in the slope and amplitude 
of the velocity dispersion profiles as a function of the evolutionary 
phase.  For example, the youngest phase \textit{Young 1} presents almost flat profiles, with  low amplitudes ($< 30$ \kms), whereas the random 
motions of more evolved stars are steeper, and with larger amplitudes in the centre (up to 100 \kms). Again choosing \textit{Young 1} as a reference 
sample, we measure that on average, \sr\ of the \textit{AGB, Young 2, BL, RGB, RRL, Young 3,} 
and \textit{RC} samples is larger by 5, 21, 24, 37, 40, 40, and 52 \kms\ , respectively. The amplitude of the combined sample is larger by 44 \kms. Similar 
mean differences are observed with the tangential component of the velocity 
dispersion. 

%======================================================================================
% Section: Bridge
%======================================================================================
\section{Magellanic Bridge and the outskirts of the Magellanic Clouds \label{sec:Bridge}}

One of the most prominent features in the outskirts of two interacting
galaxies is the formation of a bridge between them due to tidal forces
that strip gas and stars from the least to the most massive galaxy 
\citep{Toomre1972}. The relative position of the Milky Way with respect to the
Magellanic Clouds places us in the privileged position of witnessing the close
encounter between the LMC and the SMC, and of studying the Magellanic Bridge.

The stellar characterisation of the structure and kinematics of the Magellanic Bridge 
has been pursued for a long time, with simulations
\citep[e.g.][]{Besla2012,Diaz2012} and observations
\citep[e.g.][]{Irwin1985,Harris2007,Bagheri2013,Noel2013,Carrera2017}. In addition to this expected tidally induced feature, other structures such as plumes,
shells or stellar streams can be found in the outskirts of the Magellanic
Clouds \citep[e.g.][]{Deason2017,Mackey2018,MartinezDelgado2019,Navarrete2019}.  
In this section we show the quality of \egdr{3} in highlighting the
Magellanic Bridge and its kinematics, and we show several
equally interesting features in the outskirts of the Magellanic Clouds.

The Magellanic Bridge was first detected as an overdensity
in HI gas by \citet{Hindman1963}.  More recently, several studies have tried to
follow the connection between the LMC and SMC using samples of stars in different 
evolutionary phases. Because tidal forces have similar effects on stars and gas, the Bridge
would be traced by both a young stellar population with a strong correlation
with the HI distribution, and an old population made of stars stripped into the
Bridge by the tidal interaction. This is supported by dynamical simulations
\citep[e.g.][]{Guglielmo2014}.  The stellar Magellanic Bridge was first traced
by a population of young stars \citep{Irwin1985} showing in situ star
formation and a strong
correlation between the location of the stars and that of HI overdensities
\citep[e.g.][]{Skowron2014}. \citet{Casetti-Dinescu2012} selected young 
OB-type stars in a wide area between the Clouds to study the structure and 
kinematics of the Bridge using GALEX, 2MASS, and the Southern Proper Motion 4 catalogue.
\citet{Jacyszyn-Dobrzeniecka2020a} used 
Cepheids from the OGLE Collection of Variable Stars to characterise the Magellanic
Bridge with young stars, while \citet{Bagheri2013} and \citet{Noel2013} used
 RGB stars to search for an old counterpart. 
 Spectroscopic confirmation of stripped stars at the SMC side of the Bridge was
obtained by \citet{Carrera2017}.  Very recently, \citet{Grady2020} assembled a catalogue of
red giants from \gdrtwo from which the authors obtained photometric metallicities using 
machine-learning methods. Based on the metallicity structure in the Magellanic Bridge, the authors
concluded that it is composed of a mixed stellar population of LMC and SMC debris.

In this section we explore the \gaia capabilities of detecting and characterising the
Magellanic Bridge using the evolutionary phase samples described in
\secref{sec:samples}. Because the Magellanic Bridge encompasses the region in which the MC overlap, we
have to adopt a modification of the selection described in \secref{sec:samples}. This
modification takes into consideration that LMC (SMC) stars may extend farther
than 20 (11) degrees and overlap with each other spatially and in proper
motion. Our query is identical to the one described in \secref{sec:samples}, but we
queried by \textsc{HEALpix} (\texttt{NSIDE=8}) pixels that have a separation to
their centres smaller than 35 (15) degrees from the LMC (SMC) centre. The
proper motion selection described in \secref{sec:samples} was performed, but we did not
adopt any separation from either of the clouds as a membership criteria. This
produced a sample that allowed for overlap in space and velocity and also provided  a larger sky-coverage that is useful to explore stellar
structures in the outskirts of both clouds. The total number of stars and the number in
each stellar phase subsample agrees well with what we reported in
\secref{sec:samples}, but because we allowed stars to mix in PM and on sky, we obtained
numbers that are generally larger by < 1\% than in \secref{sec:samples}.

To study the Bridge, we defined two populations, one representative of the young
population, and the other the RC population in both clouds. The young population
was defined as the combination of \emph{Young 1$_{LMC}$}, \emph{Young 2$_{LMC}$},
\emph{Young 1$_{SMC}$}, and \emph{Young 2$_{SMC}$}, that is, inner-joined with the
combination of the PM-selected LMC and SMC populations. It contains $167,643$ sources. 
The RC population is defined in the same way, but using the RC subsamples. 
It contains $1,806,102$ sources.

\begin{figure*}[h]
   \begin{subfigure}[b]{0.5\textwidth} 
      \begin{center}
      \includegraphics[width=\columnwidth]{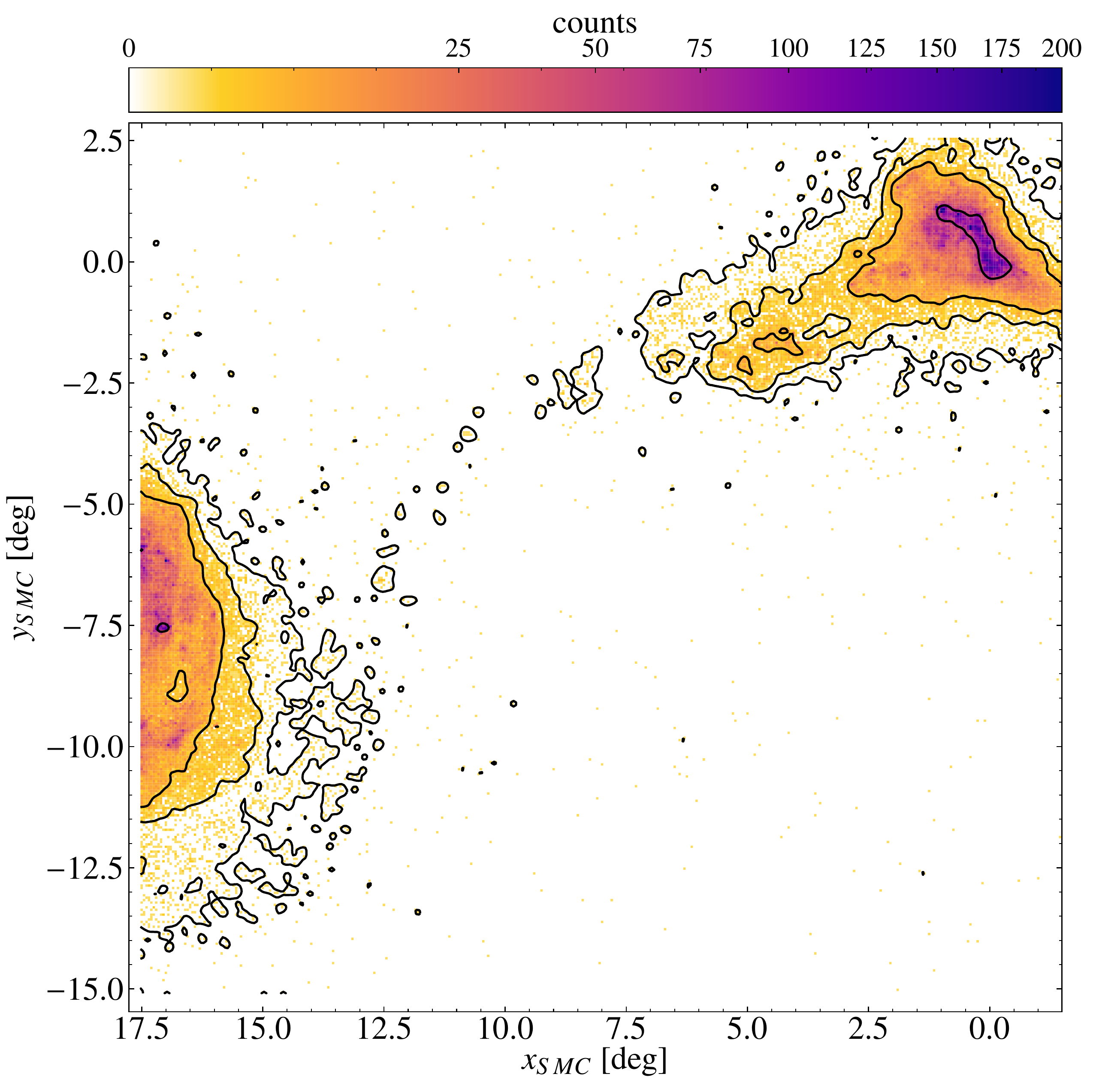}
      \end{center}
         \end{subfigure}
   \begin{subfigure}[b]{0.5\textwidth} 
      \begin{center}
      \includegraphics[width=\columnwidth]{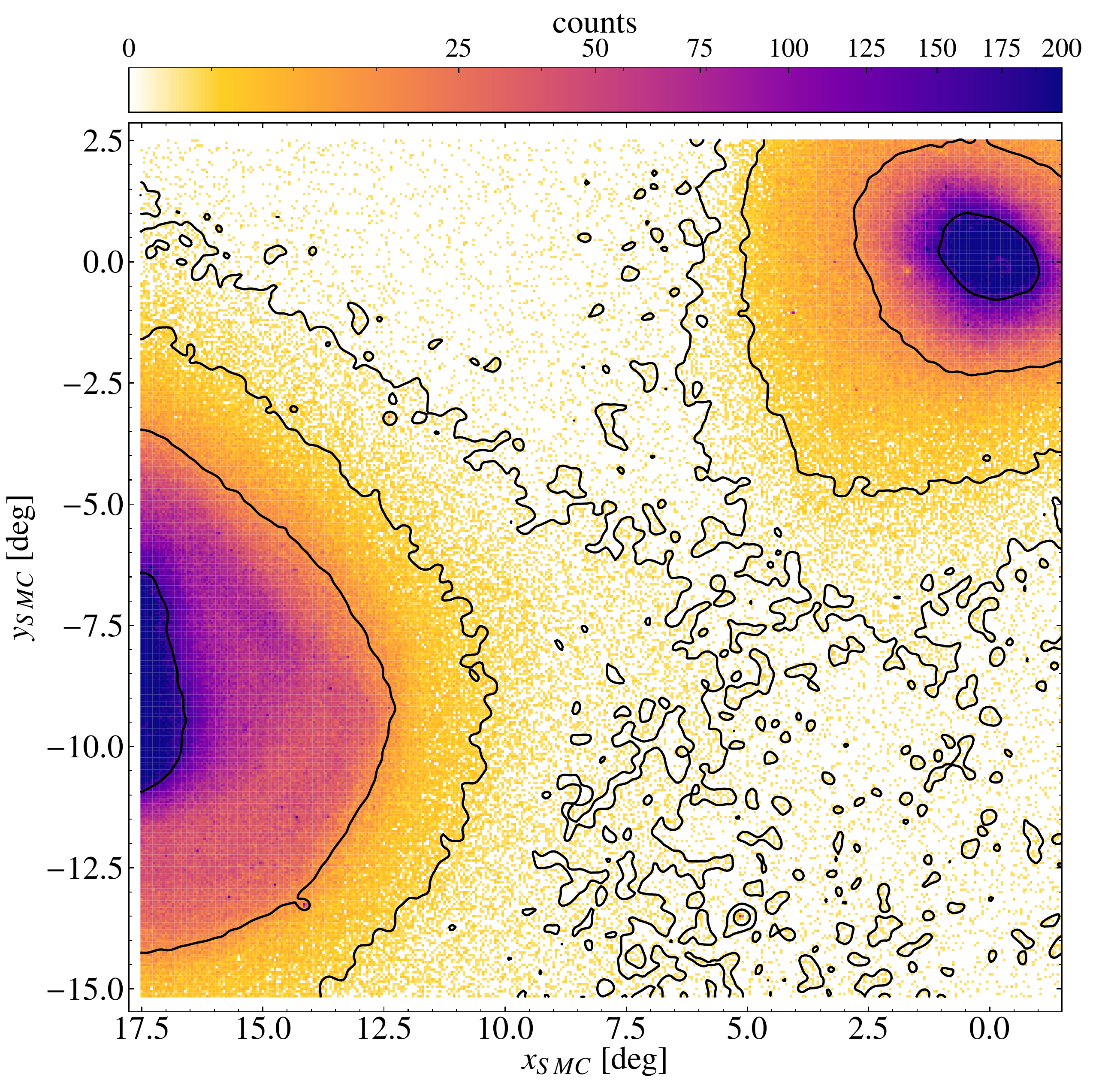}
      \end{center}
         \end{subfigure}
   \caption{Sky density plot for the Bridge region using the Young1 and Young2
       evolutionary phases (left panel) and the RC evolutionary phase (right
       panel). The bin size is $0.06$~deg in $x$ and $y$. The colour bar is in
       log scale and the black contours are at the levels $0.1$, $1$, $5,$
       and $70$ (young evolutionary phases) and $0.25$, $1.5$, $15,$ and       $165$ (RC sample). }
   \label{fig:bridge_density}
\end{figure*}

In the left panel of \figref{fig:bridge_density} we show a density plot of
the Young stellar population in the Bridge region; the 
connection between the two galaxies is obvious without applying any
statistical technique. The morphology of the young Magellanic Bridge is represented by an 
arched elongated connection between the Magellanic Clouds.  
In the right panel of \figref{fig:bridge_density} we show the density plot of
the RC sample in the Bridge region. 
The Magellanic Bridge in the \textit{RC} sample is not so clear, as
expected from a more evolved and kinematically hot population, although it has
been traced in RC stars by \citet{Carrera2017} at the near side of the SMC using
the MAGIC spectroscopic survey. In this case, it is of key importance to remove
the Milky Way foreground contamination of RC stars. This exploration has been
performed by \citet{Zivick2019} using \gdrtwo data and HST proper motions. \citet{Belokurov2019} used astrometry and broad-band photometry from \gdrtwo, and  
\citet{Schmidt2020} used data from the VISTA survey of the Magellanic Clouds 
\citep{Cioni2011} and \gdrtwo to perform a kinematic study of the region around the MC and of the Bridge region.

\begin{figure*}
   \begin{center}
   \includegraphics[width=\textwidth]{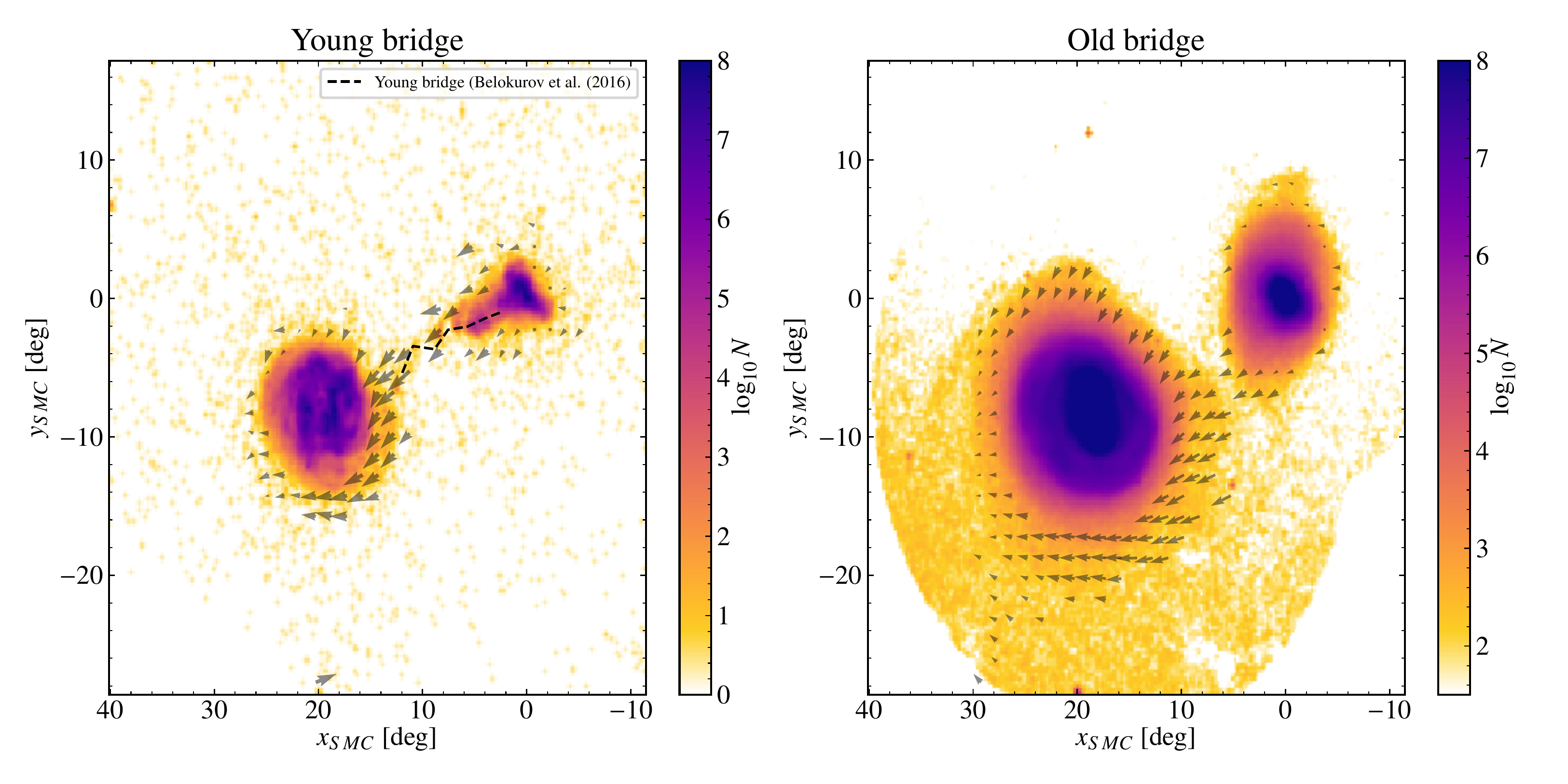}
   \end{center}
   \caption{Vector field of the proper motions in the Magellanic Clouds using
       the \textit{Young 1} + \textit{Young 2} (left panel) and \textit{RC} (right panel) 
                         samples.
       The coordinates are centred on the SMC. In the background,
       to guide the eye of the reader, we show the density in logarithmic scale. The dashed line in the left panel shows the location of the young bridge from
       \citet{Belokurov2017}. The velocity vector field is only shown for bins with
       more than 10 (200) stars in the \textit{Young 1} + \textit{Young 2} 
                        (\textit{RC}) sample.} 
   \label{fig:bridge_flow}
\end{figure*}

In \figref{fig:bridge_flow} we specifically use the proper motions included in
\egdr{3} to study the kinematic interaction
between the Magellanic Clouds. We checked the dynamical attraction of the LMC on
the SMC by plotting the vector field of the proper motion of the sources.We separately show the \textit{Young 1-2} (left panel) and the \textit{RC} evolutionary phases
(right panel). In contrast to the density plot (see
\figref{fig:bridge_density}), we clearly observe, using both evolutionaryphases, a coherent motion of stars from the SMC towards the LMC. For young stars the flow moves as we
would expect, from the SMC to the LMC along the Bridge (depicted in the background
density plot). We emphasise that the excellent quality of the \egdr{3} proper motions 
allows tracing the interaction between the Magellanic Clouds using a rather simple strategy to separate 
stars into different phases of evolution. 

The high quality of the \egdr{3} proper motions allows confirming a flow of
RC stars from the SMC towards the LMC. As mentioned above, the track of an old bridge between the LMC and SMC has recently been pursued using different tracers and strategies. In this demonstration paper we considered only the RC population, which is characteristic of  an intermediate to old population, and we did not use a typical $>10$ Gyr old population such as that of the RR Lyrae stars. Recent works have specifically targeted the RR Lyrae stars in the bridge region of the MC \citep[e.g.][]{ Belokurov2017,Clementini2019,Jacyszyn-Dobrzeniecka2020b}.
Based on their selection strategy, \citet{Belokurov2017} claimed an old RR Lyrae bridge for \gdrone RR Lyraes. The \gdrtwo bona fide RR Lyraes and those from the \egdr{3} sample (see \secref{sec:samples}) both show a smooth halo-like density distribution \citep{Clementini2019}, however. The \gdrtwo accompanying paper was confirmed by  \citet{Jacyszyn-Dobrzeniecka2020b} using the extended OGLE catalogue. \citet{Evans2018} stated in a \gdrtwo accompanying paper that a suboptimal computation affected the mean magnitude standard deviation given in \gdrone and \gdrtwo (and revised in \gdrthree \citep{EDR3-DPACP-117}), which may have affected the selection strategy of \citet{Belokurov2017} with only candidate RR Lyrae stars.
We show here that a flow of RC stars (see \figref{fig:bridge_flow}) confirms a bridge composed of an evolved population, and it would have a similar trajectory to that of \citet{Belokurov2017}. It is beyond the scope of this paper, however, to make a quantitative comparison.

In addition to the Bridge,  \citet{deVaucouleurs1972} showed a
wealth of substructure in the outskirts of the Magellanic Clouds in the 1970s. More
recently, new shells, plumes, and streams have been detected using different
surveys or photometric techniques \citep[e.g.][]{Pieres2017, Belokurov2019,
    MartinezDelgado2019}. To search for substructures around the Clouds, we
adopted a more restrictive selection using the \textit{RGB} and \textit{RC} subsamples. First, we corrected for foreground extinction using
\citet{Schlegel98} (with the correction from \citet{Schlafly11}), and we adopted a
\citep{Cardelli89} extinction curve with $R_V = 3.1$. This correction is
accurate in the outskirts of the Clouds because there is little internal
extinction from the LMC and SMC themselves. Second, we built a tighter
colour-magnitude selection polygon based on the extinction-corrected
\textit{RC} and \textit{RGB} samples; in this case, we are stricter
in the
colour range allowed for these two evolutionary phases as in the \textit{RC}
and \textit{RGB} samples described in \secref{sec:samples}.
Additionally, we applied a magnitude cut of G $<$ 19 and selected
    only stars with a parallax smaller than 0.15.  This led to a sample of stars
that is less strongly affected by Milky Way foreground, thus allowing us to explore
faint substructures in the outskirts; we call this selection \emph{LMCout}.

\begin{figure*}
   \begin{center}
       \includegraphics[width=\textwidth]{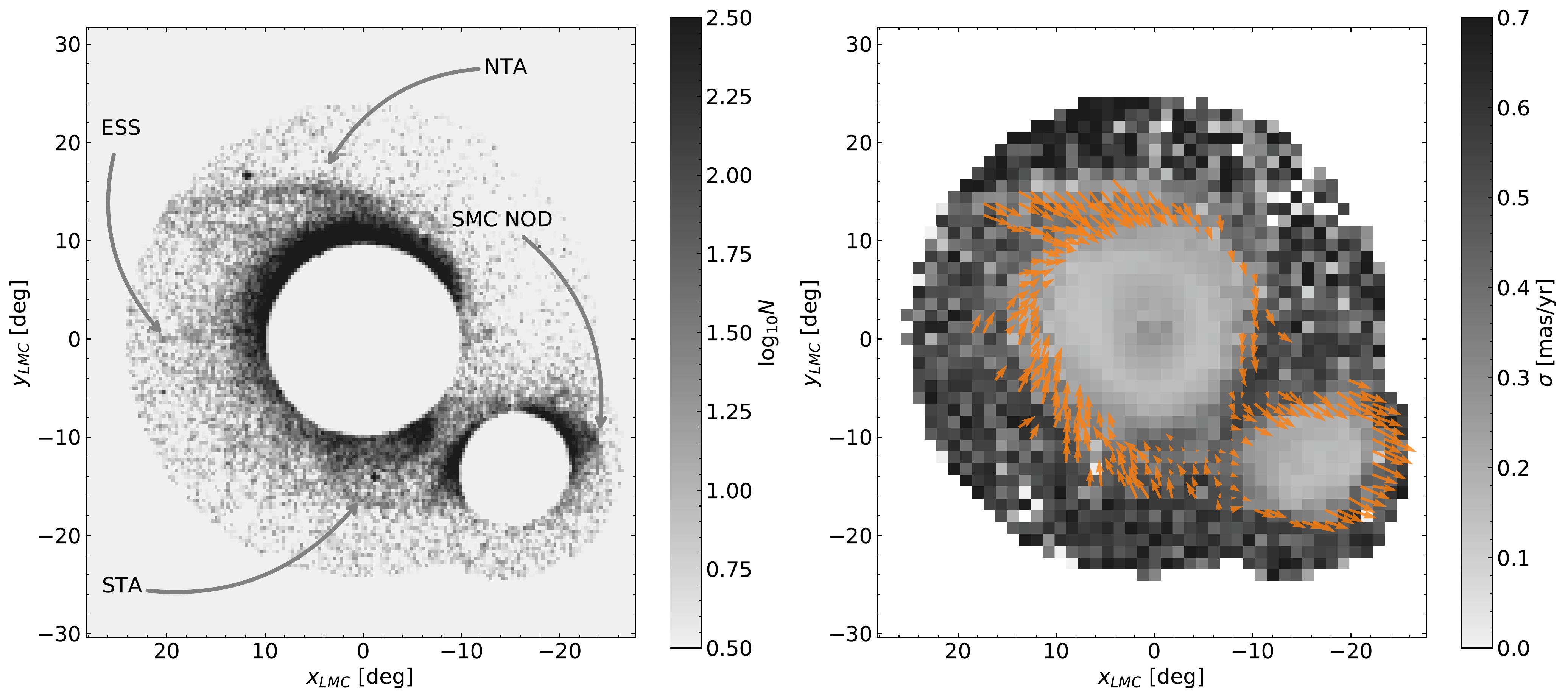}
   \end{center}
   \caption{Number count map for the \textit{LMCout} sample (left): the names of the most notable substructures 
						are given and, for a better visualisation, the inner parts of the LMC and SMC were omitted. 
						Total velocity dispersion map (right): the velocity vector field is shown as 
						orange arrows.}   \label{fig:outskirts}
\end{figure*}

In \figref{fig:outskirts} we show a star count map of the Magellanic Cloud
region to highlight the substructure found using \egdr{3}.  We also annotate a
few notable features and show the measured total dispersion and velocity vector
map (velocities and dispersion were computed using LMC-centred
coordinates). The northern tidal arm (NTA) reported initially by
\citet{Mackey16} is visible in the figure, and this structure is also visible as
a velocity low-dispersion feature, with velocities consistent with the LMC main
body.  A southern tidal arm (STA) \citep{Belokurov2019} is also evident, which
shows indications of being dynamically cold, like the NTA, and the velocities are
consistent with those of the LMC. The STA appears to be connected with the SMC through a narrow
elongation east of the SMC.  The SMC northern overdensity \citep{Pieres2017} is
also evident, and a spatially thinner structure is also seen to be connected
    to it. In addition to these known substructures, we find a faint
overdensity east of the LMC that is visible in the velocity field and
    density map. We also see a similar structure, but more conspicuous, on the
    western side of the LMC, close to $y_{LMC}=0$. We note, however, that the
    features observed near $y_{LMC}=0$ coincide with a region of elevated
    number of Gaia transits. The eastern feature is also prominent in
near-infrared maps from the VISTA Hemisphere Survey (El Youssoufi et al.,submitted).

%======================================================================================
% Section: Spiral arms
%======================================================================================
\section{Spiral arms in the Large Magellanic Cloud \label{sec:Spiral_structure}}

The LMC is a prototype of barred Magellanic spiral galaxies 
that are characterised by an off-centre bar and one prominent spiral arm. 
The dynamical interactions between the LMC and the SMC \citep[e.g.][]{Besla2016} 
are probably responsible for this and other spiral features associated with the 
galaxy \citep{deVaucouleursFreeman1972}. 
A comprehensive study of the morphology of the LMC based on  near-infrared  
observations is given in \citet{ElYoussoufi2019}, where high spatial resolution maps 
(0.03 deg$^2$) of stellar populations with different median ages show at least four 
distinct spiral features. These arms emerge predominantly from the ends of the bar, 
one in the east extending to the south, and three in the west, one of which extends north, one north-west (the most prominent arm), and the third extends south. 
The arms are well traced by stellar populations younger than a few million years, while old stellar 
populations instead show external features that may be associated with a ring-like 
structure \citep[e.g.][]{Choi2018}. 
The long-term stability of the prominent spiral arm was studied by \citet{Ruiz-Lara2020} 
using deep optical photometry to derive the star formation history throughout the galaxy. 
This structure could have formed a few million years ago at the time when the Magellanic Stream 
and the Leading Arm formed as well from a close encounter between the LMC and the SMC. 
The authors concluded that the distribution of HI gas and the coherent star formation at the 
location of the arm support this scenario. In this section we show that the spiral structure 
of the LMC can be highlighted and studied with the \egdr{3} data.

%%%%%%%%%%%%%%%%%%%
\subsection{Basic properties as a function of evolutionary phase}

\begin{figure*}[h]
                \centering
                \includegraphics[width=\textwidth]{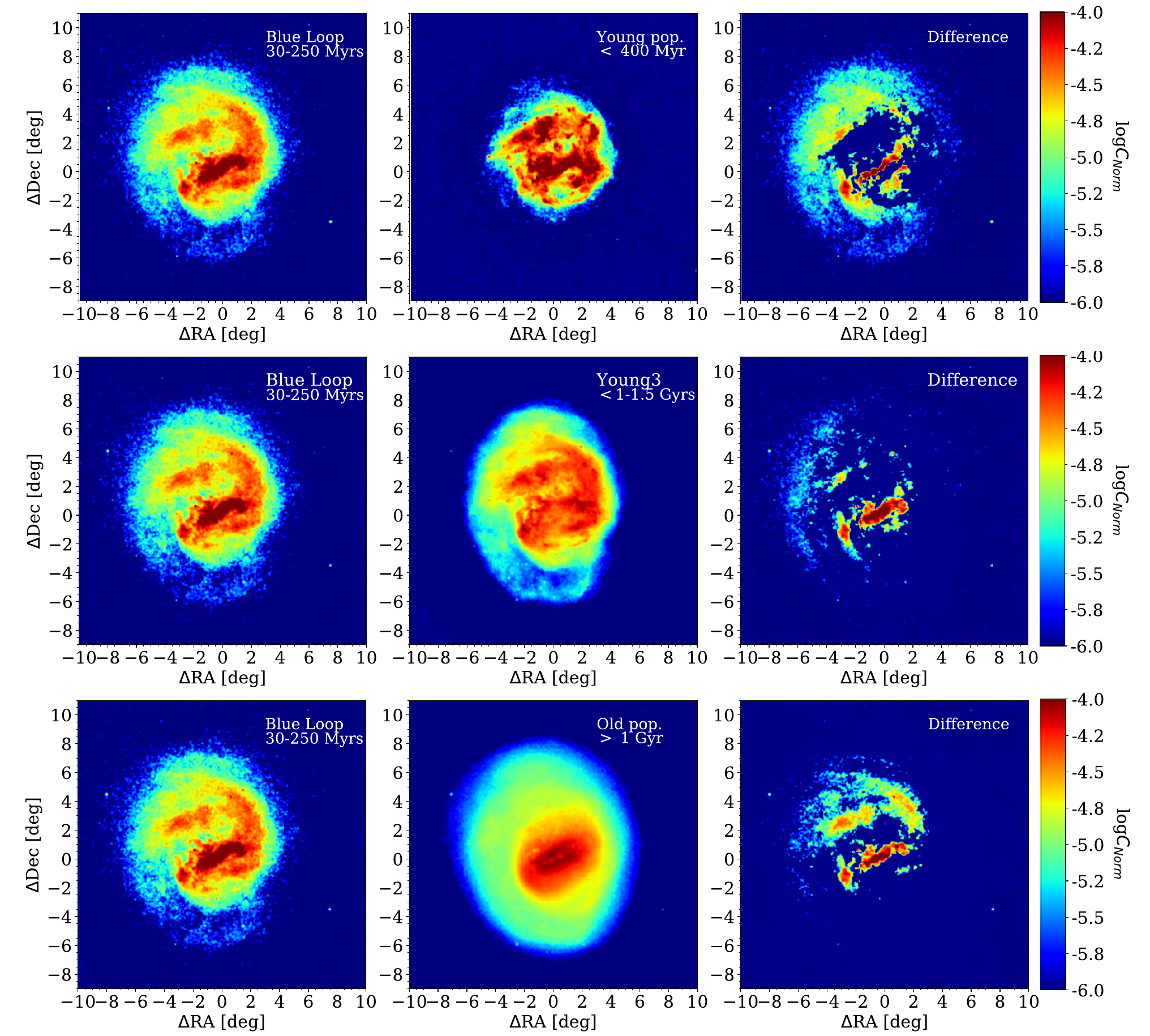}
                \caption{Differential density maps for different stellar populations in the LMC. The 
                         left column shows the density map in the \textit{BL} evolutionary phase.
                         The middle panels from top to bottom display the density maps                                             
												 of the young phase, i.e. stars with age $<$ 400 Myr, 
                         \textit{Young 3} phase, i.e. main- sequence stars with age $1-1.5$ Gyr, 
												 and old phase, i.e. stars with age $>$ 1 Gyr. 
                         The right column depicts the normalised difference between the left and 
												 middle panel maps, shown with a logarithmic stretch.}   
                \label{fig:mapDifference}
\end{figure*}

We discuss here the appearance of the spiral arms of the LMC using the 
evolutionary phase samples as proxies for age-selected samples 
(see \secref{sec:selection_evophase}). In \figref{fig:mapDifference}
we show the maps obtained in the LMC for these samples. The maps were 
constructed considering a region 
$20\times 20$ deg$^2$ around the galaxy, applying a Gaussian smoothing and 
sampling with $400\times400$ pixels, each with a dimension of $3
\arcmin \times 3\arcmin$.
Each map was normalised to the total number of objects.
The figure shows that the main structures of the LMC, that is, the bar and the 
spiral arms, are clearly outlined by \textit{BL} stars, objects with ages in the 
range of 50-350 Myr. We therefore chose this population as a reference for
the comparison of the spiral structure(s) in other stellar populations of different ages. 

Because the differential maps of the \textit{BL} 
with respect to \textit{Young 1} and \textit{Young 2} were  
similar, we merged these two evolutionary phases into a single sample. 
We refer to this merged sample as
the young population of stars with age $< 400$ Myr, which is 
shown in the middle top panel of Fig.~\ref{fig:mapDifference}.  Similar considerations 
applied to the \textit{RC}, \textit{RGB,} and  \textit{RRL} populations, all older 
than~1 Gyr, and we refer to these merged evolved populations as the old population. 
The relative differential map is shown in the lower panel of 
\figref{fig:mapDifference}. Finally, the differential map with respect to the 
\textit{Young 3} population (MS stars with ages $<1-1.5$ Gyr) is shown in the 
middle panels of the same figure. 

The analysis of the top panel in \figref{fig:mapDifference}   reveals that the 
young population is more concentrated around 
the bar and an inner northern arm, showing a clumpy structure.
The residual map with respect to the \textit{BL} shows that this latter population 
has an excess of stars along the bar, in the spiral feature at the end of the bar, 
and in an outer north-east arm (referred to as the eastern arm hereafter).

The comparison of \textit{BL} and \textit{Young 3} populations 
shows that the two populations are distributed in a very similar way, even though the \textit{BL} still displays an excess 
along the bar, especially in the eastern region, where it shows a concentration 
superior to any other population in the LMC ($\Delta {\rm RA}, \Delta {\rm DEC} \sim -3\degr, -1\degr$).
The older populations of the LMC have a homogeneous distribution; the star density decreases smoothly from 
the centre to the outskirts of the LMC. The lower density along the 
bar is caused by the \gaia incompleteness in this crowded region. The difference 
with the \textit{BL} population again shows an excess of stars along the bar and the above-mentioned clump in the eastern 
bar, but these features might in part be justified by the 
incompleteness of old populations in the more central region of the bar. In contrast, the excess of \textit{BL} stars in the inner and outer arms appears to be genuine.  

%%%%%%%%%%%%%%%%%%%
\subsection{Strength and phase of the density perturbations}
\label{sec:fftdensity}

To be more quantitative on the effect of the bar and spiral structure on the 
stellar density, we performed discrete fast Fourier transforms (FFTs) of density maps of the \textit{BL} and 
the combined samples. For this purpose, we again used 
$400\times400$ pixels maps, but at 0.04\degr\ sampling, as in 
\secref{sec:kinlmc}. This allowed us to estimate the properties of 
any asymmetries in the density maps. 

Because the apparent dominant modes of perturbations are the 
bar (second-order perturbation), the inner spiral structure starting at the end of the bar, and the 
eastern outer arm (first-order perturbation),  we present results up to the second-order 
harmonics, although the discrete FFTs yield as many orders as existing 
pixels in a vector. Therefore the analytic form equivalent to the discrete FFT applied to a 
density map is $\Sigma (R,\phi)  = \sum_k \Sigma_k (R) \cos k(\phi -\phi_k(R))$,
where $k$ is an integer, $\phi$ the azimuthal angle in the plane of the LMC with 
the reference $\phi=0$ chosen aligned to the photometric major axis of the disc 
($\Omega_{\rm phot}$), $\Sigma_0$ the axisymmetric surface density, 
and $\Sigma_k$ and $\phi_k$ are the amplitude and phase of the $k$-th 
asymmetry. 

We measured $\Omega_{\rm phot}$ by isophotal ellipse fitting 
to the stellar surface density map of the combined sample. To avoid confusion 
with Sect.~\ref{sec:kinematics}, which gives radii in a kinematic frame oriented along the kinematic position angle of  $\Omega=310\degr$, 
we refer to $R_{\rm phot}$ as the galactocentric radius measured in the photometric frame, which is aligned on $\Omega_{\rm phot}$.
With this, we find the bar semi-major axis at a position angle of $\sim 105\degr$, 
and define the one for the disc as that of the average value in the radial range 
$R_{\rm phot} =3.5-7.5$ kpc, which is $\Omega_{\rm phot} \sim 10\degr$. The photometric major 
axis therefore differs by $\sim 60\degr$ from the kinematic major axis. A similar 
discrepancy has been reported in \citet{vanderMarel2002}.

\begin{figure*}[h]
                \centering
                \includegraphics[width=\textwidth]{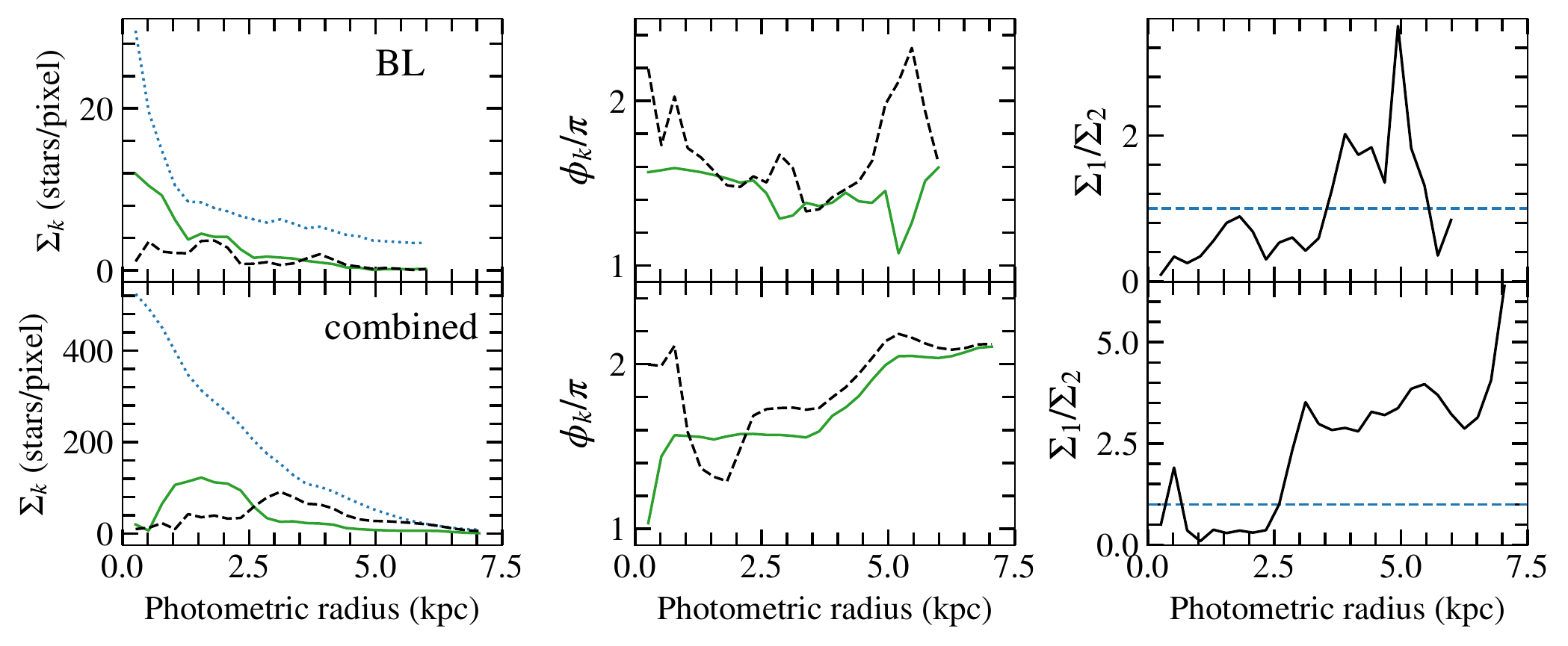}
                \caption{Results of the Fourier transform of stellar density maps of the LMC for the \textit{BL} 
                         and the combined samples (upper and bottom panels, respectively). 
                                                 The left and middle panels show the amplitude and phases of the first- and second-order 
                                                 asymmetries (dashed black and solid green lines, respectively). 
                         The axisymmetric density is shown as a dotted blue line. The right panel shows the 
                                                $k=1$ perturbation over the bisymmetry strength ratio.  
                        The galactocentric radius is given here in the photometric frame, with $\Omega_{\rm phot} = 10\degr$ as the reference photometric position angle.}   
                \label{fig:fftdensity}
\end{figure*}

\figref{fig:fftdensity} shows the results of the Fourier transforms.  We restrict the analysis to $R_{\rm phot} \le 7.5$ kpc. The axisymmetric density profile of 
\textit{BL} stars is more centrally peaked than that of the whole sample. 
At the peak of $\Sigma_1$ and $\Sigma_2$, the strengths of the lopsided outer spiral arm ($k=1$)
and bisymmetry ($k=2$) reach 60\% and 40\% the amplitude of the axisymmetric mode for 
the combined sample, and 48\%\ and 40\% for the \textit{BL} stars. At 
$R_{\rm phot} > 5.5$ kpc, the strength of the lopsided spiral is similar to the 
axisymmetric value. The lopsidedness and bisymmetry perturbations are therefore not negligible in the LMC. 

Both samples show that the dominant perturbation 
at low radius  is the bisymmetric mode ($R_{\rm phot}< 3.5$ kpc for \textit{BL} stars, $R_{\rm phot} < 2.5$ kpc 
for the combined stars), while the lopsided mode dominates at larger radii. 
In the inner kpc, the $\Sigma_2$ profile of the whole sample presents a 
dearth of stars that is lacking in the density map of younger stars. 
This is caused by the incompleteness of \gaia in this crowded region of the 
LMC disc. This dearth of stars also affects the inner profile 
of $\phi_2$ for all stars as a central phase dip. 

The orientation of the inner $k=2$
perturbation does not change much in the inner disc, with a bar 
 oriented with a phase angle of $1.6\pi$ rad (modulo $\pi$) for both samples. The 
$k=2$ spiral structure of \textit{BL} stars has a phase angle of $1.4\pi$ rad ($R_{\rm phot} > 3$ kpc), while 
that for the combined sample smoothly increases to $\sim 2.1\pi$ rad for $R_{\rm phot}=3.5-5.2$ 
kpc, then remains constant out to the last radius. 
The phase angles of the lopsided mode continuously vary as a function of radius, and  the two stellar samples present different shapes of $\phi_1$. The similar shape 
of $\phi_1$ and $\phi_2$ at $R_{\rm phot}> 2.5$ kpc for the combined sample of stars is remarkable, and the amplitudes only differ by less than 0.2$\pi$. 
The outer spiral structure in the LMC combined sample is thus made of two modes that are tightly coupled. 

 \begin{figure}[h]
                \centering
                \includegraphics[width=\columnwidth]{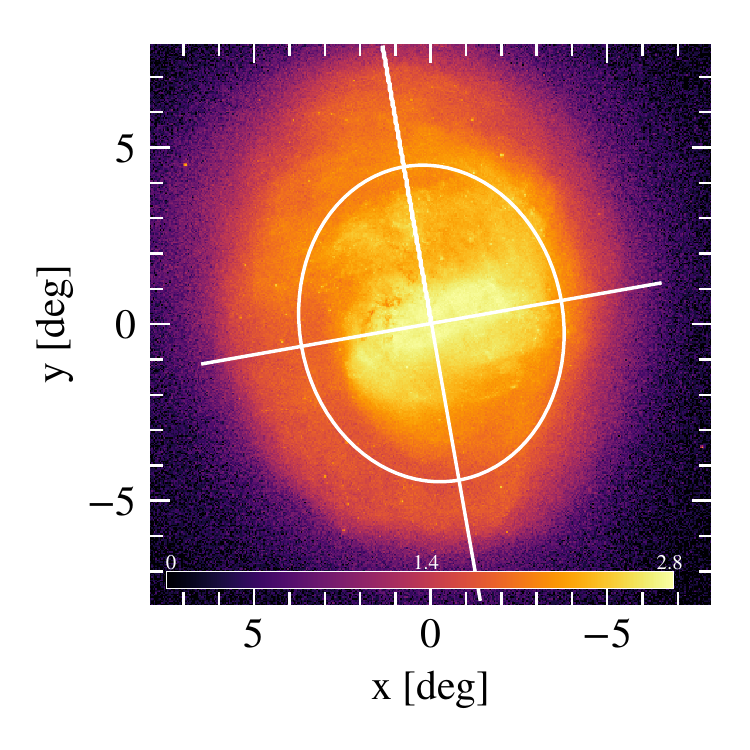}
                \caption{Stellar density map of the combined sample of the LMC. 
                The lopsided eastern spiral arm is located beyond $R_{\rm phot}=3.8$ kpc
                (shown as an ellipse). The perpendicular 
                 axes show the orientation of the photometric major and minor axes (\secref{sec:fftdensity}).
                 The density scale shown by the colour bar is logarithmic; the values indicate the star count in pixels.}   
                \label{fig:rphotdensity}
\end{figure}

%%%%%%%%%%%%%%%%%%%
\subsection{Across and along streaming motions in the eastern spiral arm}
The LMC velocity fields have been shown to exhibit variations 
stemming from the juxtaposition of an observational sawtooth-like pattern 
inherent to \textit{Gaia}, and others likely caused by intrinsic perturbations of 
the gravitational potential of the LMC (\secref{sec:kinlmc}). Here we illustrate in a simple 
way the variation of \vt\ and \vr\ along and across the dominant outer $k=1$ spiral arm to the east in the 
combined sample. 

\begin{figure*}[h]
   \centering
                \includegraphics[width=0.8\textwidth]{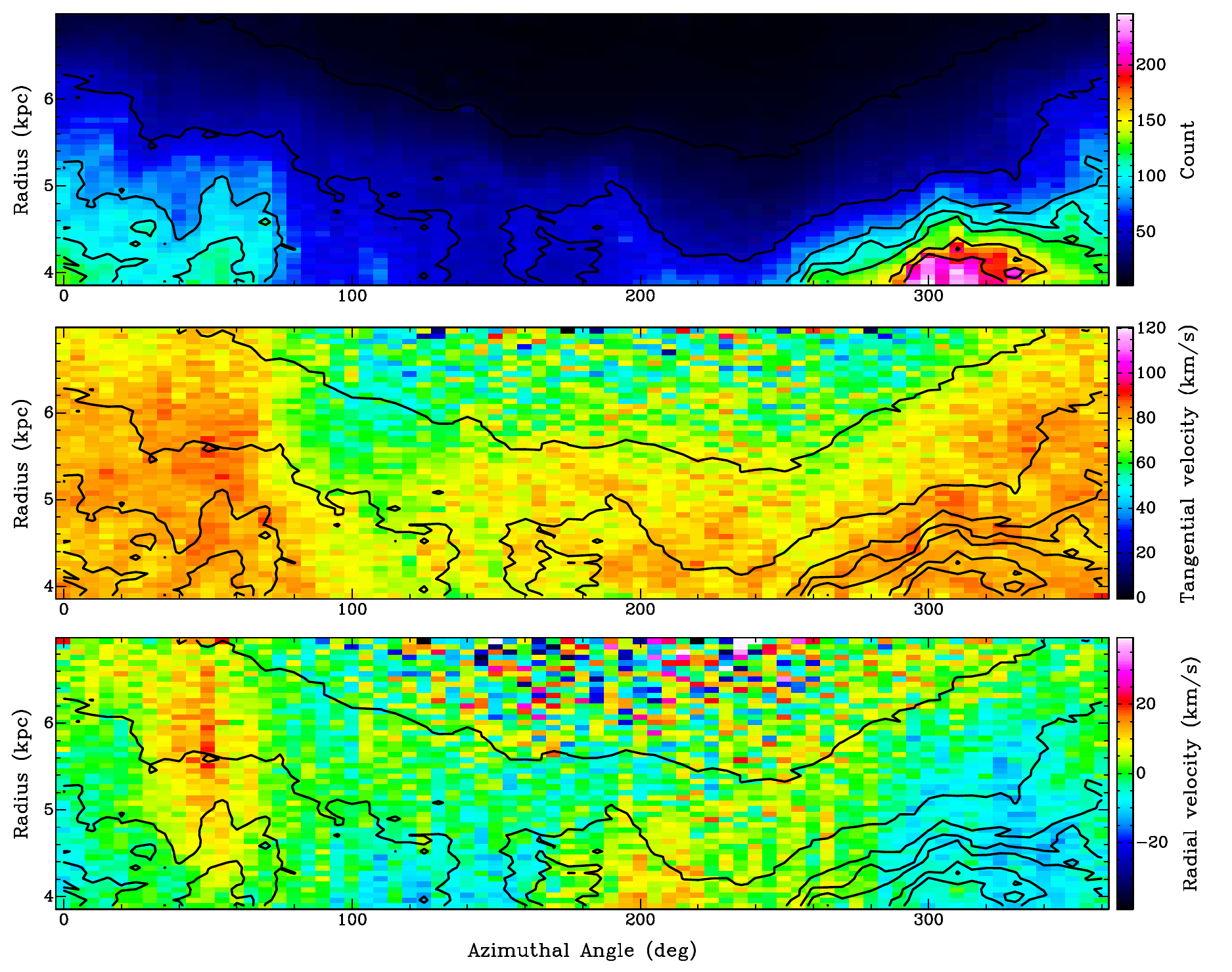}
                \caption{Streaming motions across and along the eastern spiral arm of the LMC. 
                                                 The azimuth-radius diagrams show the stellar density (top panel), the 
                                                 tangential velocity (middle panel), and radial velocity (bottom panel). 
                                                Contours represent the stellar density shown in the upper panel.}   
                \label{fig:azimraddiagram}
\end{figure*}

To isolate the  effects of the outer arm better, we only considered the region where the inner 
 $k=2$ mode becomes negligible, that is, all pixels located at $R_{\rm phot}>3.8$ kpc (\figref{fig:rphotdensity}). We 
built azimuth-radius diagrams of the stellar density and   tangential and 
radial velocities by calculating average  
star counts, \vt\ and \vr\  in bins of 5\degr\ size in azimuthal angle, and 
63 pc in radius (\figref{fig:azimraddiagram}).
The horizontal variation is thus a good proxy of the streaming of \vt\ and 
\vr\ along the eastern spiral arm, while the vertical axis is a good proxy 
for the velocity variation across the spiral arm.

The density of the spiral arm is strongly asymmetric as a function of 
azimuthal angle, caused by its lopsided nature. 
The uppermost isocontour of density (mean star count of $\sim 20$ stars) approximately delineates the maximum 
radial extent of the spiral arm, which extends to $R_{\rm phot}\sim 7$ kpc along 
the photometric major axis ($\phi = 0\degr$) to $R_{\rm phot} \sim 5.4$ kpc
($\phi = 240\degr$). The highest densities around $\phi = 300\degr$ at lower radii correspond 
to regions of the LMC that are part of the inner spiral structure, thus not strictly belonging to the outer 
lopsided eastern arm.

Along the horizontal axis,  \vt\ is maximum in higher density regions and 
minimum in lower density regions. When we consider pixels below the outermost contour, the azimuthal 
streaming in the arm is relatively constant ($60 \lesssim v_\phi 
\lesssim 90$ \kms). An exception to this occurs at $R \sim 6$ kpc owing to the lower values of \vt\ around $\phi = 100\degr$. 
As the pixels above the uppermost contour likely probe stars beyond the spiral arm, the difference in colours between pixels 
below (redder) or above (bluer, $v_\phi <60$ \kms) the uppermost contour clearly shows the effect of the arm on \vt\ in the azimuthal direction. 
The radial velocity also varies significantly with azimuth. It 
is  stronger in higher density regions around $\phi = 50\degr$ and $\phi=200\degr$
 ($v_R > -10$ \kms) 
and in lower density regions for $100 < \phi < 180\degr$ , but with inward motions ($v_R < -10$ \kms). The noise in \vr\ is higher outside the arm at large radii.

Along the vertical axis, \vt\ is observed to decline with radius across the 
spiral arm, and the decrease is not complete at the same rate for different azimuthal angles. 
This implies a wide diversity of shapes and amplitudes 
in the LMC rotation curve as a function of azimuth.
We have observed this trend in the \vt\ map of \secref{sec:kinlmc}.
The radial velocity also varies across the spiral arm, but there appears to be 
no clear rule, unlike for \vt. 
For example, 
the peak of \vr\ at $\phi = 50\degr$  occurs at $R_{\rm phot} = 5.5-6$ kpc, thus beyond 
the location of the density peak ($R_{\rm phot} \lesssim 4.5$ kpc). 
However, at an angle of for instance $\phi=200\degr$, the opposite is observed, with larger \vr\ for 
 higher density regions of this azimuthal angle ($R_{\rm phot} \lesssim 4.5$ kpc). 
 
%======================================================================================
% Conclusions
%======================================================================================
\section{Conclusions \label{sec:conclusion}}

Using the new \egdr{3} data, we studied the structure and
kinematics of the Magellanic Clouds with a new basis. The increased completeness and precision
of the new release have allowed us to improve upon previous results 
using \gdrtwo, although (by design, because this is just a demonstration
paper) we have certainly barely scratched the potential of the new data for the
study of the Clouds. 

% DR2 to DR3
In \secref{sec:DR2DR3} we compared the \gdrtwo and \egdr{3} data in
the region of the LMC and SMC, showing the improvement in the astrometry and
photometry from one release to the other. Not only the precision
has increased, but the systematic effects are significantly reduced.
The reduced crowding effects in the photometry are particularly relevant for the central regions of the Clouds.

% Spatial structure
We have explored the use of the astrometric
data for the determination of the 3D structure of the LMC. Our attempts 
to use Bayesian modelling to reconstruct the geometry of the system have shown 
that despite the significant improvements in \egdr{3}, the systematic effects 
still present on the parallaxes (the regional zero-points discussed in 
\cite{EDR3-DPACP-128}) distort most of the signal of the 3D structure of the LMC 
(in the astrometry), and that there is not enough information in the summary 
statistics used by our approximate Bayesian method to simultaneously infer the 
local parallax zero-point variations and 
the geometry of the LMC. However, we do not rule out that it may be possible 
to determine this geometry by adding additional external restrictions and/or 
finding an optimal way to include additional information of the density distribution 
of the stars in the LMC area. 

% Kinematics & spiral arms
Our kinematic modelling of the proper motions has allowed us
to derive radial and tangential velocity maps and global profiles for the LMC. This is
the first time to our knowledge that the two planar components of the ordered and 
random motions are derived for multiple stellar evolutionary phases in a galactic disc
outside the Milky Way. We show that younger stellar phases rotate faster than older ones. 
This is a clear effect of the asymmetric drift on the stellar kinematics. 
We have also been able to find the rotational centre of the stars in the LMC and showed 
that is significantly offset from the rotational centre of the H\,\textsc{i} gas.

On the other hand, we showed that the radial velocity is mostly negative 
in the inner 5 kpc, and inward radial motions are stronger in the bar region for younger stars. 
This velocity varies strongly as a function position with respect to the bar, and therefore 
to some extent reflects streaming motions along the bar.
We observed asymmetric radial and tangential motions in the disc, such as those across and 
along the large-scale outer spiral arm of the LMC, in which the tangential velocity is 
larger in higher stellar density regions of the arm, and lower at lower density. There 
appears to be no clear rule in the streaming of the radial motion as a function of the position 
in the arm, however. 

We showed that the radial and tangential random motions decrease from the 
disc centre out to the outskirts, but not at the same rate in the evolutionary phases. 
Older stars lie in a dynamically hotter disc than younger stars. While we have found higher 
velocity dispersions aligned with the bar in all stellar phases for both  components, 
we also found evidence that only the radial component of more evolved stars exhibits a 
central feature of higher amplitude, which is oriented perpendicular to the bar. 

Our analysis of the stellar density maps has shown more concentrated 
and clumpier distributions for younger stars in the bar and inner spiral structure(s) 
than older disc stars. Analysis of Fourier harmonics of density maps also revealed that 
the inner disc is perturbed predominantly by bisymmetries, the bar and spiral arms, while 
the outer disc is perturbed by a dominant lopsided spiral arm. The peak strengths of these 
perturbations can be as high as 40\% to 60\% of the axisymmetric amplitude in the inner 
region, and higher in the disc outskirts for the lopsidedness.

% Bridge
Finally, we carried out a specific study of the Magellanic Bridge. Using two different evolutionary phases 
(the young and RC samples), we were able to trace the density and velocity flow of 
the stars from the SMC towards the LMC following the Bridge. We showed that it apparently wraps around the 
LMC, connecting with the young southern arm-like structure of the galaxy. The quality of the \egdr{3} proper motions also allowed us to confirm the bridge in the \textit{RC} evolutionary phase, at a position 
slightly shifted from that of the young evolutionary
phase. Additionally, we were able to study the outskirts of both Magellanic Clouds and detected some
well-known features, such as the north and south tidal arms of the LMC and the northern overdensity 
of the SMC. Our data also suggests a faint overdensity east of the LMC, which has only recently 
been reported with the help of near-infrared maps (El Youssoufi, private communication).

%======================================================================================
% Acknowledgements
%======================================================================================
\begin{acknowledgements}

% The "standard Gaia EDR3 acknowledgements" (from https://gaia.esac.esa.int/dpacsvn/DPAC/docs/ReleaseDocumentation/GEDR3/cite.tex);

This work has made use of data from the European Space Agency (ESA) mission
{\it Gaia} (\url{https://www.cosmos.esa.int/gaia}), processed by the {\it Gaia}
Data Processing and Analysis Consortium (DPAC,
\url{https://www.cosmos.esa.int/web/gaia/dpac/consortium}). Funding for the DPAC
has been provided by national institutions, in particular the institutions
participating in the {\it Gaia} Multilateral Agreement.

%The "official ESA/Gaia/DPAC acknowledgements" (from https://gaia.esac.esa.int/dpacsvn/DPAC/docs/ReleaseDocumentation/GEDR3/acknow.tex).

This work presents results from the European Space Agency (ESA) space mission \gaia. \gaia\ data are being processed by the \gaia\ Data Processing and Analysis Consortium (DPAC). Funding for the DPAC is provided by national institutions, in particular the institutions participating in the \gaia\ MultiLateral Agreement (MLA). The \gaia\ mission website is \url{https://www.cosmos.esa.int/gaia}. The \gaia\ archive website is \url{https://archives.esac.esa.int/gaia}.

The \gaia\ mission and data processing have financially been supported by, in alphabetical order by country:

the Algerian Centre de Recherche en Astronomie, Astrophysique et G\'{e}ophysique of Bouzareah Observatory;
the Austrian Fonds zur F\"{o}rderung der wissenschaftlichen Forschung (FWF) Hertha Firnberg Programme through grants T359, P20046, and P23737;
the BELgian federal Science Policy Office (BELSPO) through various PROgramme de D\'eveloppement d'Exp\'eriences scientifiques (PRODEX) grants and the Polish Academy of Sciences - Fonds Wetenschappelijk Onderzoek through grant VS.091.16N, and the Fonds de la Recherche Scientifique (FNRS);
the Brazil-France exchange programmes Funda\c{c}\~{a}o de Amparo \`{a} Pesquisa do Estado de S\~{a}o Paulo (FAPESP) and Coordena\c{c}\~{a}o de Aperfeicoamento de Pessoal de N\'{\i}vel Superior (CAPES) - Comit\'{e} Fran\c{c}ais d'Evaluation de la Coop\'{e}ration Universitaire et Scientifique avec le Br\'{e}sil (COFECUB);
the National Science Foundation of China (NSFC) through grants 11573054 and 11703065 and the China Scholarship Council through grant 201806040200;  
the Tenure Track Pilot Programme of the Croatian Science Foundation and the \'{E}cole Polytechnique F\'{e}d\'{e}rale de Lausanne and the project TTP-2018-07-1171 'Mining the Variable Sky', with the funds of the Croatian-Swiss Research Programme;
the Czech-Republic Ministry of Education, Youth, and Sports through grant LG 15010 and INTER-EXCELLENCE grant LTAUSA18093, and the Czech Space Office through ESA PECS contract 98058;
the Danish Ministry of Science;
the Estonian Ministry of Education and Research through grant IUT40-1;
the European Commission’s Sixth Framework Programme through the European Leadership in Space Astrometry (\href{https://www.cosmos.esa.int/web/gaia/elsa-rtn-programme}{ELSA}) Marie Curie Research Training Network (MRTN-CT-2006-033481), through Marie Curie project PIOF-GA-2009-255267 (Space AsteroSeismology \& RR Lyrae stars, SAS-RRL), and through a Marie Curie Transfer-of-Knowledge (ToK) fellowship (MTKD-CT-2004-014188); the European Commission's Seventh Framework Programme through grant FP7-606740 (FP7-SPACE-2013-1) for the \gaia\ European Network for Improved data User Services (\href{https://gaia.ub.edu/twiki/do/view/GENIUS/}{GENIUS}) and through grant 264895 for the \gaia\ Research for European Astronomy Training (\href{https://www.cosmos.esa.int/web/gaia/great-programme}{GREAT-ITN}) network;
the European Research Council (ERC) through grants 320360 and 647208 and through the European Union’s Horizon 2020 research and innovation and excellent science programmes through Marie Sk{\l}odowska-Curie grant 745617 as well as grants 670519 (Mixing and Angular Momentum tranSport of massIvE stars -- MAMSIE), 687378 (Small Bodies: Near and Far), 682115 (Using the Magellanic Clouds to Understand the Interaction of Galaxies), and 695099 (A sub-percent distance scale from binaries and Cepheids -- CepBin);
the European Science Foundation (ESF), in the framework of the \gaia\ Research for European Astronomy Training Research Network Programme (\href{https://www.cosmos.esa.int/web/gaia/great-programme}{GREAT-ESF});
the European Space Agency (ESA) in the framework of the \gaia\ project, through the Plan for European Cooperating States (PECS) programme through grants for Slovenia, through contracts C98090 and 4000106398/12/NL/KML for Hungary, and through contract 4000115263/15/NL/IB for Germany;
the Academy of Finland and the Magnus Ehrnrooth Foundation;
the French Centre National d’Etudes Spatiales (CNES), the Agence Nationale de la Recherche (ANR) through grant ANR-10-IDEX-0001-02 for the 'Investissements d'avenir' programme, through grant ANR-15-CE31-0007 for project 'Modelling the Milky Way in the Gaia era' (MOD4Gaia), through grant ANR-14-CE33-0014-01 for project 'The Milky Way disc formation in the Gaia era' (ARCHEOGAL), and through grant ANR-15-CE31-0012-01 for project 'Unlocking the potential of Cepheids as primary distance calibrators' (UnlockCepheids), the Centre National de la Recherche Scientifique (CNRS) and its SNO Gaia of the Institut des Sciences de l’Univers (INSU), the 'Action F\'{e}d\'{e}ratrice Gaia' of the Observatoire de Paris, the R\'{e}gion de Franche-Comt\'{e}, and the Programme National de Gravitation, R\'{e}f\'{e}rences, Astronomie, et M\'{e}trologie (GRAM) of CNRS/INSU with the Institut National Polytechnique (INP) and the Institut National de Physique nucléaire et de Physique des Particules (IN2P3) co-funded by CNES;
the German Aerospace Agency (Deutsches Zentrum f\"{u}r Luft- und Raumfahrt e.V., DLR) through grants 50QG0501, 50QG0601, 50QG0602, 50QG0701, 50QG0901, 50QG1001, 50QG1101, 50QG1401, 50QG1402, 50QG1403, 50QG1404, and 50QG1904 and the Centre for Information Services and High Performance Computing (ZIH) at the Technische Universit\"{a}t (TU) Dresden for generous allocations of computer time;
the Hungarian Academy of Sciences through the Lend\"{u}let Programme grants LP2014-17 and LP2018-7 and through the Premium Postdoctoral Research Programme (L.~Moln\'{a}r), and the Hungarian National Research, Development, and Innovation Office (NKFIH) through grant KH\_18-130405;
the Science Foundation Ireland (SFI) through a Royal Society - SFI University Research Fellowship (M.~Fraser);
the Israel Science Foundation (ISF) through grant 848/16;
the Agenzia Spaziale Italiana (ASI) through contracts I/037/08/0, I/058/10/0, 2014-025-R.0, 2014-025-R.1.2015, and 2018-24-HH.0 to the Italian Istituto Nazionale di Astrofisica (INAF), contract 2014-049-R.0/1/2 to INAF for the Space Science Data Centre (SSDC, formerly known as the ASI Science Data Center, ASDC), contracts I/008/10/0, 2013/030/I.0, 2013-030-I.0.1-2015, and 2016-17-I.0 to the Aerospace Logistics Technology Engineering Company (ALTEC S.p.A.), INAF, and the Italian Ministry of Education, University, and Research (Ministero dell'Istruzione, dell'Universit\`{a} e della Ricerca) through the Premiale project 'MIning The Cosmos Big Data and Innovative Italian Technology for Frontier Astrophysics and Cosmology' (MITiC);
the Netherlands Organisation for Scientific Research (NWO) through grant NWO-M-614.061.414, through a VICI grant (A.~Helmi), and through a Spinoza prize (A.~Helmi), and the Netherlands Research School for Astronomy (NOVA);
the Polish National Science Centre through HARMONIA grant 2018/06/M/ST9/00311, DAINA grant 2017/27/L/ST9/03221, and PRELUDIUM grant 2017/25/N/ST9/01253, and the Ministry of Science and Higher Education (MNiSW) through grant DIR/WK/2018/12;
the Portugese Funda\c{c}\~ao para a Ci\^{e}ncia e a Tecnologia (FCT) through grants SFRH/BPD/74697/2010 and SFRH/BD/128840/2017 and the Strategic Programme UID/FIS/00099/2019 for CENTRA;
the Slovenian Research Agency through grant P1-0188;
the Spanish Ministry of Economy (MINECO/FEDER, UE) through grants ESP2016-80079-C2-1-R, ESP2016-80079-C2-2-R, RTI2018-095076-B-C21, RTI2018-095076-B-C22, BES-2016-078499, and BES-2017-083126 and the Juan de la Cierva formaci\'{o}n 2015 grant FJCI-2015-2671, the Spanish Ministry of Education, Culture, and Sports through grant FPU16/03827, the Spanish Ministry of Science and Innovation (MICINN) through grant AYA2017-89841P for project 'Estudio de las propiedades de los f\'{o}siles estelares en el entorno del Grupo Local' and through grant TIN2015-65316-P for project 'Computaci\'{o}n de Altas Prestaciones VII', the Severo Ochoa Centre of Excellence Programme of the Spanish Government through grant SEV2015-0493, the Institute of Cosmos Sciences University of Barcelona (ICCUB, Unidad de Excelencia ’Mar\'{\i}a de Maeztu’) through grants MDM-2014-0369 and CEX2019-000918-M, the University of Barcelona's official doctoral programme for the development of an R+D+i project through an Ajuts de Personal Investigador en Formaci\'{o} (APIF) grant, the Spanish Virtual Observatory through project AyA2017-84089, the Galician Regional Government, Xunta de Galicia, through grants ED431B-2018/42 and ED481A-2019/155, support received from the Centro de Investigaci\'{o}n en Tecnolog\'{\i}as de la Informaci\'{o}n y las Comunicaciones (CITIC) funded by the Xunta de Galicia, the Xunta de Galicia and the Centros Singulares de Investigaci\'{o}n de Galicia for the period 2016-2019 through CITIC, the European Union through the European Regional Development Fund (ERDF) / Fondo Europeo de Desenvolvemento Rexional (FEDER) for the Galicia 2014-2020 Programme through grant ED431G-2019/01, the Red Espa\~{n}ola de Supercomputaci\'{o}n (RES) computer resources at MareNostrum, the Barcelona Supercomputing Centre - Centro Nacional de Supercomputaci\'{o}n (BSC-CNS) through activities AECT-2016-1-0006, AECT-2016-2-0013, AECT-2016-3-0011, and AECT-2017-1-0020, the Departament d'Innovaci\'{o}, Universitats i Empresa de la Generalitat de Catalunya through grant 2014-SGR-1051 for project 'Models de Programaci\'{o} i Entorns d'Execuci\'{o} Parallels' (MPEXPAR), and Ramon y Cajal Fellowship RYC2018-025968-I;
the Swedish National Space Agency (SNSA/Rymdstyrelsen);
the Swiss State Secretariat for Education, Research, and Innovation through
 the ESA PRODEX programme,
the Mesures d’Accompagnement, the Swiss Activit\'es Nationales Compl\'ementaires, and the Swiss National Science Foundation;
the United Kingdom Particle Physics and Astronomy Research Council (PPARC), the United Kingdom Science and Technology Facilities Council (STFC), and the United Kingdom Space Agency (UKSA) through the following grants to the University of Bristol, the University of Cambridge, the University of Edinburgh, the University of Leicester, the Mullard Space Sciences Laboratory of University College London, and the United Kingdom Rutherford Appleton Laboratory (RAL): PP/D006511/1, PP/D006546/1, PP/D006570/1, ST/I000852/1, ST/J005045/1, ST/K00056X/1, ST/K000209/1, ST/K000756/1, ST/L006561/1, ST/N000595/1, ST/N000641/1, ST/N000978/1, ST/N001117/1, ST/S000089/1, ST/S000976/1, ST/S001123/1, ST/S001948/1, ST/S002103/1, and ST/V000969/1.

\end{acknowledgements}

%======================================================================================
% Bibliography
%======================================================================================
\bibliographystyle{aa} % style aa.bst

% Up to date DPAC bib file at: https://gaia.esac.esa.int/dpacsvn/DPAC/docs/ReleaseDocumentation/GEDR3/dpac.bib
\bibliography{bibliography}

%======================================================================================
% Appendix
%======================================================================================
\begin{appendix}

%----------------------------------------
\section{Sky plots of the samples used\label{sec:appendix}}

This appendix provides some additional figures complementary to the main text, 
collected here to avoid cluttering the main body and facilitate reading of the paper.

\begin{figure}[h]
   \begin{center}
     \includegraphics[width=0.9\columnwidth]{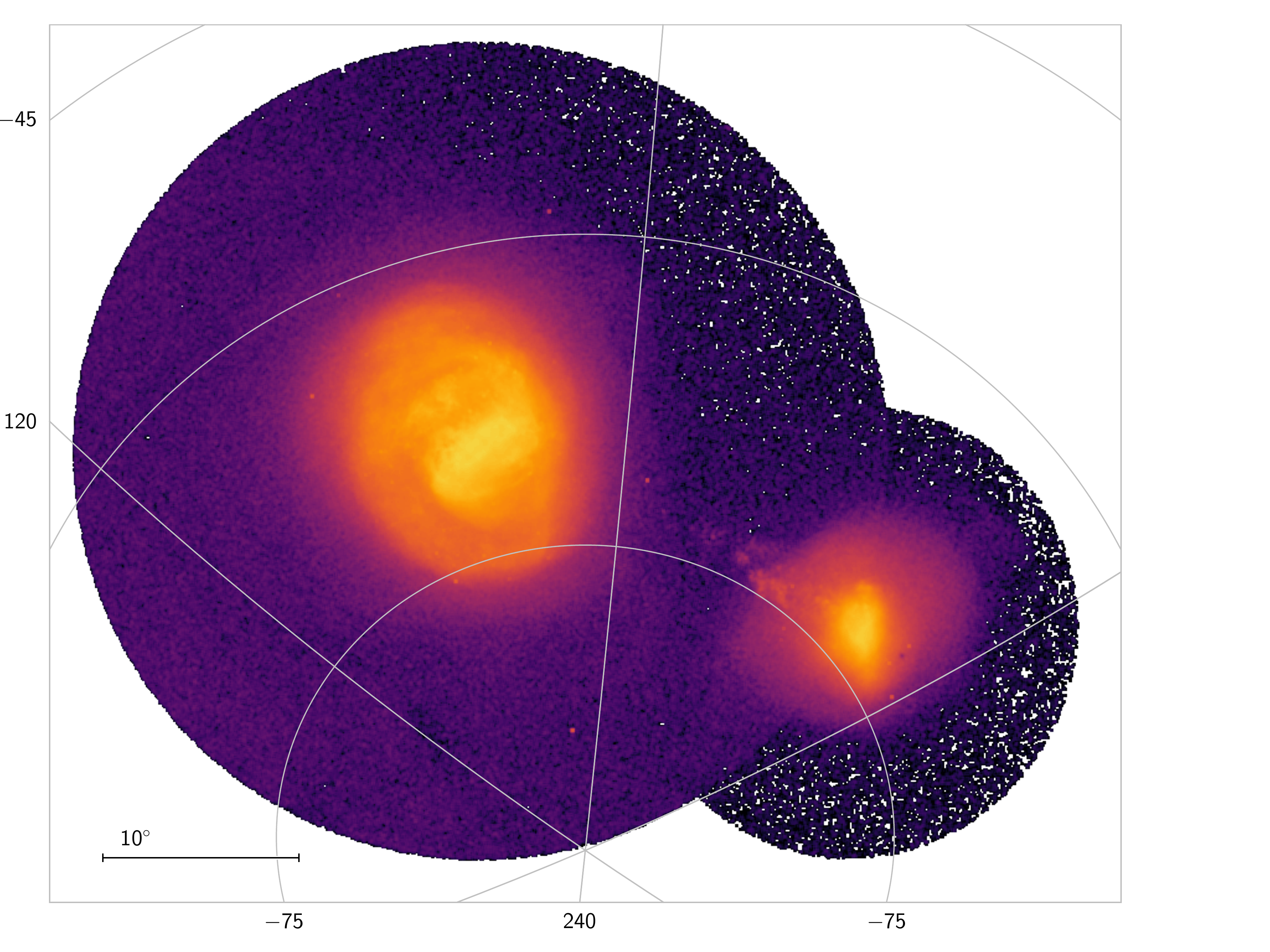}
   \end{center}
   \caption{Source density in the two circular areas with a radius of 20\degr\
   and 11\degr, centred on the LMC and SMC. Only the 12.4M sources selected
   as potential members are included, and the criteria are therefore slightly    different for the two circles. }
   \label{fig:Appendix_density}
\end{figure}

\begin{figure}[h]
   \begin{center}
     \includegraphics[width=0.9\columnwidth]{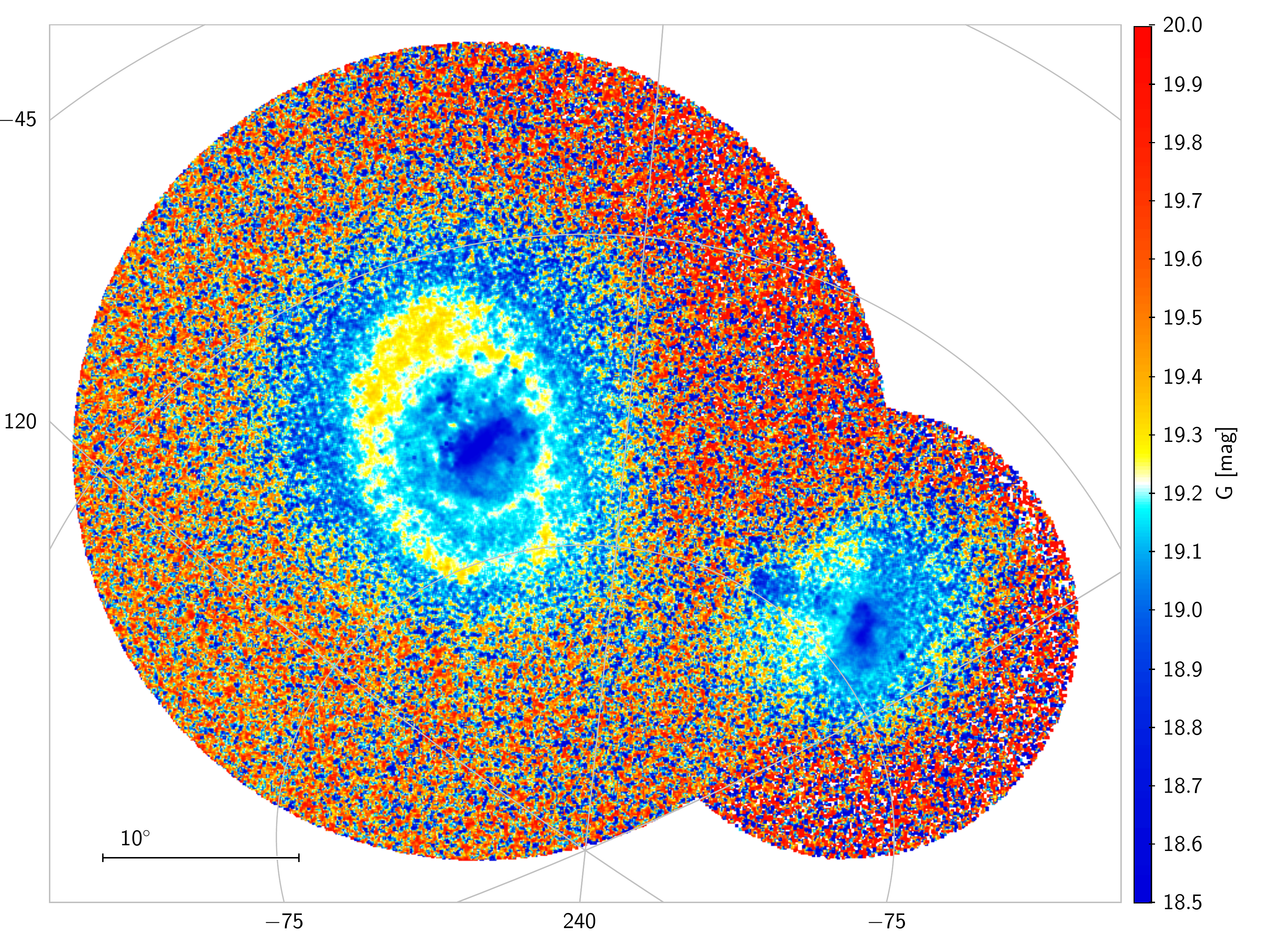}
   \end{center}
   \caption{Mean \gmag\ magnitude in the LMC and SMC.}
   \label{fig:Appendix_gmap}
\end{figure}

\begin{figure}[h]
   \begin{center}
     \includegraphics[width=0.9\columnwidth]{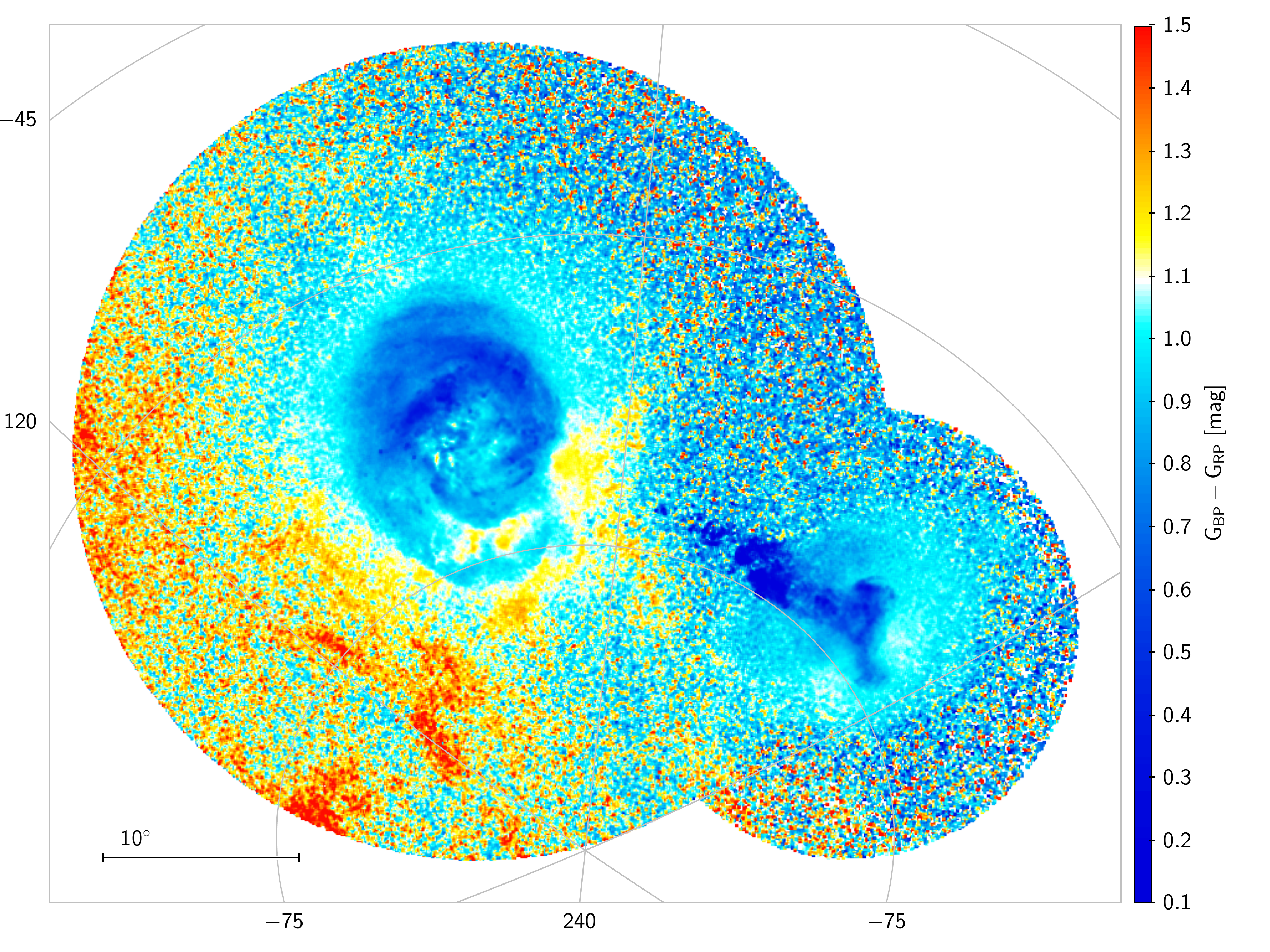}
   \end{center}
   \caption{Mean colour, $\gbp-\grp$ in the LMC and SMC. 
   }
   \label{fig:Appendix_bprp}
\end{figure}

\begin{figure}[h]
   \begin{center}
     \includegraphics[width=0.9\columnwidth]{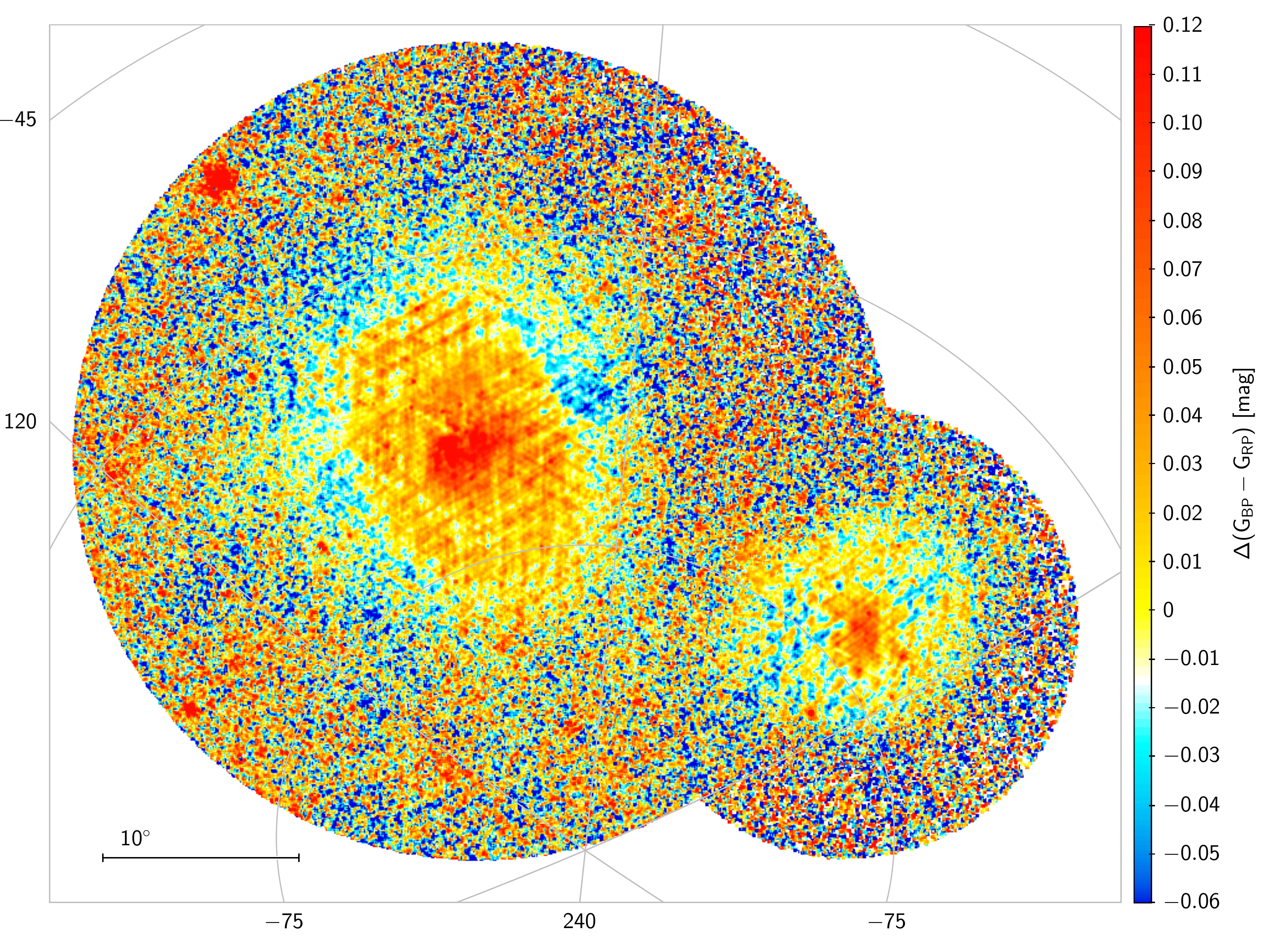}
   \end{center}
   \caption{Mean change in colour, $\gbp-\grp$, in \gdrtwo\ and \egdr{3}\
   for sources in the LMC and SMC.
   Positive values mean that the sources are now redder.
     }
   \label{fig:Appendix_delta_bprp}
\end{figure}

\begin{figure}[h]
   \begin{center}
     \includegraphics[width=0.9\columnwidth]{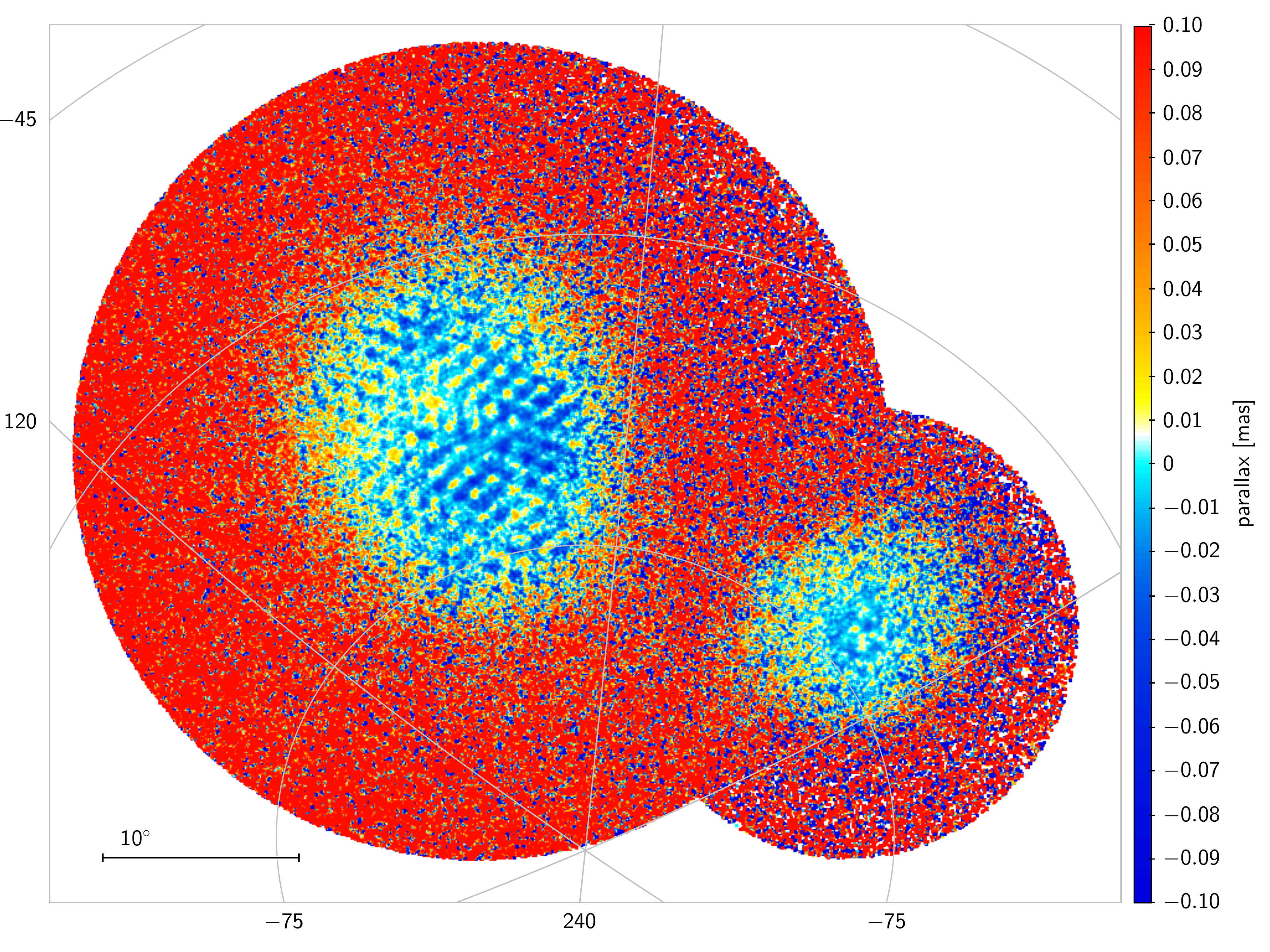}
   \end{center}
   \caption{Mean parallax for sources in the LMC and SMC. No corrections
    have been applied to the parallaxes.
    }
   \label{fig:Appendix_parallax}
\end{figure}

First, in \figref{fig:Appendix_density} we illustrate the joint sky distribution of our 
two (LMC and SMC) basic clean samples. The selection radius for both clouds is clearly 
visible. Following this map, the next figures show the distribution of the mean $G$ magnitude
(\figref{fig:Appendix_gmap}), mean $\gbp-\grp$ (\figref{fig:Appendix_bprp}), the variation
of this mean between \gdrtwo and \egdr{3} (\figref{fig:Appendix_delta_bprp}), and the mean
parallax (\figref{fig:Appendix_parallax}).

\begin{figure*}[h]
      \begin{center}
      \includegraphics[width=\textwidth]{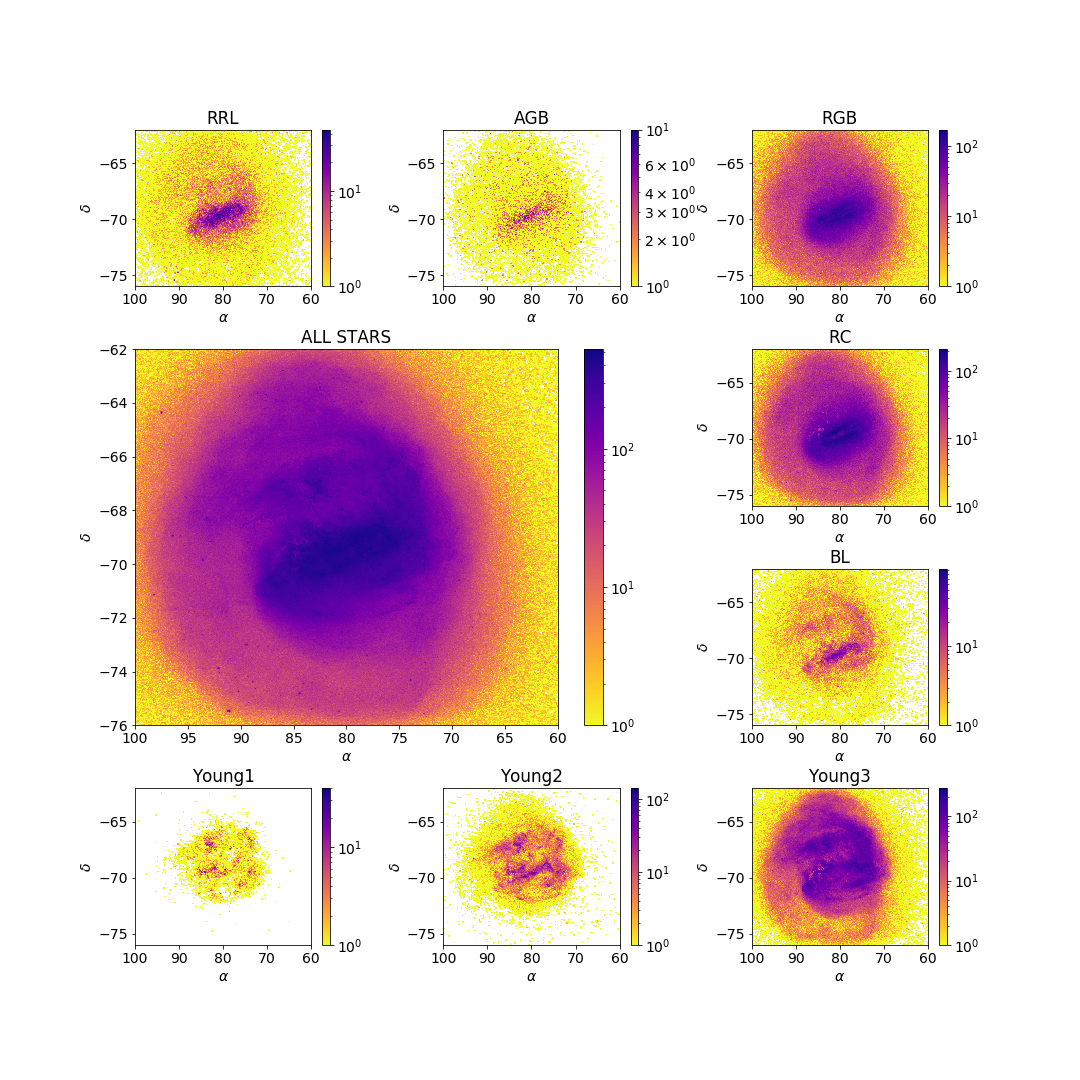}
      \end{center}
   \caption{Sky density plots for the complete LMC clean sample and the different evolutionary phase subsamples.}
         \label{fig:LMC_SkyPlots}
\end{figure*}

\begin{figure*}[h]
      \begin{center}
      \includegraphics[width=\textwidth]{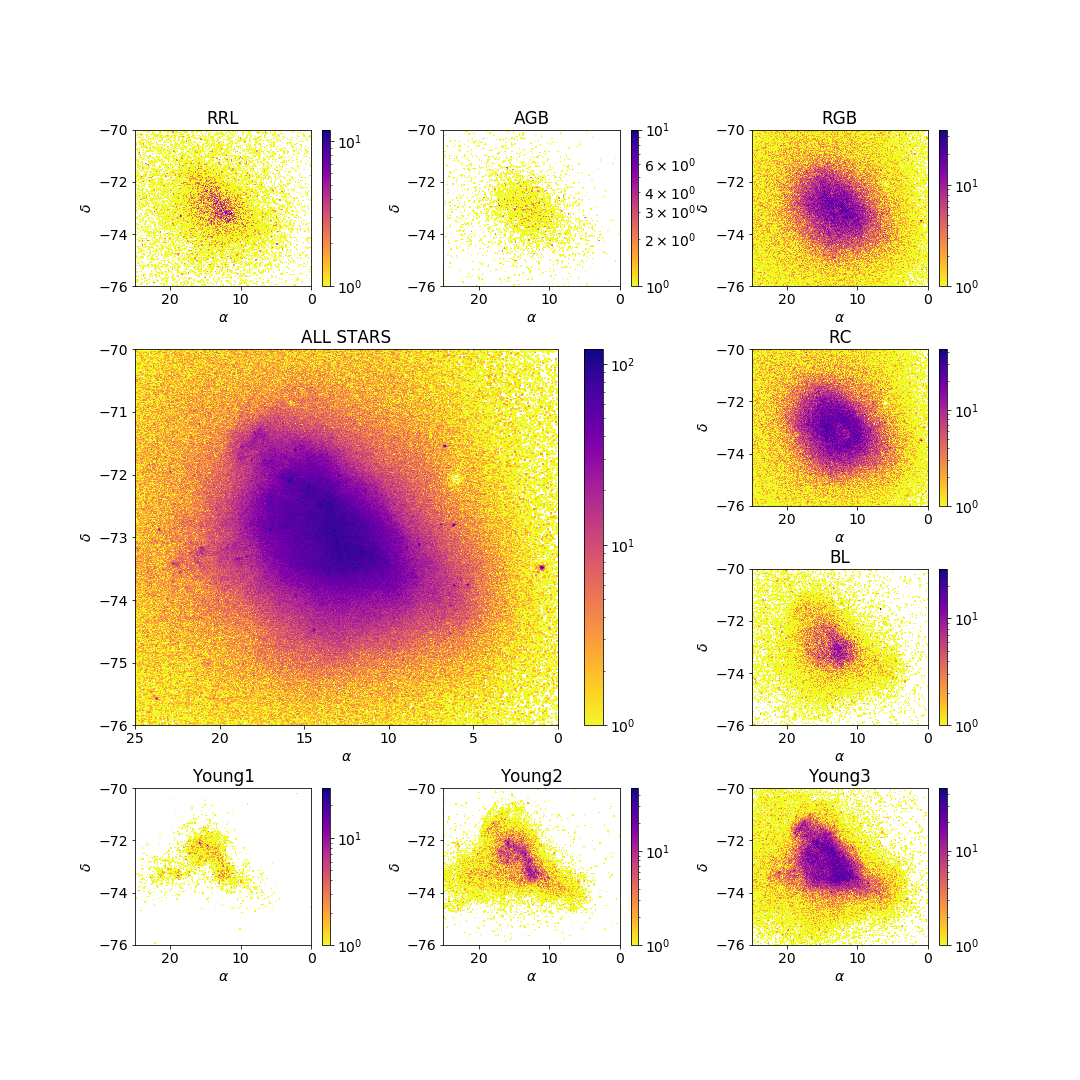}
      \end{center}
   \caption{Sky density plots for the complete SMC clean sample and the different evolutionary phase subsamples.}
         \label{fig:SMC_SkyPlots}
\end{figure*}

Next, we provide plots of the sky distribution for each one of our evolutionary phase samples;
the plots show that each sample traces different structures of the clouds. \figref{fig:LMC_SkyPlots} 
contains the plots for the LMC and \figref{fig:SMC_SkyPlots} those for the SMC.

\clearpage

%----------------------------------------
\section{LMC velocity maps and profiles}
\label{sec:appendixvelolmc}

\begin{figure*}
                \centering
                \includegraphics[width=0.95\textwidth]{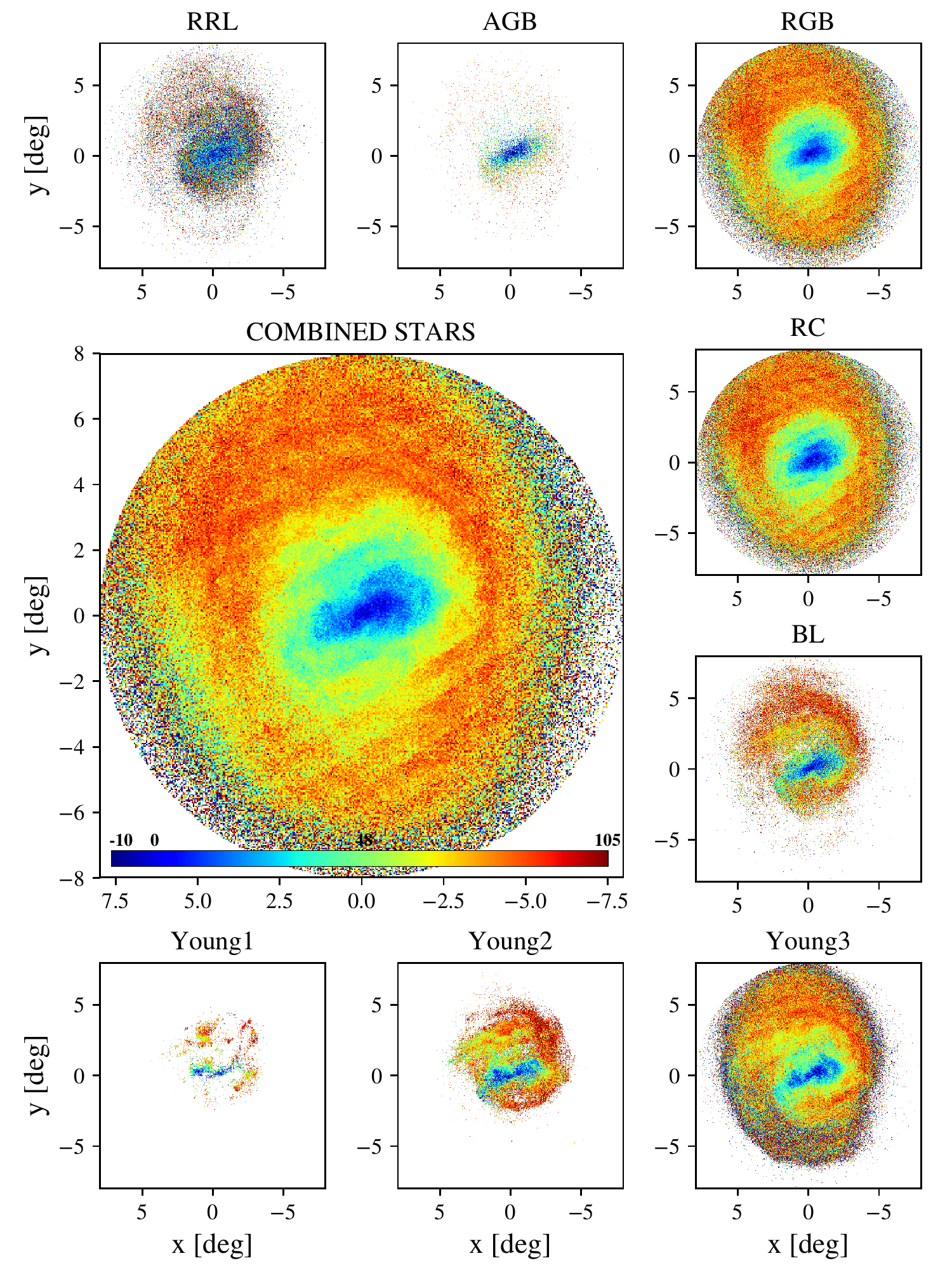}
                \caption{\label{fig:lmcvphimap} Azimuthal velocity maps $v_\phi$ of the LMC
                                                 for the combined sample (main panel) and the various evolutionary phases. The linear velocity 
                                                 scale shown by the colour bar in the main panel is the same in all subpanels and
                                                 has been chosen to highlight velocity patterns better.}
\end{figure*}

\begin{figure*}
                \centering
                \includegraphics[width=0.95\textwidth]{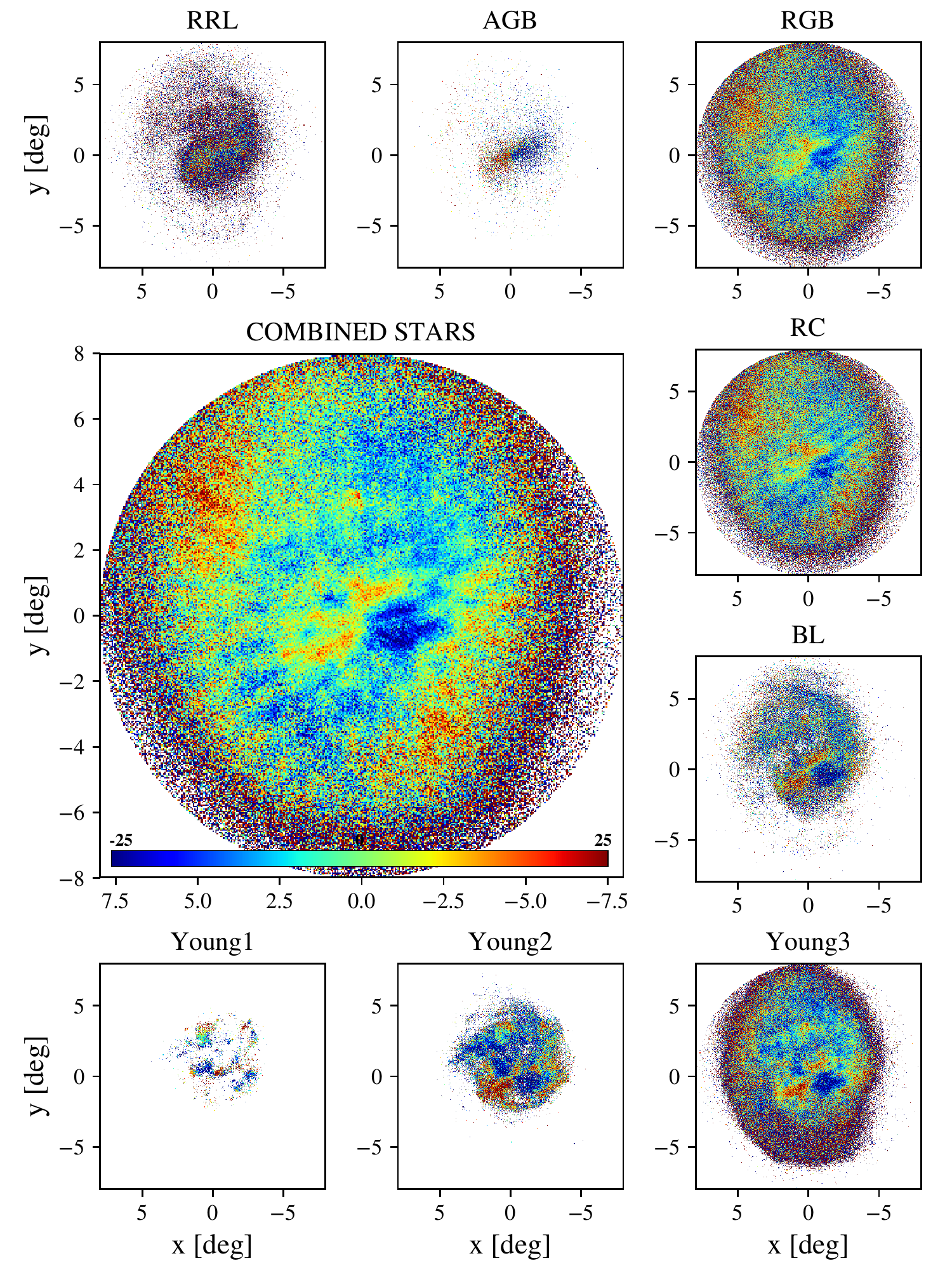}
                \caption{\label{fig:lmcvradmap} Radial velocity maps $v_R$ of the LMC 
                         for the combined sample (main panel) and the various evolutionary phases. The linear 
                                                 velocity scale shown by the colour bar in the main panel is the same in all subpanels and
                         has been chosen to highlight velocity patterns better.}
\end{figure*}

\begin{figure*}
                \centering
                \includegraphics[width=0.95\textwidth]{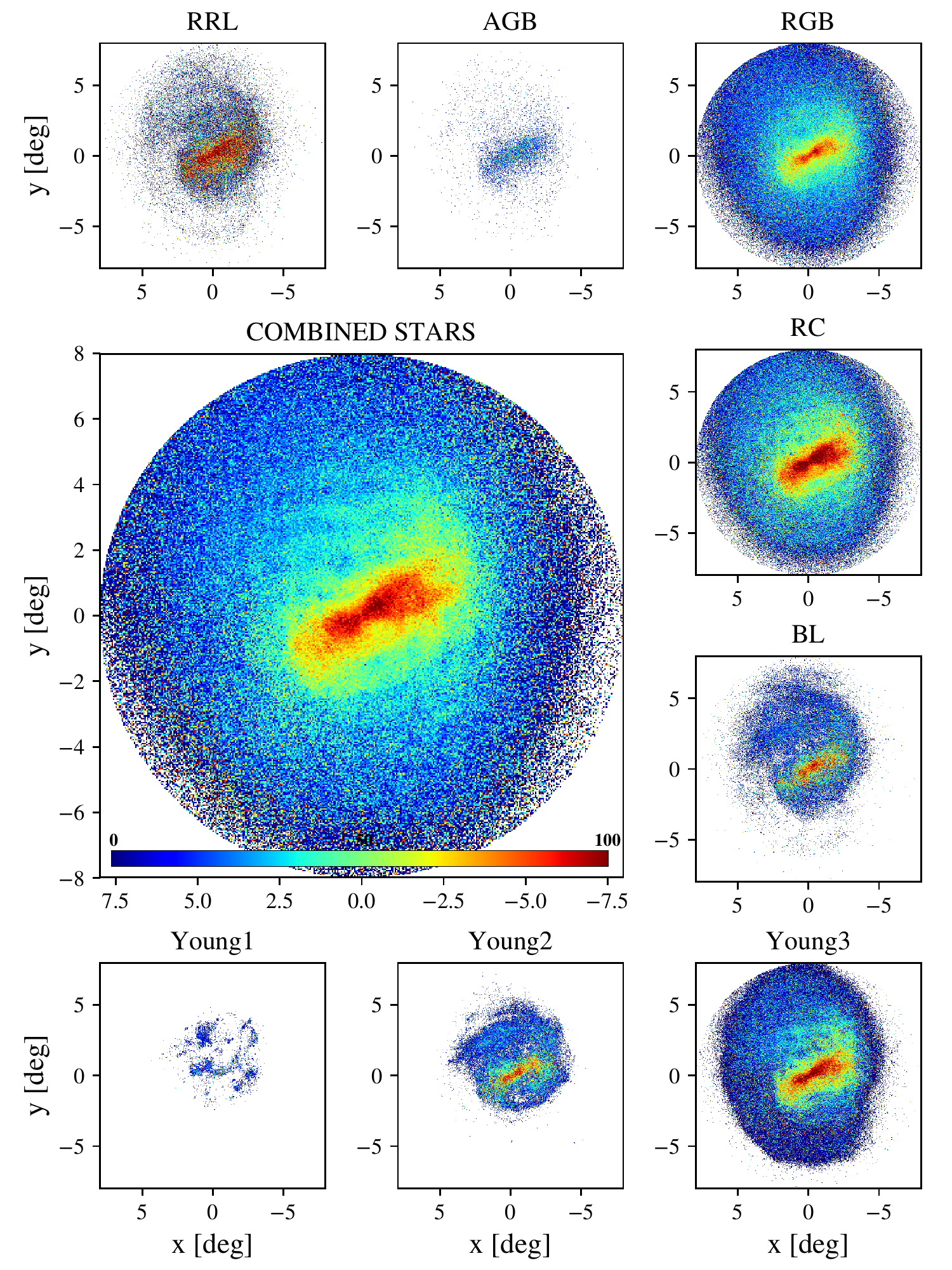}
                \caption{\label{fig:lmcsigvphimap} Azimuthal velocity dispersion maps $\sigma_\phi$ of the 
                                LMC for the combined sample (main panel) and the various 
                                                                                evolutionary phases. The linear velocity scale shown by the colour bar in the 
                                                                                main panel is the same in all subpanels and has been chosen to highlight velocity patterns better.}
\end{figure*}

\begin{figure*}
                \centering
                \includegraphics[width=0.95\textwidth]{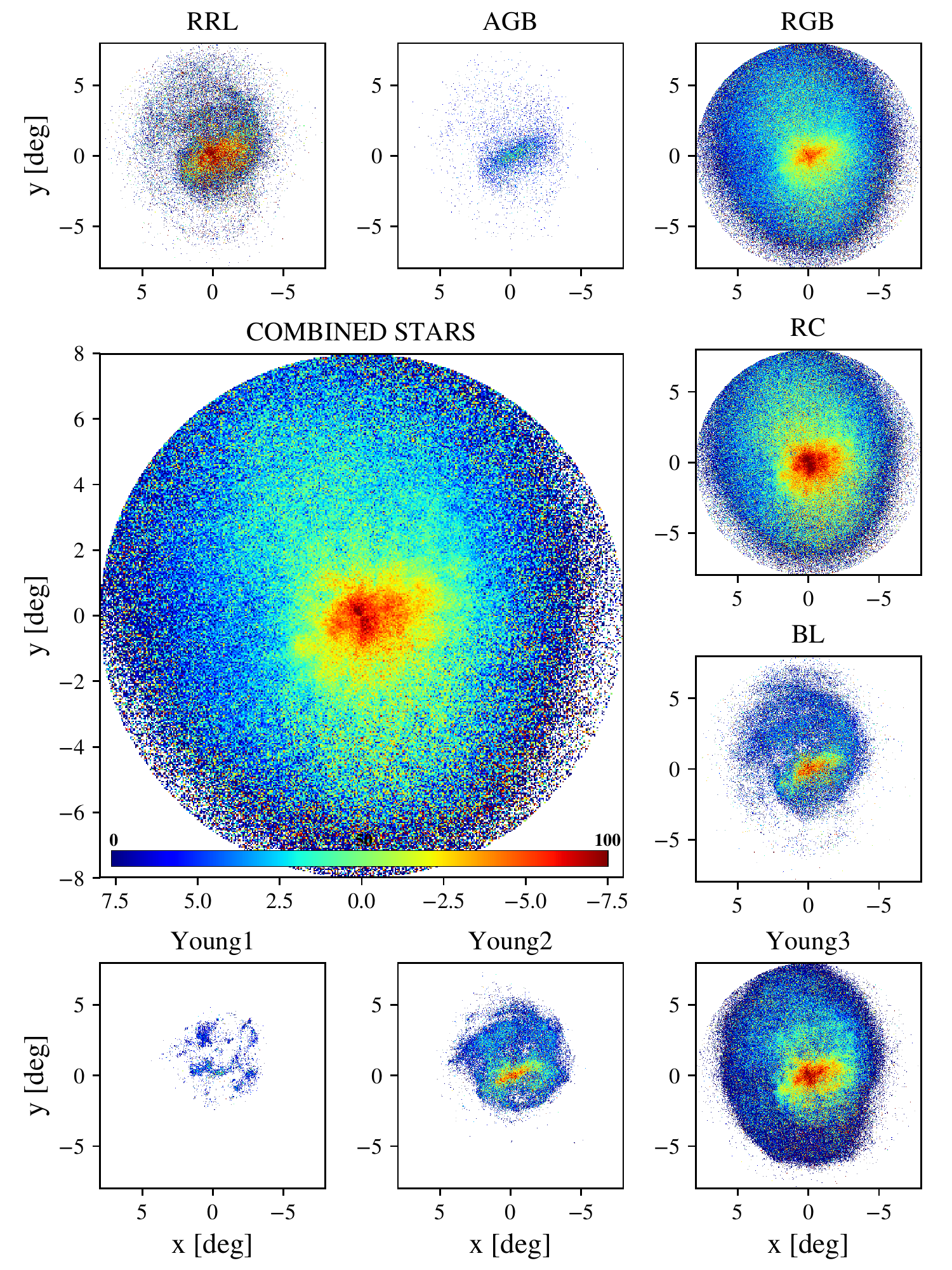}
                \caption{\label{fig:lmcsigvradmap}  Radial velocity dispersion maps $\sigma_{R}$ of the LMC for the combined sample (main panel) and the various evolutionary phases. 
                                                        The linear velocity scale shown by the colour bar in the main panel is the same in all 
                                                        subpanels and has been chosen to highlight velocity patterns better.}
\end{figure*}

\begin{figure*}
    \includegraphics[width=0.95\textwidth]{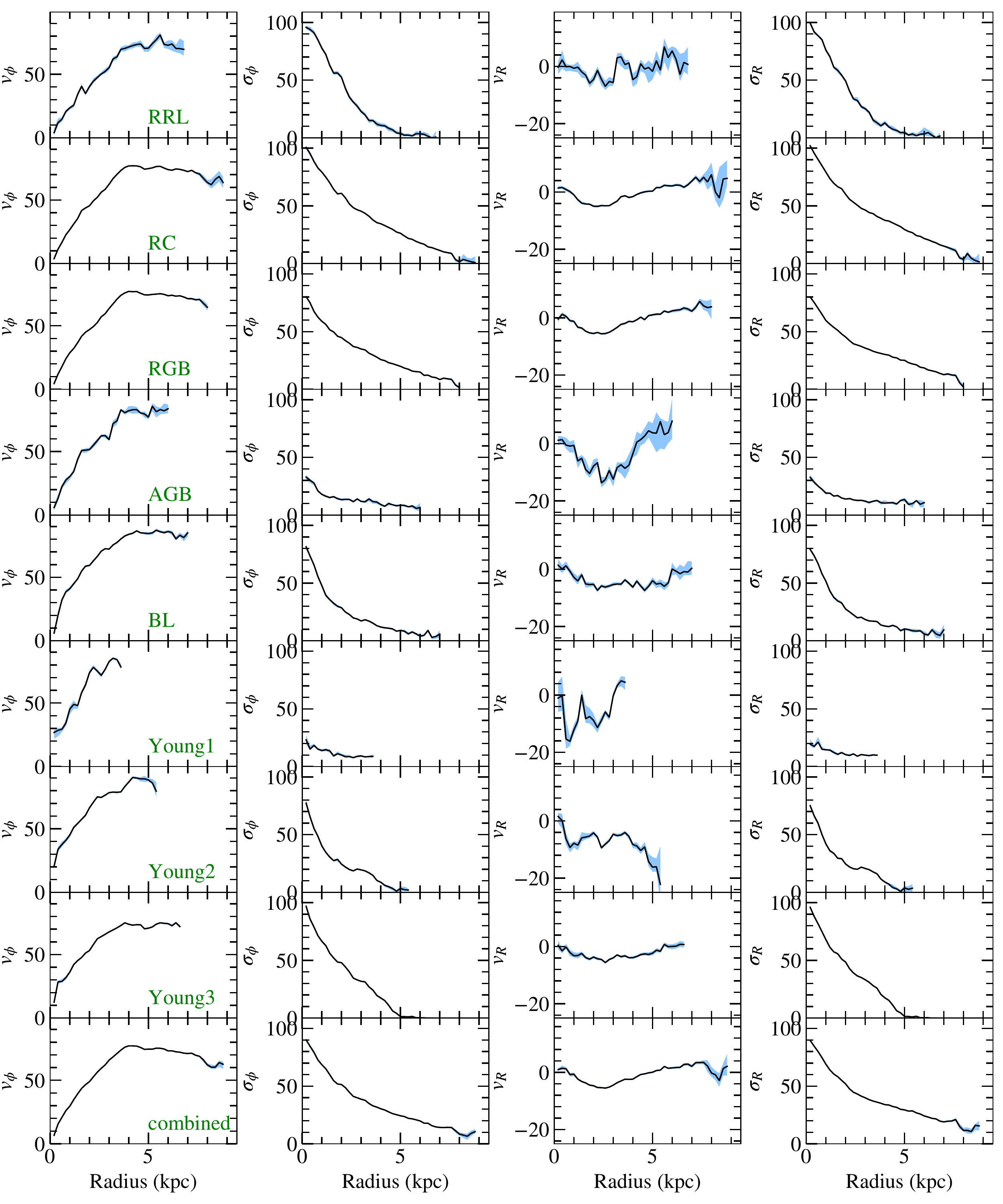}
                \caption{\label{fig:veloprof} Velocity profiles of the LMC. Rotation curve (\vt),
                         tangential velocity dispersion (\st), radial velocity (\vr), and radial 
                                                 velocity dispersion (\sr) from left to right for each
                         stellar  evolutionary phase (from top to bottom). Velocities are in \kms. 
                                                 The bottom row is the result for the combined sample. The blue shaded 
                                                 areas correspond to the uncertainties.}
\end{figure*}

We include here several figures that complement the analysis of the kinematics of the LMC presented in 
\secref{sec:kinematics}. Figures \ref{fig:lmcvphimap} to \ref{fig:lmcsigvradmap} present
the velocity maps of the LMC (azimuthal and radial velocities, mean values, and dispersion)
for the eight evolutionary phases and the combined sample of stars.
Finally, \figref{fig:veloprof} presents the velocity profiles of the LMC (as a function
of the distance to its centre) traced using the different populations defined by our 
evolutionary phases. 

\end{appendix}

\end{document}